\documentclass[a4paper,11pt]{article}
\pdfoutput=1 % if your are submitting a pdflatex (i.e. if you have
\usepackage{jcappub} % for details on the use of the package, please
\usepackage[normalem]{ulem}      
\usepackage{bm}

\usepackage[T1]{fontenc} % if needed

\newcommand{\be}{\begin{equation}}
\newcommand{\ee}{\end{equation}}
\newcommand{\beq}{\begin{equation}}
\newcommand{\eeq}{\end{equation}}
\usepackage{bm}
\newcommand{\de}{{\textrm d}}
\newcommand{\drm}{\textrm}

\newcommand{\galaxy}{\cite{xia2015, cuoco2015, fornasa2016, regis2015, xia2011, shirasaki2015, cuoco2017, ammazzalorso2018, hashimoto2020}}
\newcommand{\clusters}{\cite{branchini2017, hashimoto2019, colavincenzo2020, tan2020}}
\newcommand{\cmb}{\cite{fornengo2015, feng2017}}
\newcommand{\shear}{\cite{shirasaki2014, troster2017, shirasaki2016, shirasaki2018, ammazzalorso2020}}

\title{\boldmath Got plenty of nothing: cosmic voids as a probe of particle dark matter}

\author[a,b,c]{S. Arcari,}
\author[d,e, f]{E. Pinetti,}
\author[a,f]{N. Fornengo}

\affiliation[a]{Department of Physics, University of Torino, Via P. Giuria 1, 10125 Torino, Italy}
\affiliation[b]{Dipartimento di Fisica e Scienze della Terra, Università degli Studi di Ferrara, Via G. Saragat 1, 44122 Ferrara, Italy}
\affiliation[c]{Istituto Nazionale di Fisica Nucleare, Sezione di Ferrara, Via G. Saragat 1, 44122 Ferrara, Italy}
\affiliation[d]{Theoretical Astrophysics Department, Fermi National Accelerator Laboratory, Batavia, Illinois, 60510, USA}
\affiliation[e]{University of Chicago, Kavli Institute for Cosmological Physics, Chicago, IL 60637, USA}
\affiliation[f]{Istituto Nazionale di Fisica Nucleare, Sezione di Torino, Via P. Giuria 1, 10125 Torino, Italy}

\emailAdd{stefano.arcari@unife.it}
\emailAdd{epinetti@fnal.gov}
\emailAdd{nicolao.fornengo@unito.it}

\abstract{The search for a particle dark matter signal in terms of radiation produced by dark matter annihilation or decay has to cope with the extreme faintness of the predicted signal and the presence of masking astrophysical backgrounds. It has been shown that using the correlated information between the dark matter distribution in the Universe with the fluctuations of the cosmic radiation fields has the potential to allow setting apart a pure dark matter signal from astrophysical emissions, since spatial fluctuations in the radiation field due to astrophysical sources and dark matter emission have different features. The cross-correlation technique has been proposed and adopted for dark matter studies by looking at  dark matter halos (overdensities). In this paper we extend the technique by focusing on the information on dark matter distribution offered by cosmic voids, and by looking specifically at the gamma-ray dark matter emission: we show that, while being underdense and therefore producing a reduced emission as compared to halos, nevertheless in voids the relative size of the cross-correlation signal due to decaying dark matter vs. astrophysical sources is significantly more favourable, producing signal-to-background ratios $S/B$ (even significantly) larger than 1 for decay lifetimes up to $2 \times 10^{30}$ s. This is at variance 
with the case of halos, where $S/B$ is typically (even much) smaller than 1. We show that forthcoming galaxy surveys such as Euclid combined with future generation gamma-ray detectors with improved specifications have the ability to provide a hint of such a signal with a predicted significance up to $5.7\sigma$ for galaxies and $1.6\sigma$ for the cosmic shear. The bound on the dark matter lifetime attainable exploiting voids is predicted to improve on current bounds in a mass range for the WIMP of $25\div900$ GeV.}

\begin{document}
\maketitle
\flushbottom

\section{Introduction}
\label{sec:intro}

The evidence for the presence of dark matter in the Universe, while being overwhelming, still lacks of an understanding of the nature of what we call dark matter. A natural explanation relies on the existence of one (or more) new type(s) of elementary particle(s), which would form the dark matter. A test of the particle physics interpretation of dark matter in terms of a new elementary particle is expected to produce a variety of signals which are possible due to its particle physics nature. Among them, the production of cosmic radiation in terms of radiation (from radio to gamma rays, depending on the mass of the particle, which sets the maximal energy of the produced signal) or neutrinos is one of the most investigated channels. 

The search for a particle dark matter signal in terms of radiation produced by dark matter annihilation or decay, while offering a wide set of opportunities since the signal can be produced in every structure where dark matter is present (galaxies, clusters of galaxies, filamentary structures connecting them), nevertheless it is typically faint for most of the particle physics models, which makes it often dominated by masking astrophysical backgrounds. In order to attempt to extract the dark matter signal from the backgrounds, in \cite{camera2013, Camera:2014rja} it was proposed to look for the cross-correlation between a gravitational tracer of dark matter (like the cosmic shear or the galaxy distribution) and cosmic radiation fields (like gamma-rays, for heavy dark matter like WIMPs - see also \cite{fornengo2014}). The correlated information between where dark matter is and the ensuing fluctuations induced on the radiation fields could give a handle to separate the two signals, since spatial fluctuations in the radiation field due to astrophysical sources (which are essentially pointlike) and dark matter emission (more diffuse) have different features. 

This has been further elaborated and explored on data for the cross-correlation between dark matter gamma-ray emission with galaxies \galaxy, clusters \clusters, CMB lensing \cmb\  and cosmic shear \shear, discussed in different energy bands like X-rays \cite{zandanel2015,Caputo:2019djj} and the NIRB \cite{serra2014,Gong:2015hke, Caputo:2020msf} or extended to different gravitational tracers like the HI intensity mapping \cite{pinetti2019}.

In the previous literature, the cross-correlation technique has been proposed and adopted for dark matter studies by looking at  dark matter halos (overdensities). In this paper we investigate the possibility to extend the cross-correlation technique by using the information on dark matter distribution offered by cosmic voids, for which catalogs start to become available \cite{Pan:2012,Douglass:2022ebt,Mao:2016onb}. While the signal that originates in cosmic voids is expected to be weaker than the one produced in dark matter halos, due to the fact that voids are underdense and therefore the electromagnetic emission intensity produced by dark matter annihilation or decay is smaller in size than the one produced in dark matter halos, nevertheless we will show in the following that the relative size of the dark matter signal vs. the cross-correlation signal due to astrophysical sources like active galactic nuclei (AGN) or star-forming galaxies (SFG), which represent the relevant background for our observable of interest, is significantly more favourable in voids than in halos. Therefore, by selecting cosmic voids as large-scale structure (LSS) tracers can potentially offer a cleaner signal as compared to the signal due to overdensities: this makes the cross-correlation signal in voids an interesting counter-part to the signal in halos, with a trade off between weaker but cleaner vs. stronger but with higher background signal. 

The paper is organized as follows: in Sec. \ref{sec:cross} we summarize the formalism to derive the cross-correlation angular power spectrum of any given pair of intensity fields, and the associated variance, extended to a halo and void model of LSS \cite{voivodic2020}, which we briefly review in Sec. \ref{sec:model}, along with the prescriptions for the halo and void statistics and density distribution. Sec. \ref{sec:PS} serves as a repertory of the cross-correlation 3D power spectra needed for each case study, whilst Sec. \ref{sec:W} reports the corresponding window functions. In Sec. \ref{sec:results} we show our results and finally conclusions are drawn in Sec. \ref{sec:conclusion}. Appendix \ref{app:A} and \ref{app:B} discuss the gamma-ray luminosity function (GLF) of unresolved astrophysical sources and their mass to luminosity function, respectively. Appendix \ref{app:C} shows all contributions to the total 3D power spectrum of each source field considered in this work and its variation upon the choice of the void profile. Appendix \ref{app:D} outlines the correlation between the free parameters of our analysis.

Throughout the paper we assume a flat $\Lambda$CDM cosmology with cosmological parameters as derived by the \emph{Planck} satellite in 2018 \cite{planck2018}: $H_0=67.66\;\text{km s}^{-1}\text{ Mpc}^{-1}$, $\Omega_m=0.3111$, $\Omega_\Lambda=0.6889$.

\section{The cross-correlation signal for particle dark matter}\label{sec:cross}

The dark matter signal we consider is the statistical cross-correlation between the unresolved gamma-ray emission from particle dark matter annihilation or decay and the distribution of mass in the Universe traced through cosmic shear or through the distribution of galaxies in the Universe. In this Section we briefly review the basic elements of the formalism to cross-correlate two density fields by means of the angular power spectrum (APS), while in the next Sections we will specify in details the relevant ingredients that determine the APS for the density fields of interest. We closely follow the formalism introduced in Refs. \cite{camera2013, fornengo2014}, although we extend it to include cosmic voids.

The source intensity of an observable $i$, taken along a given direction $\vec{n}$, can be written as:
\begin{equation}
    I_i(\vec{n}) = \int \de \chi\;g_i(\chi,\;\vec{n})\tilde{W}(\chi) \; ,
\end{equation}
where $\chi$ is the comoving radial distance, $g_i(\chi,\;\vec{n})$ represents the density field of the source $i$ and $\tilde{W}(\chi)$ is a window function characterising the average intensity field as a function of distance (or equivalently of redshift). It is convenient to define a normalized window function $W(\chi) = \langle g_i(\chi,\;\vec{n})\rangle\tilde{W}(\chi)$ such that $\langle I_i\rangle=\int \de\chi\;W(\chi)$. By expanding the intensity fluctuations of two given fields $i$ and $j$ in spherical harmonics, we can compute the cross-correlation angular power spectrum (CAPS):
\begin{equation}\label{eq:CAPS}
    C_\ell^{ij} = \int \frac{\de\chi}{\chi^2}\;W_i(\chi)W_j(\chi)P_{ij}\left(k=\frac{\ell}{\chi},\;\chi\right) \; ,
\end{equation}
where the window function $W_i(\chi)$ describes how the observable $i$ is distributed in redshift and its shape strongly depends on the physics behind the chosen signal. The 3D power spectrum is defined through $\langle f_i(\chi,\;\bm{k})f_j(\chi,\;\bm{k})\rangle = (2\pi)^3\delta_D(\bm{k}-\bm{k'})P_{ij}(k,\;\chi)$ where $f_i = g_i - \langle g_i\rangle$ is the fluctuation of the density field. In Eq. \eqref{eq:CAPS} we adopt the Limber approximation \cite{limber1953, limber1992, limber1998}, which is typically valid for the relevant scales explored in cross-correlation studies involving gamma rays \cite{fornengo2014, Camera:2014rja, pinetti2019}.

In the halo model \cite{cooray2002}, the 3D power spectrum can be split into two terms, taking into account the correlations between source fields either from the same halo (1Halo) or from two different halos (2Halo). In this work we extend the CAPS formalism in a Halo-Void Model of  LSS \cite{voivodic2020}, allowing matter to lie within halos and voids. The model can be further extended to include a dust component \cite{voivodic2020}, but we leave this further level of complexity to future works. Introducing more structures in the model leads to new correlation terms, therefore the power spectrum will be decomposed not only into 1Halo (1H) and 2Halo (2H) terms, but it will also include 1Void (1V) and 2Void (2V) components as well as the Halo-Void (HV) mixed term. A detailed derivation of the relevant terms will be given in the next sections.

Under the hypothesis of gaussianity, the variance of the predicted CAPS is (see e.g. \cite{pinetti2019}):
\begin{equation}\label{eq:CAPSvariance}
    (\Delta C^{ij}_\ell)^2 = \frac{1}{(2 \ell+1)f_\text{sky}}\left[(C^{ij}_\ell)^2 + \left(C^{ii}_\ell+\frac{N^i}{(B_\ell^i)^2}\right)\bigg(C^{jj}_\ell+\frac{N^j}{(B_\ell^j)^2}\bigg)\right] \; ,
\end{equation}
where $f_\text{sky}$ is the observed fraction of the sky, $C^{ii}_\ell$ and $C^{jj}_\ell$ represent the auto-correlation angular power spectra associated to the observable $i$ and $j$, respectively, whereas $N^i$ and $N^j$ are their corresponding noises. Their beam functions $B_\ell^i$ and $B_\ell^j$ in harmonic space refer to the angular resolution of the chosen detector. Eq. (\ref{eq:CAPSvariance}) represents the uncertainty on the predicted CAPS, and will be used to determine whether the signal is detectable. 

All quantities in Eqs. \eqref{eq:CAPS} and \eqref{eq:CAPSvariance} will be explicitly specified below. In the following, $i$ labels the gamma-ray intensity, either from astrophysical sources or annihilating/decaying particle dark matter, while $j$ represents the gravitational tracer under consideration in our analysis, that is cosmic shear and galaxy distribution.

\section{The model for halos and voids}
\label{sec:model}
An overwhelming amount of theoretical and observational evidences favors the idea that structure in the Universe has risen out of a nearly homogeneous primordial cosmos through gravitational instability. The process of structure formation is hierarchical (larger clustering are formed through the continuous merging of smaller structures), as backed by most existing theories, and therefore strongly depends on initial conditions, whose knowledge is one of the primary questions in cosmology. 

Tracing back the evolution of structures retains the potential to broaden our understanding on the primordial Universe and much effort has been put into developing both analytical and numerical schemes to explain such phenomena. On one hand, linear and higher order perturbation theory descriptions of gravitational clustering \cite{bernardeau2002, baumann2012, carrasco2012, carrasco2014} from Gaussian initial conditions explain the evolution and mildly non-linear clustering of dark matter, but break in highly non-linear regimes \cite{konstandin2019} (i.e. at scales smaller than few megaparsecs) and do not provide a rigorous framework to describe the clustering of galaxies. Also, very large scales are troublesome to work with, from the observational side, because of the small amount of data available. On the other hand, smaller non-linear scales can only be described by numerical dark matter simulations of the LSS clustering \cite{teyssier2002, springel2005}. The latter show that an initially smooth matter distribution evolves into a complex web of knots, sheets and filaments. These numerical simulations provide detailed information on the distribution of mass within these structures \cite{moore1999, navarro1996} when performed at high resolution but relatively small volume, but they are also useful to constrain the abundance and spatial distribution of structures in the Universe when performed at lower resolution with large volume \cite{evrard2002, angulo2014}. The drawback is that simulations are usually computationally expensive and unable to provide an analytical description of the initial conditions, on which they are highly dependent.

Data from forthcoming large-area imaging and redshift surveys of galaxies \cite{LSST} and weak lensing \cite{DES, Euclid} will provide constraints on the dark matter distribution on large scales as well as on the galaxy formation history. At the same time, the Sunyaev–Zel’dovich effect \cite{sunyaev1980} probes the distribution of the pressure on large scales, and can be observed through wide-field surveys \cite{orlowski2021}. Moreover, dark and/or baryonic matter leave their imprints on CMB in the form of secondary temperature fluctuations on small scales \cite{sunyaev1980, sachs1967, lemarchand2019}. 

The halo model (HM) \cite{cooray2002} has been, up to today, the uttermost successful analytical description of non-linear scales and provides a self-consistent explanation of the observations discussed above. This model relies on the central assumption that all matter in the Universe lies within dark matter halos and offers a simple framework to explain the transition between non-linear and linear scales, which are dominated by the 1Halo and 2Halo terms, respectively. However, the predicted power spectrum is in accordance with N-body simulations only within 20\% around the transition and, at the same time, it either requires a normalisation of the 2Halo term or for the halo abundance to be integrated down to very low and untested masses (M $\ll 10^4\;h^{-1}\;\text{M}_\odot$), in order to account for all matter in the Universe. As discussed in \cite{voivodic2020}, there have been several attempts to modify the HM \cite{mead2015, schmidt2016, chen2019, valageas2011}. These alternative models typically either introduce new free parameters that cannot be fitted using only halo properties, or do not significantly improve the HM predictions. 

Recently, a successful self-consistent modification of the HM has been proposed by Voivodic et al. \cite{voivodic2020}, namely the halo-void model (HVM). This model is based on relaxing the central assumption of the HM and allowing matter to lie not only within halos but also within cosmic voids.

Including voids as building blocks of the Universe leads to new terms in the prediction of the cross-correlation signal: the 1Void, 2Void, and Halo-Void terms, in addition to the 1Halo and 2Halo contributions (already present in the HM). In the past decades, the scientific community has mostly focused on modeling the dark matter halos, while cosmic voids have been largely unappreciated. Nevertheless, voids constitute the dominant volume fraction of the Universe and can be used as powerful, independent probes for our theories of structure formation. Underdense regions are indeed ideal environments to constraint dark energy \cite{biswas2010, pollina2016, sahlen2016, pisani2015} and modified gravity \cite{barreira2015, cai2015, perico2019, voivodic2017}.
%, as the impact of ordinary gravitating matter is mitigated in their interior.}
In addition, thanks to their extreme sensitivity to background cosmological changes, voids are more closely related to initial conditions \cite{damico2011, chan2019, song2009} and can be valuable in order to constrain cosmology \cite{cai2016, chantavat2016, hamaus2014, hamaus2015, hamaus2016, chuang2017, lavaux2012}. Also, they are the perfect case-study for the excursion set theory\footnote{The excursion set approach provides a useful framework to describe the formation histories of gravitationally bound structures such as virialized halos or cosmic voids.} \cite{bond1991} given their rather spherical symmetry\footnote{Unlike the evolution of density peaks, primordial asphericity of negative density perturbations is quickly lost as they expand \cite{sheth2004}.}  and the existence of a refined repertory of fitting prescriptions for their density profile, mass function and linear bias.

The HVM requires the following ingredients: the halo and void density profile, the halo and void mass function and the halo and void linear bias. The model computes the halo and void statistics through the excursion set formalism \cite{bond1991} with two barriers \cite{voivodic2017, simone2011, sheth2004, jennings2013}. This approach guarantees a fully self-consistent model that takes into account the void-in-cloud, void-in-void and cloud-in-void effects, in addition to the cloud-in-cloud effect that appears in the HM. Moreover, there is no need for a normalization on large scales or for the abundances to be integrated down to very low halo masses, as the matter within smaller halos is taken into account in larger voids. Voivodic et al. \cite{voivodic2020} also show how considering both halos and voids improves the transition between the 2Halo and the 1Halo term.

Following the prescriptions of the HVM, the total matter density field is given by the sum of the halo and void contributions:
\begin{equation}\label{eq:rhoHVDM}
    \rho(\textbf{x}) = \sum_i^{\text{halos}}\rho_h(\textbf{x}-\textbf{x}_i\,|\,M_i) + \sum_j^{\text{voids}}\rho_v(\textbf{x}-\textbf{x}_j\,|\,M_j) \; ,
\end{equation}
where $\rho_h(\textbf{x}-\textbf{x}_i\,|\,M_i)$ is the density profile of a halo with mass $M_i$ centered at $x_i$ and $\rho_v(\textbf{x}-\textbf{x}_j\,|\,M_j)$ is the density profile of a void with mass $M_j$ centered at $x_j$. Note how the HM is recovered when the last term in Eq. \eqref{eq:rhoHVDM} is neglected, i.e. if we set the matter density in voids to zero across the whole Universe. We can rewrite Eq. \eqref{eq:rhoHVDM} as:
\begin{equation}\label{eq:3.2}
\begin{split}
    \rho(\textbf{x}) = \int \de M\int \de^3x'\Bigg[ &\sum_i^{\text{halos}}\delta_D(M-M_i)\delta_D(\textbf{x}'-\textbf{x}_i)\rho_h(\textbf{x}-\textbf{x}'\,|\,M)\\
    +&\sum_j^{\text{voids}}\delta_D(M-M_j)\delta_D(\textbf{x}'-\textbf{x}_j)\rho_v(\textbf{x}-\textbf{x}'\,|\,M) \Bigg] \; .
\end{split}
\end{equation}
The 2-point correlation function reads
\begin{equation}\label{eq:xi}
    \xi(\textbf{r}) = \frac{1}{\overline{\rho}_m^2}\langle\rho(\textbf{x})\rho(\textbf{x}+\textbf{r})\rangle - 1\; ,
\end{equation}
where $\overline{\rho}_m$ is the average matter density in the Universe. Plugging Eq. \eqref{eq:3.2} into Eq. \eqref{eq:xi}, we get that Eq. \eqref{eq:xi} can be written as the sum of three terms: a pure halo term (containing the correlation between particles within one or two halos), a pure void term (containing the correlation between particles within one or two voids) and a mixed term (containing the correlation between two particles within a halo and a void, respectively). In particular, the pure halo correlation reads:
\be
\begin{split}
    \xi_h(\textbf{r}) &= 
    \dfrac{1}{\overline{\rho}_m^2}\int \de M_1\de M_2\de^3x_1\de^3x_2\;
    \rho_h(\textbf{x}-\textbf{x}_1\,|\,M_1)\rho_h(\textbf{x}-\textbf{x}_2+\textbf{r}\,|\,M_2)\\
    &\times \left\langle \sum_{i,\,j}\delta_D(M_1-M_i)\delta_D(M_2-M_j)\delta_D(\textbf{x}_1-\textbf{x}_i)\delta_D(\textbf{x}_2-\textbf{x}_j) \right\rangle \; ,
\end{split}
\ee
where we can distinguish two terms: the 1Halo term for particles residing within the same halo and the 2Halo term for particles residing in different halos:
\begin{align}
    \label{eq:I1H}&I^{1H} = \delta_D(M_1-M_2)\delta_D(\textbf{x}_1-\textbf{x}_2)\frac{\de n_h}{\de M_1}\\
    \label{eq:I2H}&I^{2H} = \frac{\de n_h}{\de M_1}\frac{\de n_h}{\de M_2}[1+\xi_{hh}(\textbf{x}_1-\textbf{x}_2\,|\,M_1,\,M_2)] \; ,
\end{align}
where $\drm{d}n_h/\drm{d}M$ denotes the halo mass function, which measures the differential number density of halos in the mass range $[M,\;M + \de M]$:
\begin{equation}
    \frac{\de n_h}{\de M_1}=\left\langle \sum_{i}\delta_D(M_1-M_i)\delta_D(\textbf{x}_1-\textbf{x}_i) \right\rangle
\end{equation}
and $\xi_{hh}(\textbf{x}_1-\textbf{x}_2\,|\,M_1,\,M_2)$ is the halo-halo 2-point correlation function of halos with mass $M_1$ and $M_2$. A similar discussion applies to the other terms of the 2-point correlation function. However, for the cross-correlation the only non vanishing term is naturally the one with particles residing in two distinct structures. The terms appearing in Eq. \eqref{eq:xi} can be summarised as:
\begin{align}
    \label{eq:xi1H}\xi^{1H}(\textbf{r}) &= \frac{1}{\overline{\rho}^2_m}\int \de M\,\frac{\de n_h}{\de M}\int \de^3y\,\rho_h(\textbf{y}|M)\rho_h(\textbf{y}+\textbf{r}|M)\\[1mm]
    \xi^{2H}(\textbf{r}) &= \frac{1}{\overline{\rho}^2_m}\int \de M_1\,\frac{\de n_h}{\de M_1}b_h(M_1)\int \de M_2\,\frac{\de n_h}{\de M_2}b_h(M_2)\\
    \notag&\times\int \de^3y_1\,\rho_h(\textbf{y}_1|M_1)\int \de^3y_2\,\rho_h(\textbf{y}_2|M_2)\xi^L(\textbf{y}_1-\textbf{y}_2)\\[1mm]
    \xi^{1V}(\textbf{r}) &= \frac{1}{\overline{\rho}^2_m}\int \de M\,\frac{\de n_v}{\de M}\int \de^3y\,\rho_v(\textbf{y}|M)\rho_v(\textbf{y}+\textbf{r}|M)\\[1mm]
    \xi^{2V}(\textbf{r}) &= \frac{1}{\overline{\rho}^2_m}\int \de M_1\,\frac{\de n_v}{\de M_1}b_v(M_1)\int \de M_2\,\frac{\de n_v}{\de M_2}b_v(M_2)\\
    \notag&\times\int \de^3y_1\,\rho_v(\textbf{y}_1|M_1)\int \de^3y_2\,\rho_v(\textbf{y}_2|M_2)\xi^L(\textbf{y}_1-\textbf{y}_2)\\[1mm]
    \label{eq:xiHV}\xi^{HV}(\textbf{r}) &= \frac{1}{\overline{\rho}^2_m}\int \de M_1\,\frac{\de n_h}{\de M_1}b_h(M_1)\int \de M_2\,\frac{\de n_v}{\de M_2}b_v(M_2)\\
    \notag&\times\int \de^3y_1\,\rho_h(\textbf{y}_1|M_1)\int \de^3y_2\,\rho_v(\textbf{y}_2|M_2)\xi^L(\textbf{y}_1-\textbf{y}_2) \; ,
\end{align}
where we have used the linear approximation (tree-level) for the structure-structure 2-point correlation ($\xi_{xy}$ for $x,y=h,\,v$). 
The density contrast $\delta_x(\textbf{x}\,|\,M)$ is then obtained from the linear matter density contrast $\delta_m^L(\textbf{x})$ as (see e.g. \cite{cooray2002, voivodic2020}):
\begin{equation}
    \delta_x(\textbf{x}\,|\,M) = b^L_x(M)\delta_m^L(\textbf{x}) \; ,
\end{equation}
where $b^L_x(M)$ denotes the linear bias of $x$.
For any combination of structures $xy$ we have:
\begin{equation}
    \xi_{xy}(\textbf{r}) = b_x(M_x)\,b_y(M_y)\,\xi^L(\textbf{r}) \; ,
\end{equation}
where we have dropped the $L$ apex on the linear bias to ease the notation. Two conditions have to hold: the total matter density of the Universe has to be equal to the sum of the total matter density in halos and voids, and matter does not have to be biased with respect to itself. These two conditions lead to the two following constraints:
\begin{align}
    \overline{\rho}_h + \overline{\rho}_v = \label{eq:rhoTOT}\overline{\rho}_m \\
    \label{eq:bTOT}1-\overline{b}_h-\overline{b}_v = 0 \; ,
\end{align}
where
\begin{align}
    \label{eq:rhohtotal}&\overline{\rho}_h = \int \de M\,M\frac{\de n_h}{\de M}\\
    \label{eq:rhovtotal}&\overline{\rho}_v = \int \de M\,M\frac{\de n_v}{\de M} \; ,
\end{align}
and
\begin{align}
    \label{eq:bhtotal}&\overline{b}_h = \frac{1}{\overline{\rho}_m}\int \de M\,M\frac{\de n_h}{\de M}b_h(M)\\
    \label{eq:bvtotal}&\overline{b}_v = \frac{1}{\overline{\rho}_m}\int \de M\,M\frac{\de n_v}{\de M}b_v(M) \; .
\end{align}

Assuming spherical symmetry ($\rho_x(\textbf{r}\,|\,M)=\rho_x(r\,|\,M)$ for $x=h,\,v$) and Fourier transforming Eqs. \eqref{eq:xi1H}-\eqref{eq:xiHV} we have:
\begin{align}
    \label{eq:P1H}P^{1H}(k) &= \frac{1}{\overline{\rho}^2_m}\int \de M\;\frac{\de n_h}{\de M}|\rho_h(k\,|\,M)|^2\\[1mm]
    \label{eq:P2H}P^{2H}(k) &= \frac{1}{\overline{\rho}^2_m}\left[\int \de M\;\frac{\de n_h}{\de M}\rho_h(k\,|\,M)b_h(M)\right]^2P^L(k)\\[1mm]
    \label{eq:P1V}P^{1V}(k) &= \frac{1}{\overline{\rho}^2_m}\int \de M\;\frac{\de n_v}{\de M}|\rho_v(k\,|\,M)|^2\\[1mm]
    \label{eq:P2V}P^{2V}(k) &= \frac{1}{\overline{\rho}^2_m}\left[\int \de M\;\frac{\de n_v}{\de M}\rho_v(k\,|\,M)b_v(M)\right]^2P^L(k)\\[1mm]
    \label{eq:PHV}P^{HV}(k) &= \frac{1}{\overline{\rho}^2_m}\int \de M_1\;\frac{\de n_h}{\de M_1}\rho_h(k\,|\,M_1)b_h(M_1)
    \int \de M_2\;\frac{\de n_v}{\de M_2}\,\rho_v(k\,|\,M_2)b_v(M_2)\,P^L(k) \; ,
\end{align}
where $P^L(k)$ is the linear matter power spectrum, given by the Fourier transform of the linear 2-point correlation function $\xi^L(r)$.
Note how the HM is easily recovered when setting $\rho_v(k\,|\,M)=0$. The total matter power spectrum is now given by:
\begin{equation}\label{eq:PHVDMterms}
    P(k) = P^{1H}(k)+P^{2H}(k)+P^{1V}(k)+P^{2V}(k)+2 \, P^{HV}(k)
\end{equation}
and, using Eqs. \eqref{eq:bTOT} and \eqref{eq:rhoTOT}, it can be shown that the total matter power spectrum reduces to the linear matter power spectrum on very large scales ($k\ll1$ Mpc$^{-1}$). However, for the HM, these constraints return $\overline{b}_h = 1$ and $\overline{\rho}_h = 1$, which, for a standard halo mass function and bias, have a very slow convergence, requiring integration down to tiny and untested masses. The HVM solves this shortcoming by taking into account the matter within smaller halos in larger voids (see Sec. \ref{sec:convergence} and \cite{voivodic2020} for a brief discussion on this point).

In the following sections we discuss the  ingredients required to compute the power spectrum: halo/void density profiles, mass functions and linear biases. For the last two, we adopt the recipes provided by Voivodic et al. \cite{voivodic2020}, derived from the excursion set theory with two static barriers \cite{simone2011, voivodic2017, voivodic2020}.  Hence, the obtained mass functions and linear biases naturally incorporate the exclusion of voids inside halos and vice versa.

\subsection{Density profiles}\label{sec:profile}
In this work, we adopt for halos the standard Navarro-Frenk-White (NFW) profile \cite{navarro1996}, and for voids the Hamaus-Sutter-Wandelt (HSW) profile \cite{hamaus2014b}. Both assume a spherical density distribution of matter in the structure.

\subsection*{Halos}
The NFW density profile for a halo of virial mass $M$  can be parameterized as follows \cite{navarro1996}:
\begin{equation}\label{eq:NFW}
    \rho_{\text{NFW}}(r\,|\,M) = \frac{\rho_s}{c(M)\dfrac{r}{r_{\rm vir}} \left(1+c(M)\dfrac{r}{r_{\rm vir}}\right)^2} \; ,
\end{equation}
where $\rho_s$ is the characteristic density for which the volume-integrated profile returns the virial mass, $r_{\rm vir}$ is the virial radius\footnote{The virial radius is defined through $M=\frac{4\pi}{3}\overline{\rho}_m\Delta_{\rm vir}r_{\rm vir}^3$, where $\overline{\rho}_m$ is the average matter density of the Universe and $\Delta_{\rm vir}\sim330$ is the halo density contrast at the time of virialization for the fiducial cosmology.} and $c(M)$ is the concentration parameter \cite{duffy2008}. Its Fourier transform, truncated at $r_{\rm vir}$ is given by:
\begin{equation}
    \rho_h(k\,|\,M) =4\pi\int^{r_{\rm vir}}_0 \de r\;r^2 \, \frac{\sin{kr}}{kr} \, \rho_{\text{NFW}}(r\,|\,M) \; .
\end{equation}

\subsection*{Voids}
\begin{figure}[tbp]
\centering 
\includegraphics[width=.99\textwidth]{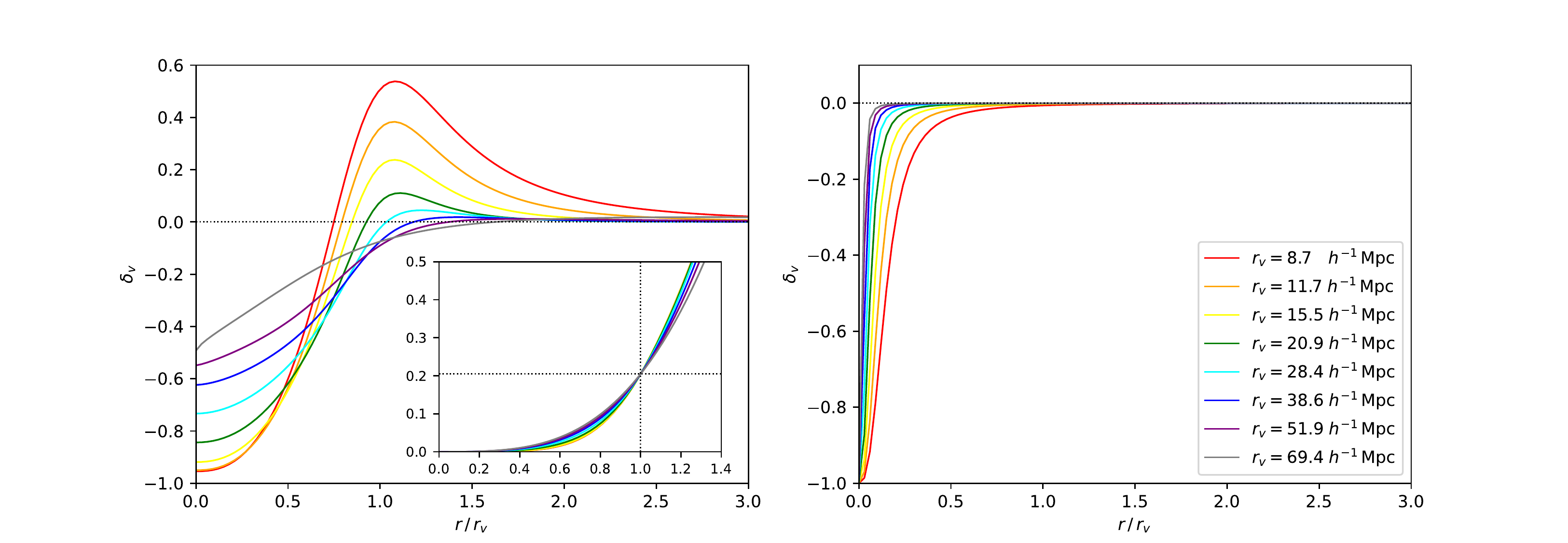}
\hfill
\caption{\label{fig:HSW} (\emph{Left}): Hamaus-Sutter-Wandelt void density contrast \cite{hamaus2014}, at $z=0$, for different void radii. Smaller voids are more underdense in their central region but show higher compensation walls in their outskirt, i.e. they tend to be overcompensated. Larger voids are less underdense in their central region but show lower compensation walls in their outskirt, i.e. they tend to be undercompensated. In the small box we show the integrated overdensity, as a function of $r/r_v$, for a normalized HSW profile, so that the void overdensity $\Delta_v \simeq 0.2047$ is reached, for any void, at its radius. This allows us to consider spherical voids with fixed void overdensity. (\emph{Right}): Void density contrast for the profile proposed in Ref. \cite{voivodic2020}, based on empty centers and no compensation walls.}
\end{figure}
Given their low density, vastness and unsphericity at some scales, voids are troublesome to study both with observations and simulations. Most void finders show them as deeply underdense in their interiors\footnote{Voids of smaller size show emptier central regions, while larger voids tend to be slightly denser in their centers \cite{hamaus2014}.}, and the profiles exhibit overdense compensation walls with a maximum located slightly outside their effective radii, shifting outwards for larger voids. The height of the compensation wall decreases with void size, causing the inner profile slope to become shallower and the wall to widen. This trend divides all voids into being either overcompensated or undercompensated, depending on whether the total mass within their compensation wall exceeds or falls behind their missing mass in the center, respectively. Ultimately, at sufficiently large distances to the void center, all profiles approach the mean background density. The HSW profile \cite{hamaus2014} is a simple empirical formula that can accurately capture the properties described above:
\begin{equation}\label{eq:HSW}
    \frac{\rho_v(r\,|\,r_v)}{\overline{\rho}_m}-1 = \delta_c\frac{1-\left(\dfrac{r}{r_s}\right)^\alpha}{1+\left(\dfrac{r}{r_v}\right)^\beta} \; ,
\end{equation}
where $r_s$ is the characteristic radius for which $\rho_v=\overline{\rho}_m$, $r_v$ is the effective radius of the considered void and $(\alpha,\;\beta,\;\delta_c)$ are free parameters.
In particular, $\delta_c$ represents the central density contrast and $\alpha$ and $\beta$ are the inner and outer slopes of the compensation wall. These parameters are usually determined through best-fits of N-body dark matter simulations but Hamaus et al. \cite{hamaus2014} also provide empirical functions, which we use in our analysis, relating them to $r_s$ and $r_v$, therefore decreasing the degrees of freedom.

Nevertheless, the choice of the profile strongly depends on the void finder, which N-body simulations should be calibrated on. Other works have used void finders that neither rely on any central particle, but on Voroni vertices, nor show any compensation wall (see e.g. \cite{voivodic2020}). These properties can be summarized in the one-parameter empirical formula proposed in \cite{voivodic2020}:
\begin{equation}\label{eq:tanh}
    \frac{\rho_v(r\,|\,r_v)}{\overline{\rho}_m}-1 = \frac{1}{2}\left[ 1 + \tanh{\left(\frac{y-y_0}{s(r_v)}\right)}\right] - 1\; ,
\end{equation}
where $y=\ln{(r/r_v)}$ and $y_0=\ln{(r_0/r_v)}$. The radius $r_0$ can be parameterized in terms of $s$, which remains the only free quantity. Ref. \cite{voivodic2020} shows, through N-body simulations, that its dependence on $r_v$ is very weak and can be safely fixed to $s=0.75$, for all void radii.

The left and right panel of Fig. \ref{fig:HSW} show the void density contrast described by Eq. \eqref{eq:HSW} and Eq. \eqref{eq:tanh}, respectively, for multiple bins in void radii, and follow efficiently all of the related properties outlined above. 

The Fourier transform of the void profile is given by:
\begin{equation}\label{eq:voidFourier}
    \rho_v(k\,|\,M(r_v)) =4\pi\int^{r_v}_0 \de r\;r^2 \, \frac{\sin{kr}}{kr} \, \rho_v(r\,|\,M(r_v)) \; ,
\end{equation}
where, in principle, $M(r_v) = 4\pi\int_0^{r_v}\de r\;r^2\,\rho_v(r\,|\,r_v)$. However, in the following discussion we will consider spherical voids with an overdensity $\Delta_v \simeq 0.2047$ (see e.g. \cite{sheth2004}), whose mass can be computed through $M(r_v)=\frac{4\pi}{3}\,\overline{\rho}_m\Delta_vr^3_v$. Hence, the profile needs to be properly normalized in order for voids of any radius to reach the chosen overdensity $\Delta_v$:
\be
    \rho_v(r\,|\,r_v) \longrightarrow \frac{\Delta_v}{\Delta(r_v)}\rho_v(r\,|\,r_v) \; ,
\ee
where
\be\label{eq:integrated_overdensity}
    \Delta(r_v) = \frac{3}{r_v^3}\int_0^{r_v} \de r\;r^2\,\frac{\rho_v(r\,|\,r_v)}{\overline{\rho}_m} \; .
\ee
Note how, by doing so, Eq. \eqref{eq:voidFourier} can be rewritten in terms of the normalized profile, but the integration should be performed up to a void radius extracted from a different $M(r_v)$ relation, in order to obtain the same value for the Fourier transform of the void profile. Calculating the latter through the normalized profile is only an expedient to ease the computation, nonetheless, the physical profile remains that shown in either panels of Fig. \ref{fig:HSW}, which follow the void properties previously discussed. The small box in the left panel of Fig. \ref{fig:HSW} shows the evolution of Eq. \eqref{eq:integrated_overdensity} as a function of the integration upper limit, for the HSW normalized profile. Independently of the void size, the required void overdensity $\Delta_v$ is reached when integrating up to the void radius. 

Studies of voids' properties are yet at their early stages and the choice of the void profile usually depends on the used void finder. However, while for definiteness in the analyses shown below we adopt the HSW profile, nevertheless we have verified that the actual choice of the profile has only a sub-percent effect on the power spectra (see Sec. \ref{sec:PS}), and we leave further discussion on the use of different profiles, in the context of our work, to future studies.

\subsection{Mass functions}
The halo/void mass functions used in the HVM are computed through the excursion set theory with two static linear barriers (2SB) \cite{voivodic2017, voivodic2020} to avoid double counting of matter. We show here a comparison of these mass functions with those computed following Sheth-Tormen \cite{sheth1999}, for halos, and standard Press-Schechter theory \cite{press1974}, for voids.

A central ingredient is the variance of the linear density field smoothed at some scale $R$ (with a top-hat window function $\widehat{W}$):
\begin{equation}\label{eq:sigma}
    \sigma^2(R) = \int \frac{\de k}{2\pi^2}\;k^2P^L(k)\widehat{W}^2(kR) \; ,
\end{equation}
where for any structure we have:
\begin{equation}
    M=\frac{4\pi}{3}\,\overline{\rho}_mR^3 \; .
\end{equation}

\subsection*{Halos}
\begin{figure}[tbp]
\centering 
\includegraphics[width=.99\textwidth]{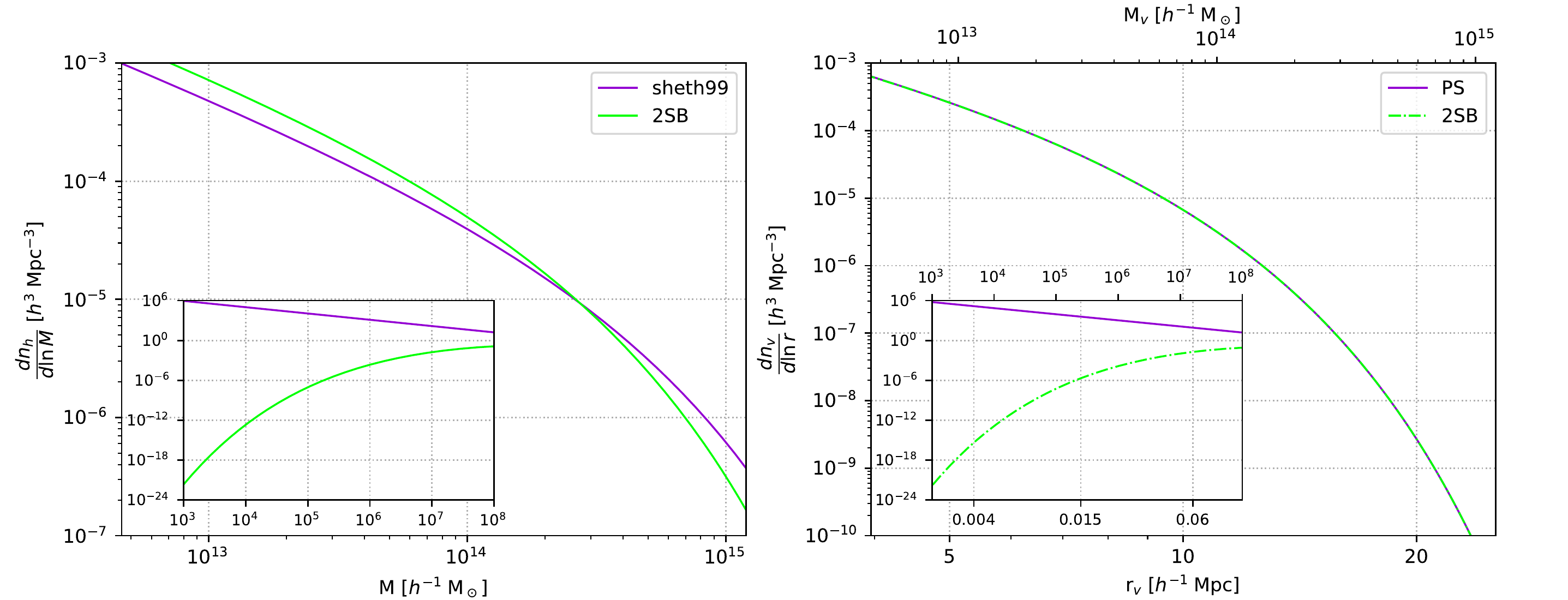}
\caption{\label{fig:massfunctions} (\emph{Left}): Halo mass function as a function of halo mass. (\emph{Right}): Void mass function as a function of void radius/mass. Both results are computed at $z=0$. We compare two different models: Sheth-Tormen for halos or Press-Schecther for voids (purple) and 2SB (green). The insets show how, at very small masses, the excursion set prediction decreases rapidly, differently from the standard Sheth-Tormen or Press-Schecther abundances.}
\end{figure}
The halo mass function can be written as:
\begin{equation}\label{eq:nh}
    \frac{\de n_h}{\de\ln{M}}=f_h(\sigma)\frac{\overline{\rho}_m}{M}\frac{\de\ln{\sigma^{-1}}}{\de\ln{M}} \; ,
\end{equation}
where $f_h(\sigma)$, the multiplicity function, determines the model.
In this work, we consider the multiplicity functions:
\begin{align}
    \label{eq:fhsheth99}f_h^{\text{sheth99}}(\sigma) &= 2A\nu\left(1+\frac{1}{(a\nu)^p}\right)\sqrt{\frac{a}{2\pi\nu}}\;\text{exp}\left(-\frac{a\nu}{2}\right)\\[1mm]
    \label{eq:fh2SB} f_h^{\text{2SB}}(\sigma) &= 2\,\sum_n\frac{n\pi}{\delta_T^2}\,\sigma^2\,\sin{\left(\frac{n\pi\delta_c}{\delta_T}\right)}\,\text{exp}\left[-\frac{n^2\pi^2}{2\delta_T^2}\,\sigma^2\right] \; ,
\end{align}
where $\delta_T=\delta_c+|\delta_v|$ with $\delta_c\simeq1.686$ and $\delta_v\simeq-2.717$ the linear critical (over and under) densities for halo and void formation, respectively. Here $f_h^{\text{sheth99}}$ is the Sheth-Tormen multiplicity function \citep{sheth1999}, with $\nu=(\delta_c^2/\sigma^2)$ and $(A, a , p)$ free parameters. $f_h^{2\text{SB}}$ is the excursion set prediction for a model with two static barriers.

Note how $f_h^{\text{sheth99}}$ is normalized to unity, whereas $f_h^{2\text{SB}}$ is not, since it already accounts for the existence of voids. As a direct consequence, the excursion set predictions with two barriers tend to decrease rapidly at small masses, as shown by the left panel of Fig. \ref{fig:massfunctions}. This directly reflects on no need to integrate Eq. \eqref{eq:rhohtotal} down to very low and untested masses.

\subsection*{Voids}
The void mass function is usually expressed in terms of its radius, rather than its mass:
\begin{equation}
    \frac{\de n_v}{\de\ln{r}}=\Delta_v\left[\frac{f_v(\sigma)}{V(r_L)}\frac{\de\ln{\sigma^{-1}}}{\de\ln{r_L}}\right]_{r_L=\,r/1.7} \; ,
\end{equation}
where $\Delta_v=({r_L}/{r_v})^{1/3}\simeq0.2$ is the void overdensity and $f_v(\sigma)$ determines the model. The last equation follows from Eq.\eqref{eq:nh} and derives from the conservation requirement of the volume density, as discussed by Jennings et al. \cite{jennings2013}.

In this work, we consider the following multiplicity functions:
\begin{align}
    \label{eq:fvpressschechter}
    f_v^{\text{PS}}(\sigma)&=\sqrt{\frac{2}{\pi}}\frac{|\delta_v|}{\sigma}\;\text{exp}^{-\frac{\delta^2_v}{2\sigma^2}}\\[1mm]
    \label{eq:fv2SB} f_v^{\text{2SB}}(\sigma) &= 2\,\sum_n\frac{n\pi}{\delta_T^2}\,\sigma^2\,\sin{\left(\frac{n\pi|\delta_v|}{\delta_T}\right)}\,\text{exp}\left[-\frac{n^2\pi^2}{2\delta_T^2}\,\sigma^2\right] \; ,
\end{align}
where $\delta_c$, $\delta_v$ and $\delta_T$ are the same parameters discussed for the halo mass function. Here $f_v^{\text{PS}}$ is the Press-Schechter multiplicity function \cite{press1974} and $f_v^{2\text{SB}}$ is the excursion set prediction for a model with two static barriers.

In the right panel of Fig. \ref{fig:massfunctions} we compare the Press-Schechter and 2SB void mass functions. Once again, we note how the excursion set prediction decreases rapidly at small masses (differently from Press-Schechter theory), avoiding the need to integrate Eq. \eqref{eq:rhovtotal} down to tiny untested masses.

\subsection{Linear bias}
We present here the linear biases used in this work. In particular, we show the excursion set predictions \cite{simone2011, sheth2004, jennings2013, voivodic2017} proposed by Voivodic et al. \cite{voivodic2020}. Using these biases makes the model fully self-consistent with the 2SB mass functions. Moreover, the combination of 2SB bias and mass function avoids double counting of matter, excluding the overlap of structures.

\subsection*{Halos}
The relevant functions for halos that we use are:
\begin{align}
    \label{eq:bhshethtormen}b_h^{\text{sheth99}}(\sigma) &= 1+\frac{a\nu-1}{\delta_c}+\frac{2p}{\delta_c\left(1+(a\nu)^p\right)}\\[1mm]
    \label{eq:bh2SB}b^{2\text{SB}}_h(\sigma) &= 1-\frac{\sum_n\frac{n\pi}{\delta_T^2}\,\sin{\left(\frac{n\pi\delta_c}{\delta_T}\right)}\,\text{exp}\left[-\frac{n^2\pi^2}{2\delta_T^2}\,\sigma^2\right]\left[\textrm{cotan}{\left(\frac{n\pi\delta_c}{\delta_T}\right)\frac{n\pi}{\delta_T}}\right]}{\sum_n\frac{n\pi}{\delta_T^2}\,\sin{\left(\frac{n\pi\delta_c}{\delta_T}\right)}\,\text{exp}\left[-\frac{n^2\pi^2}{2\delta_T^2}\,\sigma^2\right]} \; .
\end{align}
%where $\delta_c$, $\delta_v$ and $\delta_T$ are the same discussed for the halo mass function.
Here $b_h^{\text{sheth99}}$ is the Sheth-Tormen linear bias \cite{sheth2001}, with $\nu=\delta_c^2/\sigma^2$ and $(a,\,p)$ are the same free parameters of the Sheth-Tormen mass function (Eq. \eqref{eq:fhsheth99}). The quantity $b_h^{2\text{SB}}$ is the prediction for a model with two static barriers.

\begin{figure}[tbp]
\centering 
\includegraphics[width=.99\textwidth]{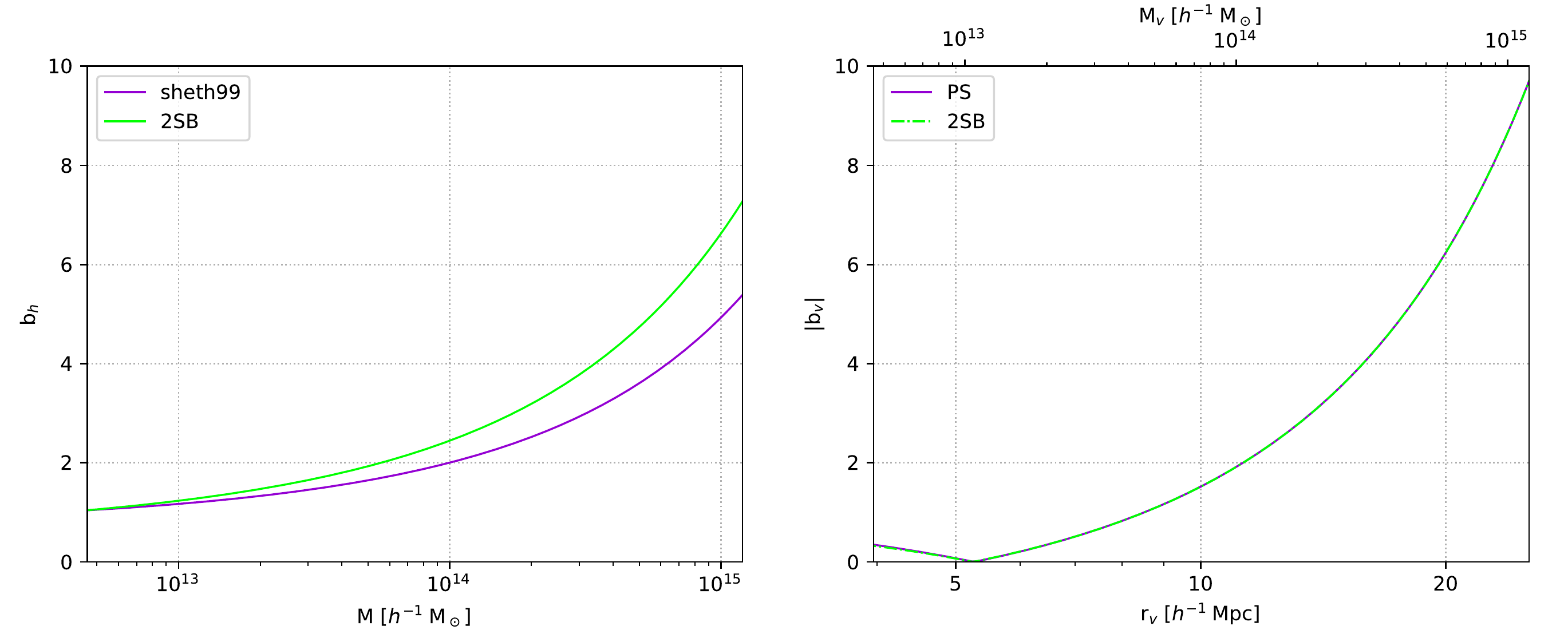}
\caption{\label{fig:linearbiases} (\emph{Left}): Halo linear bias as a function of halo mass. (\emph{Right}): Void linear bias as a function of void radius/mass. Both results are computed at $z=0$. We compare two different models: Sheth-Tormen for halos or Press-Schechter for voids (purple) and 2SB (green). For voids, we show the absolute value of the bias, as the latter is usually negative for larger voids. This is strictly connected to the absence of compensation walls for very large voids.}
\end{figure}

In the left panel of Fig. \ref{fig:linearbiases}, we compare the linear bias predictions for the functions above. Voivodic et al. \cite{voivodic2020} show how the 2SB linear bias is in very good agreement with simulations.

\subsection*{Voids}
In this work we consider the following functions:
\begin{align}
    \label{eq:bhpressschechter}b_v^{\text{PS}}(\sigma) &= 1+\frac{\nu-1}{\delta_v}\\[1mm]
    \label{eq:bv2SB}b^{2\text{SB}}_v(\sigma) &= 1 +\frac{\sum_n\frac{n\pi}{\delta_T^2}\,\sin{\left(\frac{n\pi|\delta_v|}{\delta_T}\right)}\,\text{exp}\left[-\frac{n^2\pi^2}{2\delta_T^2}\,\sigma^2\right]\left[\frac{n\pi}{\delta_T}\textrm{cotan}{\left(\frac{n\pi|\delta_v|}{\delta_T}\right)}\right]}{\sum_n\frac{n\pi}{\delta_T^2}\,\sin{\left(\frac{n\pi|\delta_v|}{\delta_T}\right)}\,\text{exp}\left[-\frac{n^2\pi^2}{2\delta_T^2}\,\sigma^2\right]} \; ,
\end{align}
where $\delta_c$, $\delta_v$ and $\delta_T$ are the same discussed for the halo mass function. Here, $b_v^{\text{PS}}$ is the Press-Schechter linear bias \cite{press1974}, with $\nu={\delta_v^2}/{\sigma^2}$. The quantity $b_v^{2\text{SB}}$ is the excursion set prediction for a model with two static barriers.

In the right panel of Fig. \ref{fig:linearbiases}, we compare the linear bias predictions for the functions above. Note that we show the absolute value of the bias since the latter is negative for larger voids in the 2SB model. This anti-correlation with the matter field is tightly connected with the compensation walls, as larger voids tend to have smaller compensation walls (see Sec. \ref{sec:profile}) and, therefore, to be under-compensated, resulting in a negative sign for the linear bias.

\subsection*{Convergence at small masses}
\label{sec:convergence}
As we know, the total power spectrum should follow linear theory on large scales. This is naturally achieved when the constraints of Eqs. \eqref{eq:rhohtotal} and \eqref{eq:bTOT} are satisfied. The former ensures that the total matter contained in all structures matches the total matter in the Universe, while the latter ensures that matter is not biased with respect to itself.

These constraints can be rewritten as:
\begin{align}
    \label{eq:Irho}I^\rho &= \frac{\overline{\rho}}{\overline{\rho}_m} = \frac{1}{\overline{\rho}_m}\int_0^\infty \de M\;M \, \frac{\de n}{\de M} = 1\\
    \label{eq:Ib}I^b &= \overline{b} = \frac{1}{\overline{\rho}_m}\int_0^\infty \de M\;M \, \frac{\de n}{\de M} \, b(M) = 1 \; ,
\end{align}
where both integrals are given by the sum of the contributions from all of the considered structures. 

Ref. \cite{voivodic2020} studies the problem by varying the lower bound of the integrals in Eqs. \eqref{eq:Irho} and \eqref{eq:Ib} for mass values $M_{\rm min}>0$. They show that the standard HM (with Sheth-Tormen-like halo mass function and linear bias), does not converge to unity even down to very low masses ($M_{\rm min}<10^4\;\text{M}_\odot$), pointing out the difficulty of the HM to recover efficiently the linear matter power on large scales and implying the need to either integrate Eqs. \eqref{eq:Irho} and \eqref{eq:Ib} down to very low masses or suitably normalize the 2Halo term. On the contrary, the sum of contributions from halos and voids, within the HVM, to Eqs. \eqref{eq:Irho} and \eqref{eq:Ib} converges to unity already for $M_{\rm min}\simeq10^9\;\text{M}_\odot$. Hence, incorporating voids in a model of LSS eliminates the need for an exotic re-normalization and saves up computational-time. 

In conclusion, allowing matter to lie within multiple structures, through the excursion set theory with two barriers \cite{simone2011, voivodic2017, voivodic2020}, offers as a fully self-consistent model, that correctly recovers the matter power on large scales, and as an efficient tool to a variety of cosmological and astrophysical studies. The foundation of such a success lies in considering the mass of smaller halos within voids of larger size.
%\vspace{-19pt}
\section{Power spectra}\label{sec:PS}
Early decoupling of dark matter allowed for its perturbations to grow undisturbed and form deep potential wells that accreted baryonic matter after recombination. As a consequence, galaxies are tightly coupled with dark matter and offer as a powerful tracer of its distribution. At the same time, galaxies host astrophysical sources that emit in several energy bands and dark matter structures bend the light, from distant sources, in the weak lensing regime. We then expect a correlation between gravitational tracers, such as i) the galaxy distribution and ii) the cosmic shear, and gamma-ray emitters, such as a) annihilating/decaying dark matter and b) unresolved astrophysical sources. The latter, while being an interesting signal by itself in the study of the unresolved component of the gamma-ray sky, nevertheless represent an irreducible background for pure dark matter studies.

In this section we generalize the formalism introduced in the previous sections to adapt the 3D power spectrum to the cross-correlation signal between two different source fields. In particular, we are interested in correlating i-ii) with a-b) within the HVM. For any couple of observables $i$ and $j$, the power spectra are:
\begin{align}
    \label{eq:1H}P^{1H}_{ij}(k) &= \int \de M\;\frac{\de n_h}{\de M}f_i^{h*}(k\,|\,M)f_j^h(k\,|\,M)\\
    \label{eq:2H}P^{2H}_{ij}(k) &= \int \de M_1\;\frac{\de n_h}{\de M_1}f_i^{h*}(k\,|\,M_1)b_{h}(M_1)\int \de M_2\;\frac{\de n_h}{\de M_2}f_j^h(k\,|\,M_2)b_{h}(M_2)P^L(k)\\
    \label{eq:1V}P^{1V}_{ij}(k) &= \int \de M\;\frac{\de n_v}{\de M}f_i^{v*}(k\,|\,M)f_j^v(k\,|\,M)\\
    \label{eq:2V}P^{2V}_{ij}(k) &= \int \de M_1\;\frac{\de n_v}{\de M_1}f_i^{v*}(k\,|\,M_1)b_{v}(M_1)\int \de M_2\;\frac{\de n_v}{\de M_2}f_j^v(k\,|\,M_2)b_{v}(M_2)P^L(k)\\
    \label{eq:HV}P^{HV}_{ij}(k) &= \int \de M_1\;\frac{\de n_h}{\de M_1}f_{(i}^{h*}(k\,|\,M_1)b_h(M_1)
    \int \de M_2\;\frac{\de n_v}{\de M_2}\,f_{j)}^v(k\,|\,M_2)b_v(M_2)\,P^L(k) \; ,
\end{align}
where we have generalized Eqs. \eqref{eq:P1H}-\eqref{eq:PHV} for any pair of source fields $f_i, f_j$. The Halo-Void term in Eq. \eqref{eq:HV} is given by the sum of the symmetric permutations of the considered fields. In the notation, we have omitted the redshift dependence of the 3D power spectrum for simplicity, but it has been considered throughout the computation.

The main idea of this work relies on the observational ability to identify cosmic voids: this would then allow to perform the cross-correlations between the relevant gravitational tracers (related to halos or voids) and the corresponding gamma-ray emission (due to dark matter annihilation/decay or the astrophysical sources) from those structures. In this case, we can split the CAPS \eqref{eq:CAPS} into two terms, each one depending on the 3D power spectrum generated only from halos or voids:
\begin{align}
    \label{eq:Ph}P_{ij}^h(k,\;z) &= P_{ij}^{1H}(k,\;z) + P_{ij}^{2H}(k,\;z)\\
    \label{eq:Pv}P_{ij}^v(k,\;z) &= P_{ij}^{1V}(k,\;z) + P_{ij}^{2V}(k,\;z) \; .
\end{align}
In order to calculate the power spectra of Eqs. \eqref{eq:Ph}, \eqref{eq:Pv} and the window functions of Eq. \eqref{eq:CAPS}, we need to specify the relation between gravitational tracers and the underlying large-scale structure as well as how the DM particles and astrophysical sources hosted in halos and voids produce the gamma-ray emission observed by our telescopes. In our work we consider cosmic shear and galaxy catalogs as our reference gravitational tracers. Clearly, in order to calculate the total signal, we need to add also the HV term.

Concerning the relation between the fields that define our signals with the underlying halo or void mass distribution, let us start with weak lensing, which directly depends on the distribution of dark matter across the line of sight; therefore the cosmic shear source field is intuitively proportional to the density profile of dark matter structures. Similarly, decaying dark matter is sourced by the distribution of dark matter itself, as it requires a single particle to occur in the decaying process, and therefore it directly traces the mass density. In both cases we have:
\begin{equation}
    f^x_l(k,\;z\,|\,M) = f^x_d(k,\;z\,|\,M)= \frac{\mathcal{F}[\rho_x](k,\;z\,|\,M)}{\overline{\rho}_m(z)} \; ,
\end{equation}
where $l$ stands for "lensing", $d$ for "decaying dark matter", $x = (h,\;v)$ denotes the considered structure and $\mathcal{F}[\varphi]$ denotes the Fourier transform of the field $\varphi$, which in our case can be either the density $\rho_x$ for decaying dark matter or the density squared $\rho^2_x$ for annihilating dark matter.

Also the galaxy distribution is sourced by the density of dark matter structures, however the latter must be weighted through the halo occupation distribution (HOD) \cite{cooray2002, berlind2002, berlind2003, zheng2007, zheng2005, cuoco2015}, namely $\langle \tilde N_g\rangle$, which is an indicator of the number of galaxies in each region of a dark matter halo:
\begin{equation}
    f^x_g(k,\;z\,|\,M) = \frac{\langle \tilde N_g^x\rangle}{\overline{n}_g^x(z)} = \frac{\langle N_\text{cen}(M)\rangle + \langle N_\text{sat}(M)\rangle \mathcal{F}[\rho_x](k,\;z\,|\,M)/M}{\overline{n}_g^x(z)} \; ,
\end{equation}
where $\overline{n}_g^x = \int \de M\;\dfrac{\de n_x}{\de M}\left(\langle N_\text{cen}\rangle + \langle N_\text{sat}\rangle\right)$ the average number of galaxies. Following \cite{cuoco2015}, and references therein, the average number of central galaxies $\langle N_\text{cen}\rangle$ and satellite galaxies $\langle N_\text{sat}\rangle$ in halos can be modeled as:
\begin{align}
    \langle N_\text{cen}\rangle &= \frac{1}{2}\left[1+\drm{erf}\left(\frac{\log M-\log M_{\rm th}}{\sigma_{\log M}}\right)\right]\\
    \langle N_\text{sat}\rangle &= \left(\frac{M}{M_*}\right)^\alpha\drm{exp}\left[-\frac{M_{\rm cut}}{M}\right] \; ,
\end{align}
where $M_{\rm th}$ denotes the approximate halo mass required to populate the halo with the considered type of galaxies and $\sigma_{\log M}$ governs the width of the transition between $0$ and $1$ for the central galaxy. The satellite distribution is described by a power law (of order $\alpha$) with an exponential cutoff $M_{\rm cut}$ at low masses. In this paper we consider the galaxy distribution from the 2MASS catalog \cite{2MASS}, as it is one of the most extended almost-all-sky catalogs. Exploiting the HOD results of Ref. \cite{zehavi2005}, we adopt a step function for central galaxies ($\langle N_\text{cen}\rangle = 0$ for $M<10^{12.1}\;\text{M}_\odot$ and $\langle N_\text{cen}\rangle = 1$ for $M\ge10^{12.1}\;\text{M}_\odot$) and take the following parameters for satellite galaxies: $\alpha=1.2$, $M_*=10^{13.5}\;\text{M}_\odot$ and $M_{\rm cut}=0$.

\begin{figure}[tbp]
\centering 
\includegraphics[width=.9\textwidth]{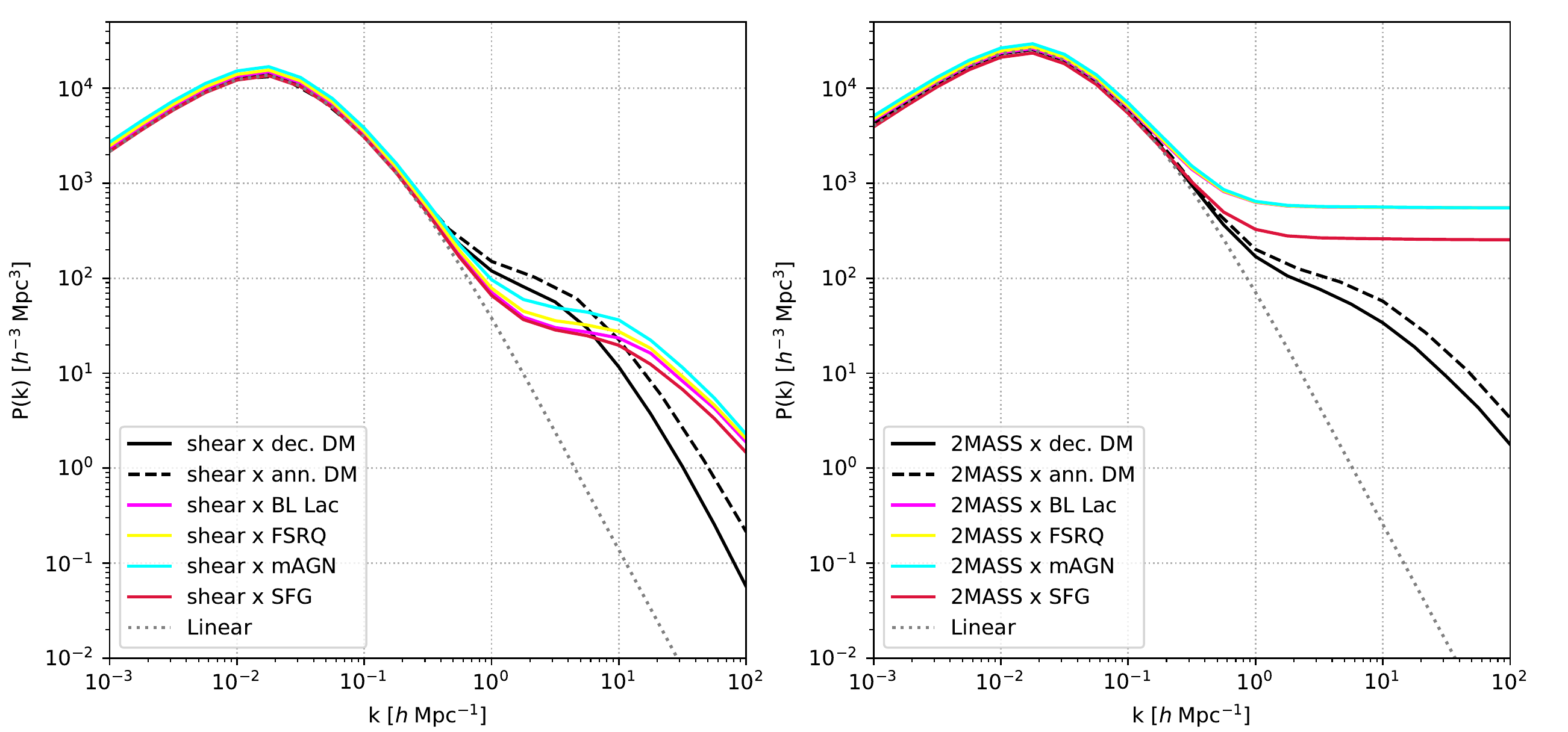}
\caption{\label{fig:PScross} Comparison of the total cross-correlation 3D power spectrum between gamma-ray sources and gravitational tracers (at $z=0.5$): cosmic shear (\emph{Left}) and the 2MASS galaxy catalog (\emph{Right}). The shown gamma-ray sources are DM decay (solid black), DM annihilation (dashed black), BL Lac (magenta), FSRQ (yellow), mAGN (cyan) and SFG (crimson red).}
\end{figure}

In the case of voids, we need to adapt the HOD to a void occupation distribution (VOD). In this regard, we assume that the total number of galaxies is the same within halos and voids of the same mass. Intuitively, if the local density fractions of the Universe components remain unchanged in voids, we expect the fraction of matter contained in galaxies to be the same as in halos. Ref. \cite{patiri2006} shows that voids usually contain bluer galaxies, however, the general properties (such as color distribution, bulge to total ratios, and concentrations) are remarkably similar to those in halos. The main difference resides in their distribution within the structure; in particular, the number of galaxies tends to increase with the radius of the void (see e.g. \cite{hamaus2014}), while void centers are expected to be extremely underdense, leaving arguably no space for central galaxies\footnote{Indeed, void finders are usually based on tessellation methods around void centers, found as the minima in the density distribution, thus representing points sensibly far from the closest galaxies.}. This can be accounted for by re-scaling the void density profile with a parameterization of the galaxy number density contrast (in voids) shown in \cite{patiri2006}. We define the number density contrast of galaxies as $1+\delta_g(r)=n_g(r)/\overline{n}_v$ (where $n_g(r)$ is the radial number density of galaxies and $\overline{n}_v$ is the average density of the considered void). The density of galaxies within voids can be rewritten as
\be
    \rho_v^{\rm gal}(r)=\rho_v(r)\frac{n_g(r)}{\overline{n}_v}\equiv\rho_v(r)(1+\delta_g(r)) \; ,
\ee
where the values for $\delta_g(r)$ have been extrapolated by the data of Ref. \cite{patiri2006}. By plugging this expression in the HOD formula shown above and setting $\langle N_\text{cen}(M)\rangle=0$ for any void we obtain the needed VOD.

In the case of annihilating dark matter, the process requires both a particle and an antiparticle to occur. The source field is therefore proportional to the square of the density profile of dark matter structures:
\begin{equation}
    f^x_a(k,\;z\,|\,M) = \frac{1}{\Delta_x^2(z)}\frac{\mathcal{F}[\rho_x^2](k,\;z\,|\,M)}{\overline{\rho}^2_m(z)} \; ,
\end{equation}
where $a$ stands for "annihilating dark matter" and $\Delta_x^2(z)$ is the clumping factor, which acts as a transfer function between $\langle\rho^2\rangle$ and $\langle\rho\rangle^2$:
\begin{equation}\label{eq:clumping}
    \Delta^2_x(z) = \frac{\langle\rho_x^2\rangle}{\overline{\rho}^2_m} = \int \de M\;\frac{\de n_x}{\de M}\int \de^3x\;\frac{\rho_x^2(\textbf{x}\,|\,M)}{\overline{\rho}^2} \; .
\end{equation}

When considering astrophysical sources we have to replace the structure mass $M$ with the luminosity $L$ of the source, the mass function $\de n_x/\de M$ with the GLF $\phi_x(L,\;z) = \de n_x/\de L$ and the linear bias with $b_s^x(M(L),\; z)$, following \cite{fornengo2014}. In this work we consider four different classes of unresolved gamma-ray astrophysical sources: BL Lacertae objects (BL Lacs), flat-spectrum radio quasars (FSRQs), misaligned AGN (mAGN) and star-forming galaxies (SFGs). The GLFs and mass-to-luminosity functions $M(L)$ of each astrophysical class are shown in Appendix \ref{app:A} and \ref{app:B}, respectively.  The source field can be expressed as:
\begin{equation}
    f_s^x(z\,|\,L) = \frac{L}{\langle g^x_s\rangle} \; ,
\end{equation}
where $\langle g^x_s(z)\rangle = \int \de L\;L\,\phi_x(L,\;z)$ is the average luminosity density of sources. 

Gamma-ray astrophysics has deeply improved in the last decade, with thousands of sources detected, allowing for a better understanding of the most violent phenomena of the Universe. However, the unresolved gamma-ray background (UGRB) still remains to be fully understood. Moreover, previous works mostly focus on unresolved sources within dark matter halos, while their presence and properties within the most underdense regions of the Universe remain somewhat foggy and require further studies. Works on the properties of galaxies \cite{patiri2006} and AGN \cite{constantin2008, furniss2015} in voids show that the environment makes very little impact on general properties. AGN are slightly more common in underdense regions, but only for the most luminous galaxies. At the same time, voids are generally younger than halos, due to the hierarchy in structure formation, showing bluer galaxies and increased rates of structure formation (the effect being evident only in the rare very massive galaxies). Nevertheless, the accretion rates do not show particular changes and we expect mostly negligible variations between the luminosity properties of sources hosted by galaxies in halos and those in voids of the same mass. Given these considerations, we assume the number of sources within voids and halos of the same mass to be constant, which is intuitively similar to what we have done for the void occupation distribution. Under this assumption, we can write the void GLF as the one of halos re-scaled by the volume fraction at fixed mass $M_*$\footnote{Dimensionally speaking, the GLF is a number density per unit of luminosity; assuming no dependence of the luminosity properties on the environment and the number of sources to be constant among different structures of the same mass, the GLF goes as the inverse of a volume.}:
\begin{equation}
    \phi_v(L,\;z)=\frac{V_h(M_*)}{V_v(M_*)}\,\phi_h(L,\;z)\sim 6.2\times10^{-4}\,\phi_h(L,\;z) \; ,
\end{equation}
where $V_h(M_*)/V_v(M_*)=\Delta_v/\Delta_{h,\,\text{vir}}$.

In the left panel of Fig. \ref{fig:PScross}, we show the total 3D power spectrum of the cross-correlation between cosmic shear and gamma-ray emitters (decaying/annihilating DM and the four classes of unresolved astrophysical sources discussed above) at $z=0.5$, computed within the HVM through the sum of contributions from Eqs. \eqref{eq:1H}-\eqref{eq:HV}. We discuss, with more detail, the importance of each term in the Appendix. The same result is shown in the right panel of Fig. \ref{fig:PScross} for the cross-correlation of the 2MASS galaxy catalog and the unresolved gamma-ray sky.

\section{Window functions of gamma-ray emission, galaxies and cosmic shear}\label{sec:W}

\begin{table}[t]
\centering
\begin{tabular}{l|ccc} 
 \hline\rule{0mm}{5mm}
 ~& $\Gamma$ & $L_{\rm min}$ $[$erg s$^{-1}]$ & $L_{\rm max}$ $[$erg s$^{-1}]$\\[1mm]
 \hline
 \rule{0mm}{5mm}BL Lacs \cite{ajello2014} & $2.11$ & $7\times 10^{43}$ & $10^{52}$\\
 
 FSRQ \cite{ajello2012} & $2.44$ & $10^{44}$ & $10^{52}$\\
 
 mAGN \cite{willot2001, mauro2014} & $2.37$ & $10^{40}$ & $10^{50}$\\
 
 SFG \cite{gruppioni2013} & $2.7$ & $10^{37}$ & $10^{42}$\\[1mm]
 \hline
 
\end{tabular}
 \caption{Spectral index $\Gamma$, minimum and maximum luminosities and related references for the classes of unresolved gamma-ray astrophysical sources considered in this paper.}
 \label{tab:sourceclass}
\end{table}

The last needed ingredient to compute the CAPS in Eq. \eqref{eq:CAPS} is the window function for the different observables considered here.

The window function of weak lensing takes the form (see e.g. \cite{bartelmann2010}):
\begin{equation}\label{eq:Wl}
    W_l(\chi) = \frac{3}{2}\frac{H_0^2}{c^2}\Omega_m(1+z)\chi\int_\chi^\infty \de\chi'\;\frac{\chi'-\chi}{\chi'}\frac{\de N}{\de\chi'}(\chi') \; ,
\end{equation}
where $\de N/\de\chi$ denotes the redshift distribution of the background sources (see \cite{euclid2020}), normalized to unit area.

For galaxies, the window function simply reduces to their redshift distribution (e.g. see \cite{xia2015}):
\begin{equation}\label{eq:Wg}
    W_g(\chi) = \frac{\de N_g}{\de z}\frac{\de z}{\de\chi} = \frac{\de N_g}{\de z}\frac{H(z)}{c} \; .
\end{equation}
As mentioned above, we consider the galaxy distribution from the 2MASS catalog \cite{2MASS}.

In the case of decaying dark matter we have:
\begin{equation}\label{eq:Wd}
    W_d(E,\;z) = \frac{1}{4\pi}\frac{\Omega_\text{DM}\rho_c}{m_\text{DM}\tau_d}\frac{\de N_d}{\de E}\left[E(1+z)\right]\text{e}^{-\tau\left[E(1+z),\;z\right]} \; ,
\end{equation}
where $\Omega_\text{DM}$ is the cosmological abundance of DM and $\rho_c$ the critical density of the Universe today. $m_\text{DM}$ and $\tau_d$ denote the mass and decay lifetime of the DM particle (here we consider $m_\text{DM} = 100\;\text{GeV}$ and $\tau_d=3\times 10^{27}\;\text{s}$, from \cite{fornengo2014b, Blanco:2018esa}, for definiteness, about at its conservative lower bound) and $\de N_d/\de E$ is the number of photons emitted by the decay in the energy band $[E,\;E+ \de E]$, here for the decaying channel $b\overline{b}$ (see e.g. \cite{cembranos2011}). Finally, $\tau$ is the optical depth for absorption on the line of sight (taken here from \cite{razzaque2009}). The latter is relevant for gamma-rays mainly due to pair production on the extra-galactic background light emitted by galaxies in the ultraviolet, optical, and infrared bands. 

\begin{figure}[tbp]
\centering 
\includegraphics[width=.99\textwidth]{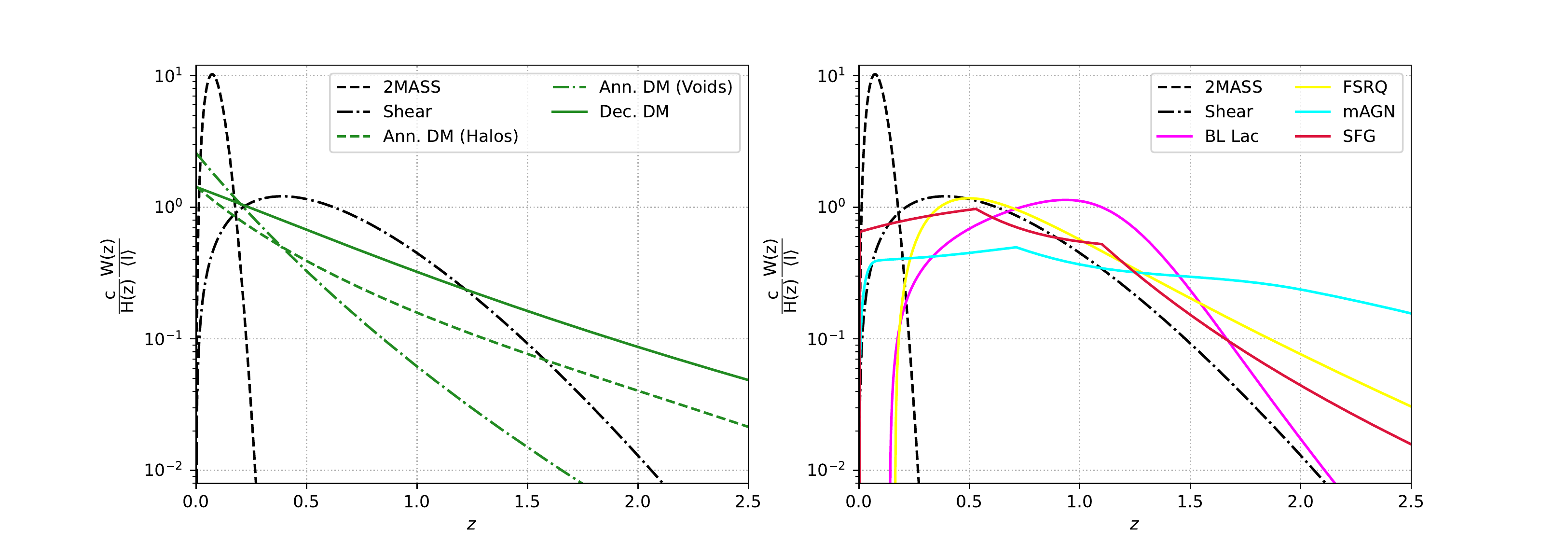}
\hfill
\caption{\label{fig:window} (\emph{Left}): Comparison between the window functions of decaying/annihilating dark matter with those of the 2MASS galaxy distribution and weak lensing. (\emph{Right}): Comparison between the window functions of each class of gamma-ray astrophysical sources and those of gravitational tracers. In both cases the functions are computed for gamma-rays with $E_\gamma=5$ GeV, and normalized to the redshift integrated intensity.}
\end{figure}

For the case of annihilating dark matter, since the source field depends on $\rho^2$, the clumping factor of Eq. \eqref{eq:clumping} enters the window function, leading to a dependence on the structure $x$, differently from the decaying case:
\begin{equation}\label{eq:Wa}
    W_a^x(E,\;z) = \frac{\left(\Omega_\text{DM}\rho_c\right)^2}{4\pi}\frac{\langle\sigma_a v\rangle}{2m_\text{DM}^2}(1+z^3)\Delta_x^2(z)\frac{\de N_a}{\de E}\left[E(1+z)\right]\text{e}^{-\tau\left[E(1+z),\;z\right]} \; ,
\end{equation}
where $\langle\sigma_a v\rangle$ is the velocity averaged annihilating cross section (we consider the thermal value of $3\times10^{-26}\;\text{cm}^{3}\text{s}^{-1}$ for a WIMP dark matter candidate with a mass around $100\;\text{GeV}$, see \cite{fornengo2014b}) and $\de N_a/\de E$ is the number of photons emitted by annihilations in the energy band $[E,\;E+dE]$, we consider the channel $b\overline{b}$ \cite{cembranos2011}.

The window function of astrophysical sources, within a structure $x$, yields (see e.g. \cite{pinetti2019}):
\begin{equation}\label{eq:Ws}
    W_s^x(E,\;z) = \left(\frac{d_L(z)}{1+z}\right)^2\int_{L_{\rm min}}^{L_{\rm max}(z)}\de L\;\frac{\de F}{\de E}(E,\;L,\;z)\phi_x(L,\;z) \; ,
\end{equation}
where $\phi_x$ is the GLF, $d_L(z)=(1+z) \, \chi(z)$ the luminosity distance and $\de F/\de E$ the spectral energy distribution (SED) which, assuming a power law for the number of gamma-rays in the energy interval $(E,\;E+ \de E)$, reads\footnote{The normalization factor in front of the energy power law comes from defining the luminosity in the energy range $(0.1,\;100)$ GeV.}:
\begin{equation}
    \frac{\de F}{\de E}(E,\;L,\;z)=\frac{L}{4\pi d_L^2(z)}(1+z)(2-\Gamma)\left[100^{2-\Gamma}-0.1^{2-\Gamma}\right]^{-1}\left(\frac{E}{\text{GeV}}\right)^{-\Gamma}\;\text{GeV}^{-2} \; ,
\end{equation}
where $\Gamma$ is the spectral index of the considered source class (see Table \ref{tab:sourceclass}). The minimum and maximum luminosities in Eq. \eqref{eq:Ws} depend on the intrinsic properties of the source class and are shown in Table \ref{tab:sourceclass}. However, since we are dealing with the unresolved sky, $L_{\rm max}$ shall never be larger than the luminosity corresponding to the sensitivity of the detector. In this work we consider the \emph{Fermi}-LAT telescope for gamma-ray surveys and assume a detector flux sensitivity $F_{\rm sens} = 10^{-10}\;\text{cm}^{-2}\text{s}^{-1}$, for photons in the energy band $1-100$ GeV, which is well compatible with 8 years of data taking\footnote{The constraint on $L_{\rm max}$ is more relevant at low redshift, where $L_{\rm sens}<L_{\rm max}$ and the window function tends to fall rapidly.} and a slightly better sensitivity for a future improved detector, discussed below. Note that Eqs. \eqref{eq:Wd}, \eqref{eq:Wa} and \eqref{eq:Ws} must be integrated in the considered energy band, before being plugged into the CAPS of Eq. \eqref{eq:CAPS}.

In Fig. \ref{fig:window} we show the normalized (to the average intensity $\langle I\rangle=\int \de \chi\;W(\chi)$) window functions for gamma-rays with $E_\gamma=5$ GeV. We compare the gravitational tracers (galaxy distribution from the 2MASS catalog and the cosmic shear) with the gamma-ray emitters: annihilating/decaying dark matter in the left panel and astrophysical sources in the right panel. We see how the latter peaks around redshifts $0.5-1$, while the former, being completely unresolved, peaks at very low redshifts and then rapidly decays.

\section{Results}\label{sec:results}
\begin{table}
\centering
\begin{tabular}{ccccccc} 
 \hline\rule{0mm}{5mm}
 Bin & $E_\text{min}\;[\text{GeV}]$ & $E_\text{max}\;[\text{GeV}]$ & $N^\gamma\;[\text{cm}^{-4}\text{s}^{-2}\text{sr}^{-1}]$ & $f_\text{sky}$ & $\sigma_0^\text{Fermi}\;[\text{deg}]$ & $E_b\;[\text{GeV}]$\\[1mm]
 \hline
 \rule{0mm}{5mm}1 & $0.5$ & $1.0$ & $1.056\times10^{-17}$ & $0.134$ & $0.87$ & $0.71$\\
 
 2 & $1.0$ & $1.7$ & $3.548\times10^{-18}$ & $0.184$ & $0.50$ & $1.30$\\
 
 3 & $1.7$ & $2.8$ & $1.375\times10^{-18}$ & $0.398$ & $0.33$ & $2.18$\\
 
 4 & $2.8$ & $4.8$ & $8.324\times10^{-19}$ & $0.482$ & $0.22$ & $3.67$\\
 
 5 & $4.8$ & $8.3$ & $3.904\times10^{-19}$ & $0.549$ & $0.15$ & $6.31$\\
 
 6 & $8.3$ & $14.5$ & $1.768\times10^{-19}$ & $0.574$ & $0.11$ & $11.0$\\
 
 7 & $14.5$ & $22.9$ & $6.899\times10^{-20}$ & $0.574$ & $0.09$ & $18.2$\\
 
 8 & $22.9$ & $39.8$ & $3.895\times10^{-20}$ & $0.574$ & $0.07$ & $30.2$\\
 
 9 & $39.8$ & $69.2$ & $1.576\times10^{-20}$ & $0.574$ & $0.07$ & $52.5$\\
 
 10 & $69.2$ & $120.2$ & $6.205\times10^{-21}$ & $0.574$ & $0.06$ & $91.2$\\
 
 11 & $120.2$ & $331.1$ & $3.287\times10^{-21}$ & $0.597$ & $0.06$ & $199.5$\\
 
 12 & $331.1$ & $1000$ & $5.094\times10^{-22}$ & $0.597$ & $0.06$ & $575.4$\\[1mm]
 \hline
\end{tabular}
 \caption{Gamma-ray energy bins used in this analysis, adherent with 8 years of data taking from \emph{Fermi}-LAT Pass 8 (see \cite{ackermann2018}). $N^\gamma$ is the auto-correlation noise, $f_\text{sky}$ the observed fraction of the sky outside the combined Galactic and point-source masks and $\sigma_0^\text{Fermi}$ the $68\%$ containment angle of the PSF, referred to the geometric center of each energy bin $E_b=\sqrt{E_\text{min} \, E_\text{max}}$.}
 \label{tab:fermi}
\end{table}
As previously discussed, we underline the potential of cross-correlating gamma-ray emission from particle dark matter (either through annihilations or decays) with the most underdense regions of the Universe. Previous works have proposed \cite{fornengo2014, camera2013, pinetti2019} and used \cite{xia2015, branchini2017, troster2017} the cross-correlation formalism between dark matter emission and gravitational tracers within dark matter halos, and have shown how the gamma-ray signal generated by unresolved astrophysical sources typically surpasses that of dark matter, making the signal-to-noise ratio somehow unfavourable. As discussed in Sec. \ref{sec:PS}, the astrophysical signal depends on the GLF which can be more than three orders of magnitude lower in voids than in halos, assuming the number of sources to be constant for halos and voids with the same mass and no relevant changes in their general properties. For this reason, we show here that, while the dark matter signal from voids is reduced in size as compared to halos, nevertheless in comparison to its astrophysical background the cross-correlation signal can be favored with respect to astrophysical sources, ticking the potential usefulness of exploiting voids to research on the dark sector.

On the observational side, the most efficient gamma-ray detector today is the \emph{Fermi}-LAT telescope, which has contributed strongly in the advancement of our knowledge on extremely powerful events in the Universe. With its excellent angular and energy resolutions and more than 10 years of service, it has been used to determine the composition of the UGRB \cite{ackermann2018, ackermann2015, ackermann2012, fornasa2016}. In this work we use, as a point of reference, the specifications adopted for the analysis of the UGRB performed in \cite{ackermann2018}, based on 8 years of data taking and a selection of events with optimal angular resolution and background rejection. Ref. \cite{ackermann2018} is currently the most up-to-date analysis of the statistical fluctuations of the UGRB, and we therefore adopt the specifications (sensitivity, energy binning, angular resolution, noise) as a reference for this analysis. We will then forecast the reach of a future gamma-ray detector (called {\it Fermissimo} for definiteness and in continuity with previous analyses \cite{Camera:2014rja, pinetti2019}), as specified in \cite{pinetti2019} and which is modeled on improved specifications (like e.g. \cite{Topchiev:2017xfp}).

We compute the cross-correlations on the 12 energy bins of \cite{ackermann2018} and reported in Table \ref{tab:fermi}. For each energy bin we display the measured photon noise $N^\gamma$, the observed portion of the sky $f_\text{sky}$ and the $68\%$ containment angle of the \emph{Fermi}-LAT point spread function (PSF) around the geometric center of the bin. These parameters are necessary to calculate the variance on the CAPS (see Eq. \eqref{eq:CAPSvariance}). The beam function of the telescope depends on the photon event class and the energy spectrum and is available through the \emph{Fermi} tools. An overall good analytical approximation is given by (see \cite{pinetti2019}):
\begin{equation}\label{eq:beam}
    B_\ell^\gamma = \text{exp}\left[-\frac{\sigma_b^2(\ell,\;E)\ell^2}{2}\right] \; ,
\end{equation}

where the dispersion angle of each energy bin evolves as (see \cite{pinetti2019}):
\begin{equation}
    \sigma_b(\ell,\;E) = \sigma_0^\text{Fermi}(E)\left[1+0.25\,\sigma_0^\text{Fermi}(E)\ell\right]^{-1} \; ,
\end{equation}
for which $\sigma_0^\text{Fermi}(E)$ is the $68\%$ containment angle of the \emph{Fermi}-LAT PSF and can be defined in terms of a referenced value $\sigma_0^\text{Fermi}(E_{\rm ref}=0.5\;\text{GeV})=1.2$ deg (see \cite{pinetti2019}):
\be
    \sigma_0^\text{Fermi}(E) = \sigma_0^\text{Fermi}(E_{\rm ref})\left[\frac{E}{E_{\rm ref}}\right]^{-0.95}+0.05\;\text{deg} \; .
\ee

\begin{figure}[tbp]
\centering 
\includegraphics[width=.9\textwidth]{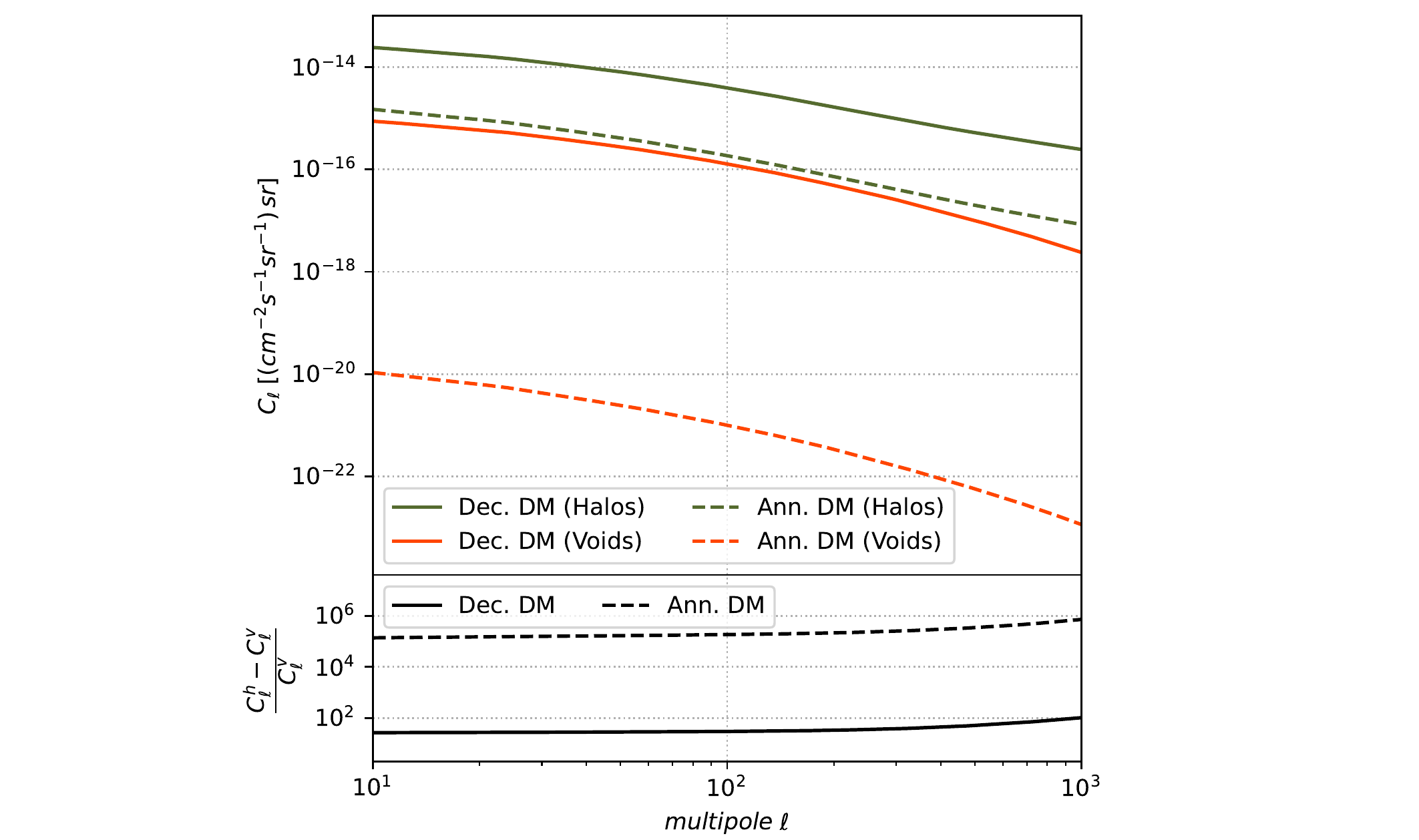}
\hfill
\caption{\label{fig:shearXDM} (\emph{Top}): CAPS of decaying (solid)/annihilating (dashed) dark matter and cosmic shear both in halos (green) and voids (orange). (\emph{Bottom}): Relative difference between the cross-correlation signal in halos and voids, for the two production cases. Note that in the case of voids, the annihilating signal is largely disfavored, being its relative difference much higher than for decaying dark matter. The dark matter particle is $m_\text{DM}= 100$ GeV, the production channel is $b\bar b$, the annihilation rate is $\langle \sigma v \rangle = 3\times10^{-26}\;\text{cm}^{3}\text{s}^{-1}$, the decay lifetime is $\tau_d=3\times 10^{27}\;\text{s}$.}
\end{figure}
On the cosmic shear side the currently operating survey is DES \cite{DES}, which provides weak lensing shape catalogs with more than $10^8$ galaxies \cite{DESb, DESc}. However, Euclid \cite{Euclid} will be the next generation galaxy survey and its forecasts \cite{euclid2020} are extremely promising, especially for cosmic shear as the telescope is expected to be very sensitive to this effect. Galaxy surveys have achieved an incredibly high angular resolution and their beam function in harmonic space $B_\ell$ can be considered equal to unity in the multipole range of interest here, while the noise associated to the auto-correlation is, for Euclid: $N^\text{Euclid}=\sigma^2_\epsilon/\overline{\rm N}_g$, where $\sigma_\epsilon=0.3$ is the intrinsic ellipticity and $\overline{\rm N}_g=30\;\text{arcmin}^{-2}$ is the average number of galaxies per steradian (see e.g \cite{Camera:2014rja}). Similarly, for galaxies the beam function can be set to unity and the noise simply reduces to the average number of galaxies.

In the gaussian error estimates on the signals of Eq. \eqref{eq:CAPSvariance}, the cross-correlation term is largely sub-dominant, while the gamma-ray auto-correlation term is dominated by the noise\footnote{This is only true when considering the specifications of \emph{Fermi}-LAT, while when considering an improved future detector, namely {\it Fermissimo}, the noise and beam function are largely reduced and their contribution to the error is comparable with that of the gamma-ray auto-correlation, and therefore we include them in the error estimate.} (see e.g. \cite{pinetti2019}). The variance of the cross-correlation between gamma-rays and gravitational tracers (namely "$t$") can therefore be safely approximated by:
\be\label{eq:errorapprox}
     (\Delta C^{t\gamma}_\ell)^2 \simeq \frac{1}{(2\ell+1)f_\text{sky}}\left[\frac{N^\gamma}{(B_\ell^\gamma)^2}\left(C^{tt}_\ell+N^t\right)\right] \; .
\ee

For definiteness, as a representative case of WIMP, in the following we will consider a dark matter particle of 100 GeV mass and annihilating or decaying into the $b\bar b$ channel with a canonical-thermal-relic annihilation rate of $\langle \sigma_a v \rangle = 3\times10^{-26}\;\text{cm}^{3} \, \text{s}^{-1}$ or a decay lifetime of $\tau_d=3\times 10^{27}\;\text{s}$ (close to its conservative lower bound \cite{fornengo2014b, Blanco:2018esa}), respectively.

Let us turn first to the discussion of the cross-correlation signal between dark matter gamma-ray emission and gravitational tracers. In particular, we first compare the decaying signal with the annihilating one (we chose to use only the cosmic shear as a reference for gravitational tracers, for simplicity). The CAPS of Eq. \eqref{eq:CAPS} depends on the spectra and window functions for dark matter and cosmic shear, whose prescriptions are shown in Secs. \ref{sec:PS} and \ref{sec:W}, respectively. In the top panel of Fig. \ref{fig:shearXDM} we show the CAPS of dark matter decay (solid lines) and annihilation (dashed lines), separately for halos (green) and voids (orange). The signal produced by decaying dark matter is higher than that of annihilating dark matter across the whole range of multipoles and for both types of structures. While both annihilating and decaying signals can be modified by changing $\langle \sigma v \rangle$ or $\tau$ (which we have set here close to their bounds), nonetheless, as a general property, the relative difference between the signal in voids and halos is highly disfavored for annihilations. In fact, as shown by the bottom panel of Fig. \ref{fig:shearXDM}, in the case of decay the signal in voids is about a factor of 50 smaller than the corresponding signal in halos, while for annihilations the ratio is significantly depressed.
This behavior is intuitively connected to the dependence of the source function on the square of the density, enhancing the already conspicuous ratio between the density of halos and voids. We conclude that underdense regions are not an efficient probe of annihilating dark matter and we will focus only on decaying dark matter for the following discussion.

\begin{figure}[tbp]
\centering 
\includegraphics[width=.9\textwidth]{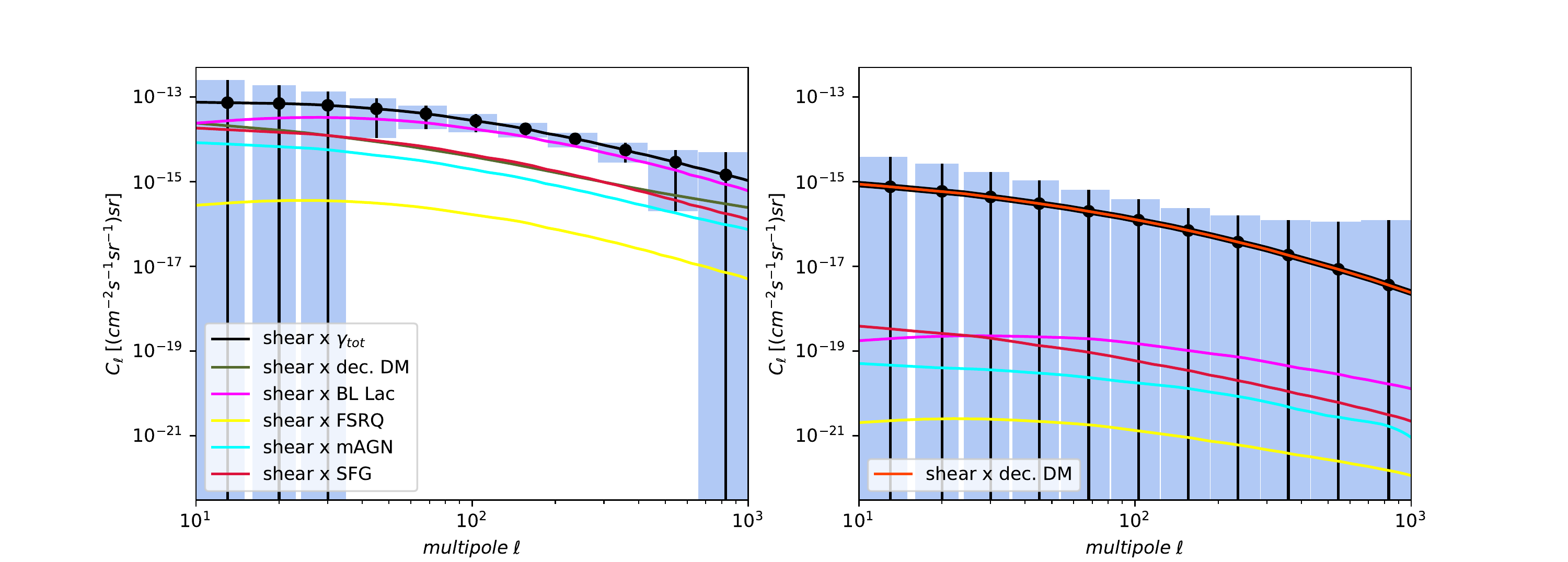}
\hfill
\caption{\label{fig:shearXgamma} Cross-correlation angular power spectrum between cosmic shear and the unresolved gamma-ray sky. The total signal (solid black) is given by the sum of contributions from all gamma-ray sources: astrophysical sources and dark matter decays (we consider a DM mass $m_\text{DM}=100\;\text{GeV}$ and decay lifetime $\tau_d=3\times10^{27}\;\text{s}$). In the left panel we show the signals computed within halos and in the right panel those within voids. The result refers to the sum of contributions from all energy bins of Table \ref{tab:fermi} and the error bars are obtained from the Gaussian estimate of the variance of the signals. Error bars are calculated for the {\sl Fermi}-LAT configuration discussed in the text.}
\end{figure}

\begin{figure}[tbp]
\centering 
\includegraphics[width=.9\textwidth]{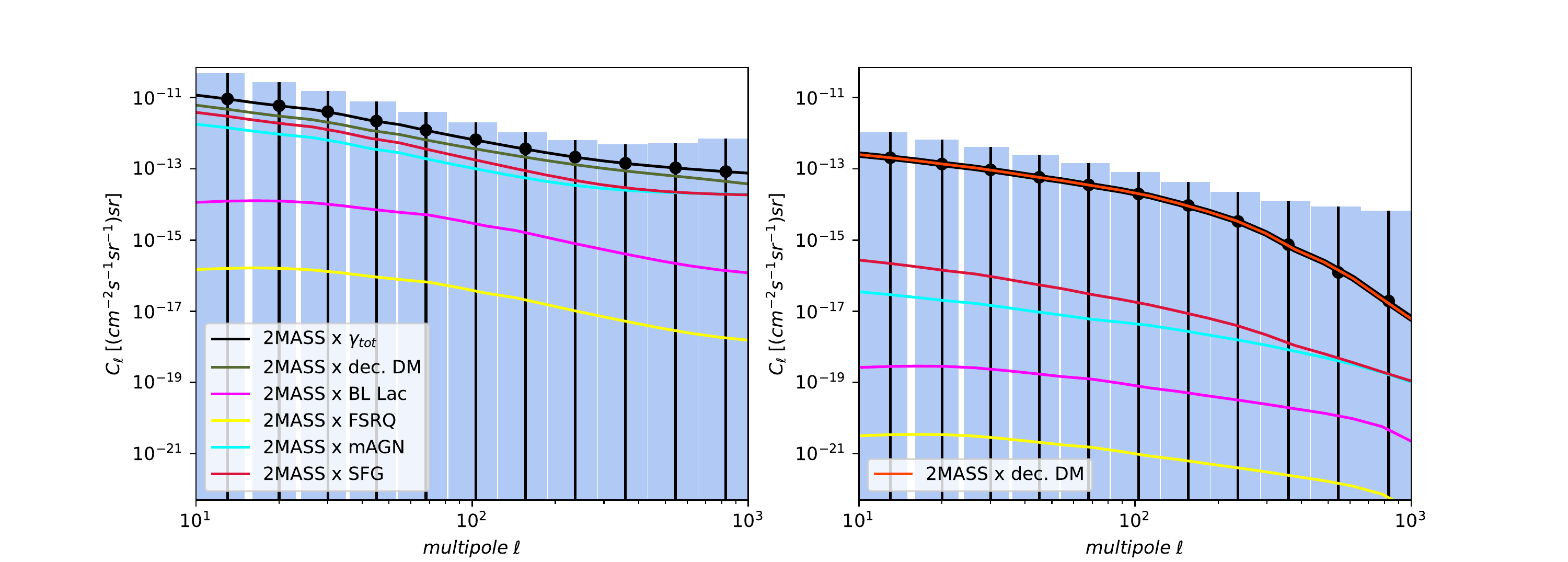}
\hfill
\caption{\label{fig:galXgamma} The same as Fig. \ref{fig:shearXgamma} for galaxies $\times$ gamma-rays.}
\end{figure}

Remaining in the realm of decaying dark matter, the relevant angular power spectra are shown in Fig. \ref{fig:shearXgamma} for the cross-correlation between cosmic shear and gamma-ray emitters, and in Fig. \ref{fig:galXgamma} for that between the galaxy distribution and gamma-ray emitters. In each figure, we display the contribution from the four classes of unresolved astrophysical sources (BL Lac, FSRQ, mAGN and SFG) and from decaying dark matter, both for halos (in the left panel) and voids (in the right panel). The results refer to the sum of contributions from all the energy bins of Table \ref{tab:fermi}. Among astrophysical sources, for the cosmic shear case, BL Lacs and SFGs provide for the dominant terms, while FSRQs are strongly sub-leading. In the case of galaxies, however, BL Lacs give way to mAGN, which climb up to the dominant contribution, together with SFGs, for high multipoles. This derives from a complex interplay between the redshift and energy dependence of the CAPS. Intuitively, mAGN have a greater impact when cross-correlating with galaxies as the 2MASS window function is peaked at lower redshift (where blazars tend to have extremely low power, due to the detector sensitivity cutoff) than the cosmic shear one, which, on the contrary, peaks right around the maximum of the window function for BL Lacs, that is strongly suppressed at low redshift (these behaviors are trivial by looking at the right panel of Fig. \ref{fig:window}).

The main observation relative to Fig. \ref{fig:shearXgamma} and \ref{fig:galXgamma} is that in the case of halos, the dark matter signal is either completely dominated by astrophysical sources or too close to their signal to be effectively distinguished by the present-day detectors. However, as shown by the right panel of Figs. \ref{fig:shearXgamma} and \ref{fig:galXgamma}, in voids decaying DM provides the dominant contribution, even almost completely setting the total signal, dominating astrophysical sources by more than three orders of magnitude for the cosmic shear and from three to one order of magnitude (for increasing multipoles) for galaxies. This behavior is due to the expected much lower density of astrophysical sources in underdense environments, as discussed in Sec. \ref{sec:PS} when examining the difference between the GLF in halos and voids. 

In summary, though the UGRB signal is expected to be much lower in voids than in halos, the signal from decaying dark matter is expected to dominate over the astrophysical background by a sizeable amount, contrary to the case of halos, making this option (if accessible, given the sensitivity of gamma-ray detectors) a background-free signal for dark matter lifetimes 3-4 orders of magnitude smaller than current bounds. 
\begin{figure}[tbp]
\centering 
\includegraphics[width=.9\textwidth]{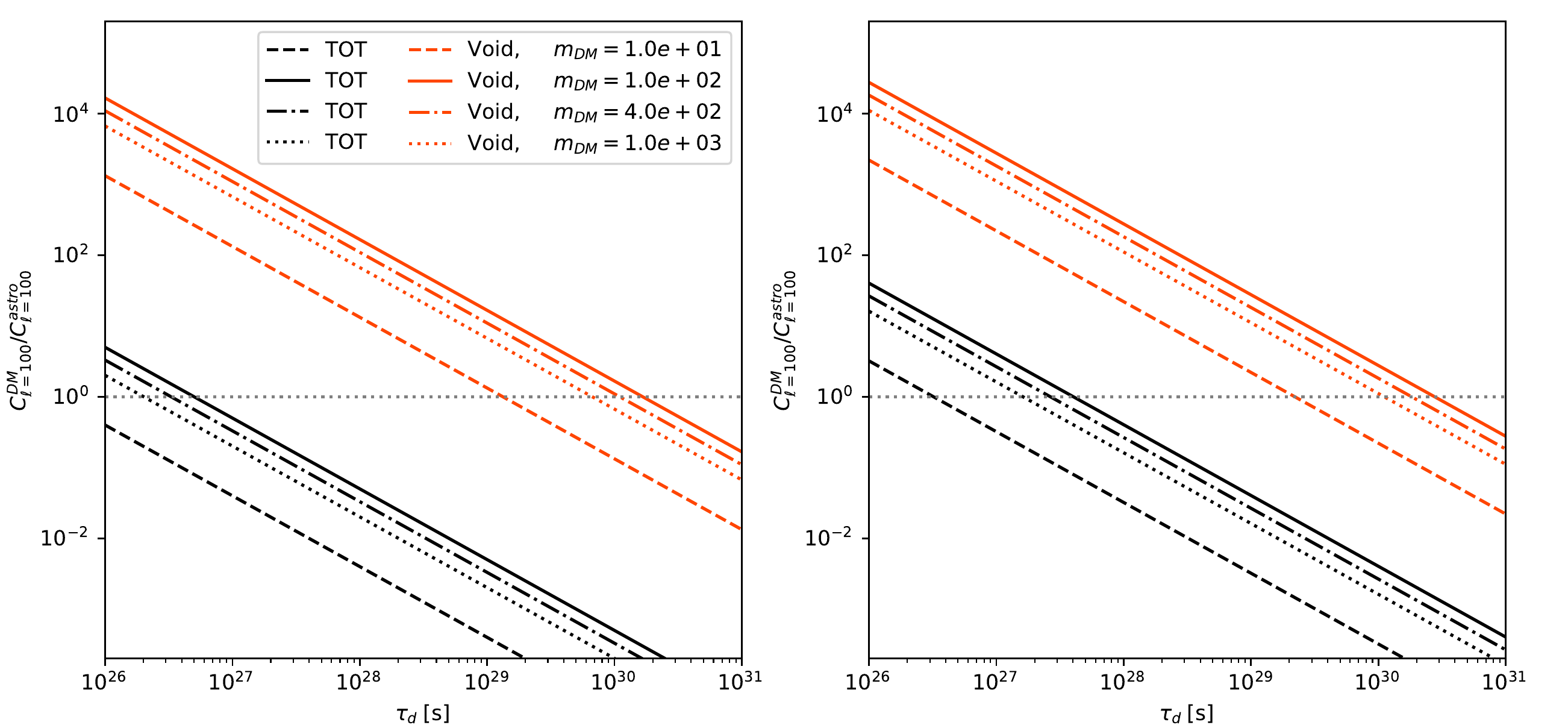}
\hfill
\caption{\label{fig:signal_to_background} Signal-to-background ratio (at $\ell=100$) as a function of the dark matter decay lifetime. In the left panel we show the result for the cross-correlation with cosmic shear, in the right panel that with galaxies. In both cases we report values for four different values of the WIMP mass. The signal-to-background ratio is shown in black for the all-sky cross-correlation (given by the sum of contributions from halos and voids) and in oranges for the correlation signal computed within cosmic voids only.}
\end{figure}
This can be seen by looking at Fig. \ref{fig:signal_to_background}, where we report the ratio between the dark matter signal and the astrophysical background as a function of the decay lifetime and for different WIMP masses. In the left panel we show the result for cosmic shear, in the right panel that of galaxies. The grey lines refer to the total contribution from both halos and voids and lead to a signal-to-background ratio (even significantly) lower than unity, if not for small values of the decay lifetime (basically already outside current bounds). On the contrary, the signal-to-background ratio within voids (orange lines) is way larger than unity also for very large values of the decay lifetime. This "background-free" situation for a relatively large fraction of the dark matter particle parameter space is analogous to the case of the galactic antideuteron signal, where a large signal-to-background ratio is present at low $\bar D$ energies \cite{Donato:2008yx}. These two channel ($\bar D$ and voids cross-correlation) share the feature of being potentially offering promising opportunities, though requiring high sensitivities \cite{Aramaki:2015pii}. 

The two gravitational tracers used here provide similar results. However, Figs. \ref{fig:shearXgamma} and \ref{fig:galXgamma} are helpful to determine which of them might be a better probe for decaying dark matter. In particular, the relative difference between the dark matter signal and that of astrophysical sources in voids, as discussed above, is on average higher for the cross-correlation with the cosmic shear, while being greater for galaxies for central multipoles as shown in Fig. \ref{fig:signal_to_background}. At the same time, the difference between the total signal in halos and in voids is lower for galaxies, especially at low multipoles where is set around one order of magnitude, compared to the two orders of magnitude for the cosmic shear.
\begin{figure}[tbp]
\centering 
\includegraphics[width=.9\textwidth]{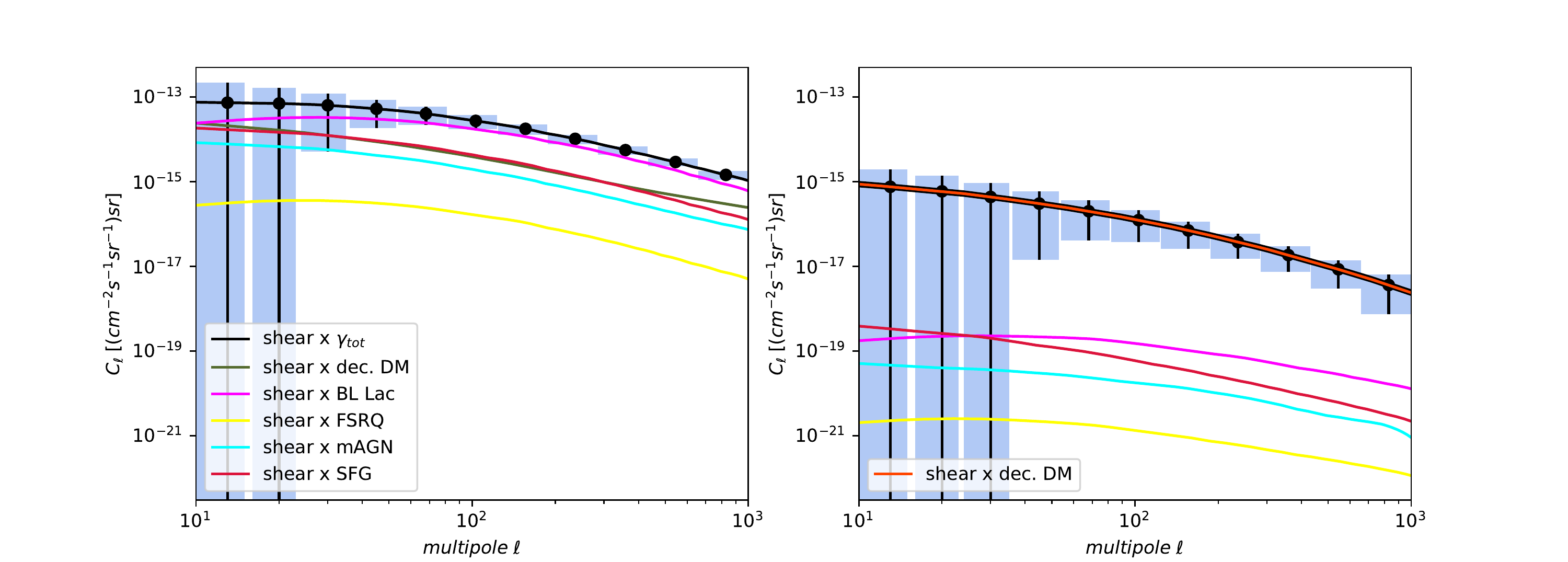}
\hfill
\caption{\label{fig:shearXgamma_fermissimo} The same as Fig. \ref{fig:shearXgamma} using the {\it Fermissimo} configuration to compute error bars.}
\end{figure}
\begin{figure}[tbp]
\centering 
\includegraphics[width=.9\textwidth]{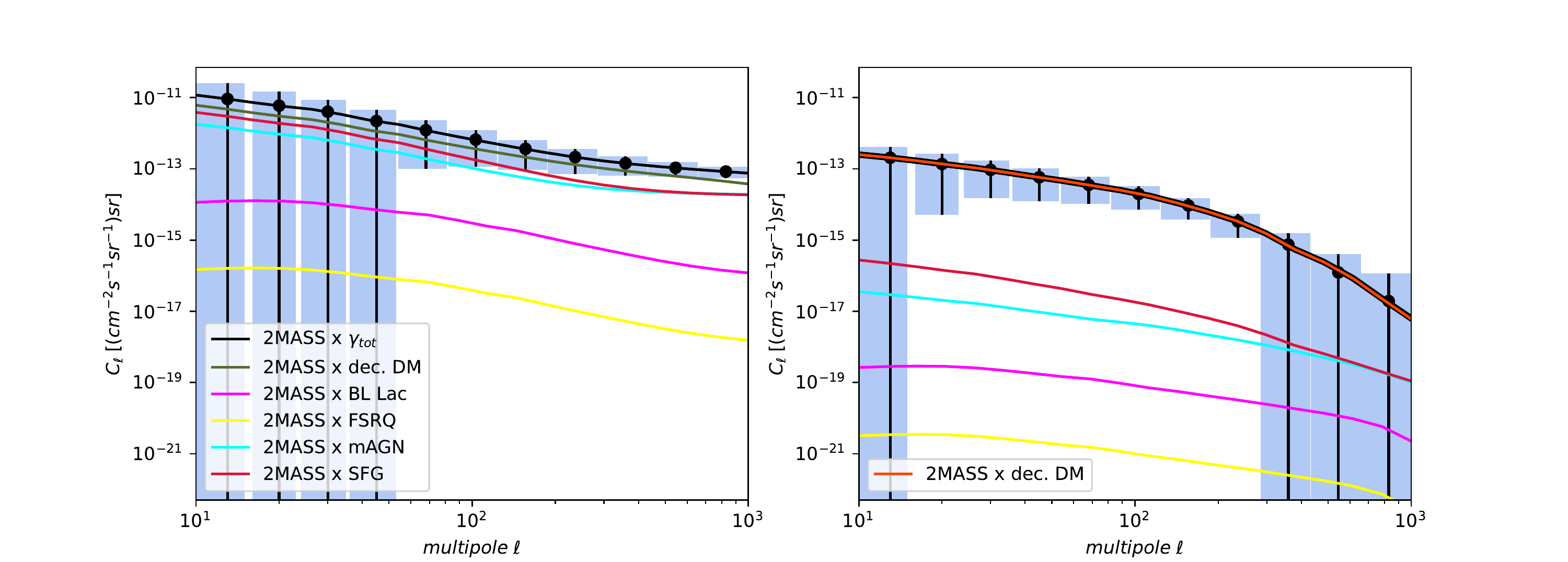}
\hfill
\caption{\label{fig:galXgamma_fermissimo} The same as Fig. \ref{fig:galXgamma} using the {\it Fermissimo} configuration to compute error bars.}
\end{figure}

Concerning detectability of a signal, we need to consider that the errors on the signals computed through Eq. \eqref{eq:CAPSvariance}
with the {\it Fermi}-LAT configuration are large at single multipoles, as well for the individual energy bins of Figs. \ref{fig:shearXgamma} and \ref{fig:galXgamma}. However, we can exploit the full range of multipoles accessible with the {\it Fermi}-LAT PSF, i.e. ranges between about $\ell \simeq 50$ (below which the large-scale galactic foreground is a limiting factor) and $\ell \simeq$ (200$ \div 1000)$ depending on the energy bin \cite{ackermann2018}. This can be traced to the strong domination of the gamma-ray noise/beam term from \emph{Fermi}-LAT, as mentioned above (see Eq. \eqref{eq:errorapprox}). 

In order to investigate the potentiality of the cross-correlation signals to probe dark matter, we consider a future gamma-ray detector with improved specifications, following \cite{pinetti2019}. First, we assume the exposure of the detector to be larger by a factor of 2 as compared with the current \emph{Fermi}-LAT specification here considered. Second, we assume that the detector PSF can be improved, adopting the same behaviour of the beam function expressed in Eq. \eqref{eq:beam} but with a better angular resolution (see \cite{pinetti2019}):
\be
    \sigma_0(E) = \alpha_\sigma\times\sigma_0^\text{Fermi}(E_{\rm ref})\left[\frac{E}{E_{\rm ref}}\right]^{-0.95}+0.001\;\text{deg} \; ,
\ee
for which we assume $\alpha_\sigma=0.2$, for definiteness. Finally, thanks to the better angular resolution, which means smaller mask, we adopt a (somewhat optimistic) larger sky-fraction coverage of $f_\text{sky} = 0.8$, allowing to slightly reduce the impact of noise. Using the same energy bins of Table \ref{tab:fermi}, we scale the noise as the inverse of the exposure, as in Ref. \cite{pinetti2019}. 
We refer to this configuration as \emph{Fermissimo} and show its effect of reducing the variance on the cross-correlation signal in Figs. \ref{fig:shearXgamma_fermissimo} and \ref{fig:galXgamma_fermissimo} (for the cosmic shear and the 2MASS galaxy distribution, respectively), when compared to the \emph{Fermi} results of Figs. \ref{fig:shearXgamma} and \ref{fig:galXgamma}. With the statistical technique discussed below, we find that the fiducial model we are adopting in the analysis ($m_\textrm{DM}=100$ GeV and $\tau_d=3\times10^{27}$ s) would lead to a statistical significance of $5.7\sigma$ in the case of galaxies (being $1.6\sigma$ in the case of cosmic shear). The corresponding numbers for the {\sl Fermi}-LAT configuration are $1.3\sigma$ and $0.6\sigma$, respectively.

Let us in fact now quantify the reach in terms of detectability of a signal or setting bounds on the DM properties that the cross-correlation technique with voids can lead to. To this aim we adopt the Fisher matrix formalism \cite{Jeffrey:1961,Vogeley:1996xu,Tegmark:1996bz}, which is a likelihood-based statistics suitable to forecast the capabilities of future experiments. Let us therefore consider our observable, namely the $C_\ell$'s, modeled through a defined set of free parameters $\theta_{a\,=\,1,\,\dots\,,\,n}$ and with an associated covariance $\Gamma_{\ell\ell'}$; we can write the Fisher matrix (a representation of the covariance matrix for the model parameters, associated to a maximum-likelihood estimate) as:
\be
    F_{ab}=\sum_{\ell\ell'}\frac{\partial C_\ell}{\partial \theta_a} \, \Gamma_{\ell\ell'}^{-1} \, \frac{\partial C_{\ell'}}{\partial \theta_b} \; .
\ee
For any free parameter $\theta_a$, its error can be estimated as $\sigma(\theta_a)=\sqrt{(F^{-1})_{aa}}$ and therefore we can set a bound at $n$-$\sigma$ significance level through:
\be\label{eq:bound}
    \theta_a^\text{bound}=n\times\sigma(\theta_a) \; .
\ee
In our case, we model the total CAPS at fixed WIMP mass with seven free parameters:
\be
\begin{split}
    C_\ell &= p\,\tilde{C}_\ell^\text{DM}+A_\text{1V}\,\tilde{C}_{\ell,\,\text{1V}}^\text{astro}(\Gamma_i)+A_\text{2V}\,\tilde{C}_{\ell,\,\text{2V}}^\text{astro}(\Gamma_i) \; ,\\
    \tilde{C}_{\ell,\,\text{1V/2V}}^\text{astro}(\Gamma_i) &= \sum_\text{i=1}^4 \tilde{C}_{\ell,\,\text{1V/2V}}^i(\Gamma_i) \; ,
\end{split}
\ee
where $p=(3\times10^{27}\;\text{s}/\tau_d)$ is the decay lifetime-normalization with respect to the fiducial cross-correlation for DM computed with $\tau_d=3\times10^{27}$ s and $A_\text{1V}$ and $A_\text{2V}$ are free normalizations for the 1V and 2V contributions to the total astrophysical signal, computed with our fiducial model (see Appendix \ref{app:A}). The latter are given by the sum of the contributions from the four unresolved astrophysical sources (BL Lac, FSRQ, mAGN and SFG), each computed with the fiducial value for the related spectral index $\Gamma_i$ from Table \ref{tab:sourceclass}. The two normalizations account for variations in shape related to the multipoles, while the spectral indexes carry all of the information about the photon's energy dependence of the spectra (see Sec. \ref{sec:W}). With our gaussian theoretical assumptions, the covariance matrix can be assumed to be diagonal: $\Gamma_{\ell\ell'}(E)=\delta_{\ell\ell'}(\Delta C_\ell(E))^2=\delta_{\ell\ell'}\sigma^2$. Therefore the Fisher matrix takes the form:
\be\label{eq:fisher}
F = \begin{pmatrix}
\sum_{\ell,E}\frac{(\tilde{C}_\ell^\text{DM})^2}{\sigma^2} & \sum_{\ell,E}\frac{\tilde{C}_\ell^\text{DM}\tilde{C}_{\ell,\,\text{1V}}^\text{astro}}{\sigma^2} & \cdots\\[3mm]
\sum_{\ell,E}\frac{\tilde{C}_\ell^\text{DM}\tilde{C}_{\ell,\,\text{1V}}^\text{astro}}{\sigma^2} & \sum_{\ell,E}\frac{(\tilde{C}_{\ell,\,\text{1V}}^\text{astro})^2}{\sigma^2} & \cdots\\[3mm]
\vdots & \vdots & \ddots\\[3mm]
\end{pmatrix} \; ,
\ee
where the sums are over all considered multipoles and the energy bins of Table \ref{tab:fermi}. It is a 7x7 symmetric square matrix since we are considering seven parameters: the DM normalization $p$, the two astrophysical normalizations $A_\text{1V}$ and $A_\text{2V}$ and the four spectral indexes $\Gamma_\text{BL}$, $\Gamma_\text{FSRQ}$, $\Gamma_\text{mAGN}$ and $\Gamma_\text{SFG}$. From Eq. \eqref{eq:bound} we can then compute the bound on $p$ and consequently on the decay lifetime $\tau_d$, with respect to varying values of the WIMP mass. The Fisher formalism allows to efficiently consider all uncertainties on the chosen models and, in particular, to consider the impact of the astrophysical background noise on the detection of the dark matter signal.
\begin{figure}[tbp]
\centering 
\includegraphics[width=.99\textwidth]{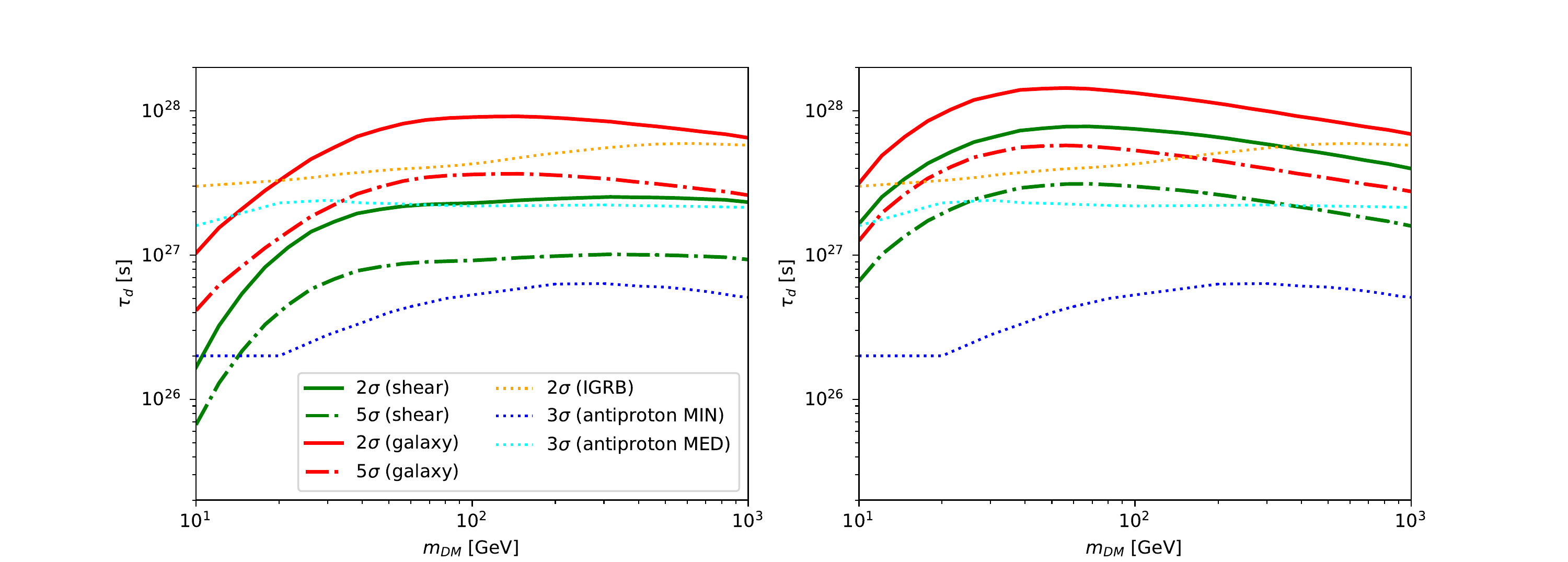}
\hfill
\caption{\label{fig:bound} Forecast of the bounds on the DM properties (decay lifetime $\tau_d$ vs. WIMP mass $m_{DM}$) attainable through the study of the cross-correlation between the unresolved gamma-ray sky and either cosmic shear (green) or the galaxy distribution (red) within cosmic voids. We show both the $2\sigma$ bound and the $5\sigma$ detection reach, considering a highly efficient gamma-ray detector, namely {\it Fermissimo}. We also report previously derived bounds from cosmic-ray antiprotons \cite{fornengo2014b} and from the isotropic gamma-ray background \cite{Blanco:2018esa}. (\emph{Left}): conservative bound. (\emph{Right}): the bound accounts for priors on the non-DM parameters; $\pm\;60\%$ for astrophysical normalizations and $\pm\;0.4$ for spectral indexes.}
\end{figure}

In Fig. \ref{fig:bound} we show the $2\sigma$ (95\% C.L. bound reach) and $5\sigma$ (detection limit) reach for the {\it Fermissimo} configuration, attainable with voids (cosmic shear in green and galaxies in red). We compare these forecasts with bounds obtained with cosmic antiprotons \cite{fornengo2014b} and the isotropic gamma-ray background (IGRB) \cite{Blanco:2018esa}. In the left panel we report the bound for the marginalized case (conservative). In the right panel we show the bounds obtained by accounting for priors on the non-DM parameters (determined from the measured uncertainties on the fiducial model parameters): for the astrophysical normalizations $A_\text{1V}$ and $A_\text{2V}$ we consider variations of $\pm\;60\%$, while for spectral indexes we take a prior of $\pm\,0.4$ on the fiducial values of $\Gamma_i$.  In general, the galaxies would allow to reach tighter bounds than cosmic shear. The forecasted bounds are competitive with previous works only for galaxies in the conservative scenario and for both galaxies and shear in the non-conservative scenario, when accounting for priors on the parameters. In particular, the current bounds could be improved (in the case of galaxies) for $m_\textrm{DM}\sim25\div900$ GeV in the conservative case (and in the whole range of considered masses for the non-conservative case), while being generally worse (conservative) or slightly worse (non-conservative) for cosmic shear. Let us notice that the drop in sensitivity toward lower masses is due to the fact that we are considering photon energies starting from 500 MeV (to conform to the specifications of Ref. \cite{ackermann2018}, from which we adopt relevant information, like the photon noise, see Table \ref{tab:fermi}): by decreasing the lower energy around 100 MeV would allow us to improve the forecasted bounds for lighter dark matter, making the lines flatter, since an energy threshold at 500 MeV cuts out a relevant fraction of the signal for dark matter masses close to 10 GeV, especially for the
$b \bar b$ decay channel we are considering here. The detector PSF at those energies would be worse, though.

In conclusion, although voids lead to a much weaker cross-correlation signal than halos, their signal-to-background ratio is much larger than 1, for decay lifetimes up to $2 \times 10^{30}$ s. This represents a situation in which a detection would point toward a dark matter interpretation, contrary to the case of a detection in halos, for which the signal-to-background ratio is unfavourable. The signal in voids is nevertheless difficult to achieve and requires improved sensitivities of gamma-ray detectors: with improved specifications like those discussed in the text, the reach on bounds on the dark matter lifetime can exceed current bounds obtained through different techniques in a mass range between 25 GeV and 900 GeV.

\section{Conclusion}\label{sec:conclusion}
In this paper we have discussed the idea of using cosmic voids as probes for particle dark matter. We considered the cosmic shear and the galaxy distribution as gravitational tracers of the matter distribution in the Universe to investigate the nature of the unresolved gamma-ray background, which can hide a signal due to dark matter annihilation or decay. Large-scale structures are responsible for the bending of light in the weak lensing regime as well as for gamma-ray emission either from unresolved astrophysical sources or particle dark matter annihilations and decays. At the same time, the distribution of galaxies is tightly connected to the distribution of dark matter as well as to that of astrophysical sources hosted by them. We thus expect a positive correlation between these gravitational tracers and the unresolved gamma-ray emission.

We have therefore computed the cross-correlation angular power spectrum between the fluctuations due to the inhomogeneous distribution of matter in the Universe, traced by either the cosmic shear or the galaxy distribution, and those induced in the unresolved gamma-ray emission, by both astrophysical sources and dark matter annihilations/decays. We separated the signals into the contributions coming from voids and halos, respectively, to investigate the main differences among their properties within the two types of structures. We show that in voids, the cross-correlation signal generated by dark matter annihilation is strongly disfavored with respect to dark matter decay, as it depends on the square of the structure's density, leading to a much greater difference between the signal in halos and the signal in voids than the case of decaying dark matter. 

On the other hand, in the case of voids and for decaying dark matter, we find that the dark matter cross-correlation signal can largely exceed the astrophysical counterpart, for an interestingly large section of the dark matter parameter space. For decay lifetimes up to $2 \times 10^{30}$ s the signal-to-background ratio is  (even significantly) larger than 1. This is at variance with the cross-correlation from halos, where the astrophysical signal dominates over the dark matter signal. This makes the signal from voids a potentially "background-free" option for dark matter searches for decay lifetimes about 3 orders of magnitude larger than current bounds.

The size of signal is nevertheless small, which makes observational opportunities to require next generation detectors. We found in fact that the combination of forthcoming galaxy surveys, such as Euclid, and the gamma-ray \emph{Fermi}-LAT telescope is not able yet to detect the signal. However, potential detectability can be achieved by considering an improved gamma-ray detector, with better angular resolution and slightly larger exposure: we assumed a factor 2 larger exposure as compared to the {\it Fermi}-LAT, and an angular resolution a factor of 5 better than  {\it Fermi}-LAT. In this case, the cross-correlation which uses galaxies as a dark matter tracer, would reach a 5.7$\sigma$ stastistical significance for the fiducial dark matter models we adopted in the analysis ($m_\text{DM}\sim100\;\text{GeV}$ and $\tau_d=3\times 10^{27}\;\text{s}$) and will be able to improve on current bounds on the dark matter decay lifetime for masses in the range $m_\textrm{DM}\sim25\div900$ GeV.

\acknowledgments
We acknowledge support from: {\sl Departments of Excellence} grant awarded by the Italian Ministry of Education, University and Research (MIUR); Research grant {\sl The Dark Universe: A Synergic Multimessenger Approach}, Grant No. 2017X7X85K funded by the Italian Ministry of Education, University and Research (MIUR); {\sl Research grant The Anisotropic Dark Universe}, Grant No. CSTO161409, funded by Compagnia di Sanpaolo and University of Torino; Research grant {\sl TAsP (Theoretical Astroparticle Physics)} funded by Istituto Nazionale di Fisica Nucleare (INFN); {\sl Fermi Research Alliance, LLC} under Contract No. DE-AC02-07CH11359 with the U.S. Department of Energy, Office of High Energy Physics.

\appendix
\section{Gamma-ray luminosity functions}
\label{app:A}
The astrophysical sources are characterised by their GLF $\phi_\gamma = \de N_{\gamma}/(\de L \, \de V)$, which specifies the number of gamma-ray sources per unit of luminosity $L$ and comoving volume $V$.
In this appendix we provide the GLFs for each astrophysical class considered in our analysis. 

\subsection{Blazars}

Following the LDDE model of Ref. \cite{ajello2014}, the GLF of BL Lacs and FSRQs can be parameterized as a broken power-law in luminosity and redshift:

\begin{align}
\phi_\gamma
\left(L, \, z \right) = \dfrac{A}{\ln (10) \, L} \left[ \left(\dfrac{L}{L_\star} \right)^{\gamma_1} + \left(\dfrac{L}{L_\star} \right)^{\gamma_2} \right]^{-1} \times \left[\left(\dfrac{1+z}{1 + z_c(L)} \right)^{-p_1} + \left(\dfrac{1+z}{1+z_c(L)} \right)^{-p_2} \right]^{-1} \, ,
\end{align}
where $z_c = z_\star \, (L/10^{48} \textrm{erg s}^{-1})^\beta$ and the other parameters are specified in Table \ref{tab:blazar_par}.

\begin{table}[t]
\centering
\begin{tabular}{l|cccccccc}
\hline\rule{0mm}{5mm}
~& $A$ [Mpc$^{-3}$] & $ L_\star$ [erg s$^{-1}$] & $\gamma_1$ & $\gamma_2$ & $p_1$ & $p_2$ & $z_\star$ & $\beta$ \\[1mm]
%\hline\hline
\hline\rule{0mm}{5mm}BL Lacs & $9.20 \times 10^{-11}$ & $2.43 \times 10^{48}$  & $1.12$ & $3.71$ & $4.50$ & $-12.88$ & $1.67$ & $4.46 \times 10^{-2}$ \\[1mm]
FSRQs & $3.06 \times 10^{-9}$ & $0.84 \times 10^{48}$ & $0.21$ & $1.58$ & $7.35$ & $-6.51$ & $1.47$ & $0.21$ \\[1mm]
\hline
\end{tabular}
\caption{Parameters of the gamma-ray luminosity function for BL Lacs and FSRQs.}
\label{tab:blazar_par}
\end{table}

\subsection{Misaligned active galactic nuclei}

\begin{table}[t]
\centering
\begin{tabular}{ccccccccc}
\hline\rule{0mm}{5mm}
$\beta_l$ & $ L_{l \star}$ [W/Hz] & $\rho_{l\star}$ [Mpc$^{-3}$] & $z_{l\star}$ & $k_l$ & $\beta_h$ &  $ L_{h \star}$ [W/Hz] & $\rho_{h}$ [Mpc$^{-3}$] & $z_{h \star}$ \\[1mm]
\hline
\rule{0mm}{5mm}$0.586$ & $10^{26.48}$  & $10^{-7.523}$ & $0.71$ & $3.48$ & $2.42$ & $10^{27.39}$ & $10^{-6.757}$ & $2.03$ \\[1mm]
\hline
\end{tabular}
\caption{Parameters of the radio luminosity function for mAGN.}
\label{tab:mAGN_par}
\end{table}

The GLF for mAGN can be derived from the radio luminosity function (RLF) through
\be 
\phi_\gamma (L, \, z) = \dfrac{k \, \eta}{(1+z)^{2- \Gamma}} \, \dfrac{1}{\ln(10) \; L^\drm{151 MHz}_\drm{tot}} \dfrac{\de L^\drm{151 MHz}_\drm{tot}}{\de L} \; \rho_r (L^\drm{151 MHz}_\drm{tot} (L),\,z) \; ,
\label{eq:GLF_radio}
\ee 
where $k = 3.05$, $\Gamma=2.37$ and
\be 
\eta = \dfrac{\de^2 V_W / \de z \de \Omega}{\de^2 V / \de z \de \Omega} \; .
\ee 
The comoving volume $\dfrac{\de^2 V_W}{\de z \, \de \Omega}$ used by \cite{willot2001} and the one in the standard $\Lambda$CDM cosmology $\dfrac{\de^2 V}{\de z \, \de \Omega}$ are 
\begin{eqnarray} 
\dfrac{\de^2 V_W}{\de z \, \de \Omega} &=& \dfrac{c^3 \, z^2 \, (2+z)^2}{4 \, H^3_{0, W} \, (1+z)^3} \; , \\[3mm]
\dfrac{\de^2 V}{\de z \, \de \Omega} &=& \dfrac{c \, d^2_L(z)}{H_0 \, (1+z)^2 \sqrt{(1-\Omega_\Lambda - \Omega_m)(1+z)^2 + (1+z)^3 \Omega_m + \Omega_\Lambda}} \; .
\end{eqnarray}
with $H_{0, W} = 50$ $\drm{km}\,\drm{s}^{-1}\,\drm{Mpc}^{-1}$.
The relation between core radio luminosity and gamma-ray luminosity is provided in \cite{mauro2014}, while \cite{Lara:2004ee}  derived the correlation between core and total luminosities:
\begin{eqnarray}
    \log L &=& 2 + 1.008 \log L_{r, \drm{core}} \\[3mm]
    \log L^\drm{5 GHz}_{r, \drm{core}} &=& 4.2 + 0.77 \log L^\drm{1.4 GHz}_{r, \drm{tot}} \; .
\end{eqnarray}
The reference radio frequency in Eq. \eqref{eq:GLF_radio} is 151 MHz thus, following \cite{Inoue:2011bm}, we consider the power-law scaling  %
\be 
\dfrac{L_r}{\nu} \propto v^{-\alpha_{r}}
\ee 
with $\alpha_r=0.80$.
The radio luminosity function \cite{willot2001} can be expressed as the sum of two components
\be 
\rho_r (L_r,z)  = \rho_l  (L_r,z) + \rho_h  (L_r,z) \; ,
\label{eq:rho_radio}
\ee
where
\begin{align}
    \begin{cases}
    \rho_l = \rho_{l\star} \left(\dfrac{L_r}{L_{l \star}} \right)^{-\beta_l} \, \exp \left(-\dfrac{L_r}{L_{l \star}} \right) \, \left(1+z \right)^{k_l} \quad \quad \; \; &\drm{for} \quad z < z_{l \star} \\[5mm]
    \rho_l = \rho_{l\star} \left(\dfrac{L_r}{L_{l \star}} \right)^{-\beta_l} \, \exp \left(-\dfrac{L_r}{L_{l \star}} \right) \, \left(1+z_{l \star} \right)^{k_l} \quad &\drm{for} \quad z \geq z_{l \star}
    \end{cases}
    \label{eq:rho_l}
\end{align}
and 
\be
\rho_h = \rho_{h \star} \left(\dfrac{L_r}{L_{h\star}} \right)^{-\beta_h} \exp \left(- \dfrac{L_{h \star}}{L} \right) \, \exp \left \{ - \dfrac{1}{2} \left(\dfrac{z-z_{h \star}}{z_{h0}} \right)^2 \right \} \, .
\label{eq:rho_h}
\ee 
For $z< z_{h \star}$ we adopted $z_{h0}=0.568$, while for $z \geq z_{h \star}$ we used $z_{h0}=0.956$. All the parameters included in Eqs.\eqref{eq:rho_l} and \eqref{eq:rho_h}  are specified in Table \ref{tab:mAGN_par}.

\subsection{Star-forming galaxies}
From the infrared luminosity function (ILF) $\phi_\drm{IR} = \dfrac{\de N_\drm{IR}}{\de \log_{10}(L_\drm{IR}) \, \de V}$ we can obtain the GLF of SFGs:
\be 
\phi_\gamma(L, \, z) = \phi_\drm{IR} \, \dfrac{\de \log_{10} (L_{8-1000 \, \mu\drm{m}})}{\de L_{0.1-100 \, \drm{GeV}}} \, .
\ee 
The luminosity $L_{0.1-100 \, \drm{GeV}}$ between 0.1 GeV and 100 GeV and the luminosity $L_{8-1000 \, \mu\drm{m}}$ between 8 $\mu$m and 1000 $\mu$m are related via \cite{Fermi-LAT:2012nqz}  
\be 
\log_{10} \, \left(\dfrac{L_{0.1-100 \drm{GeV}}}{\drm{erg s}^{-1}} \right) = \alpha_\drm{IR} \; \log_{10} \, \left(\dfrac{L_{8-1000 \, \mu\drm{m}}}{10^{10} \, L_\odot} \right) + \beta_\drm{IR}
\label{eq:L_IR}
\ee 
with coefficients $\alpha_\drm{IR} = 1.09$ and $\beta_\drm{IR} = 39.19$. Following \cite{gruppioni2013}, the ILF can be written as the sum of quiescent spiral galaxies, starburst galaxies and SFG hosting a concealed or low-luminosity AGN:
\be 
\phi_\drm{IR} = \phi_\drm{spiral} + \phi_\drm{starburst} + \phi_\drm{SF-AGN} \; .
\label{eq:phi_IR}
\ee 
The ILF of each sub-class can be modelled as
\be 
\phi_i = \phi_{0, \, i} (z) \left(\dfrac{L_{8-1000 \, \mu\drm{m}}}{L_{0, \, i}} \right)^{1- \gamma_i} \, \exp \left[- \dfrac{1}{2 \sigma^2_i} \log^2_{10} \left(1 + \dfrac{L_{8-1000 \, \mu\drm{m}}}{L_{0, \, i}} \right)\right] \; ,
\label{eq:phi_SFG}
\ee 
where $i$ = \{\textrm{spiral, starburst, SF-AGN}\}. The radio luminosity can be written as a function of the gamma-ray luminosity using Eq. \eqref{eq:L_IR} and the normalization $\phi_{0, i}$ reads
\be
\phi_{0, i} = \begin{cases}
\phi_{\star, \, {i}} \left(\dfrac{1+z}{1.15} \right)^{k_{R1, \, i}} & \drm{for} \;  z \leq z_{\star, i} \\[3mm]
\phi_{\star, \, {j}} \left(\dfrac{1+z_{\star, i}}{1.15} \right)^{k_{R1, \, i}} \, \left(\dfrac{1+z}{1+z_{\star,i}} \right)^{k_{R2}, \, \drm{i}} & \drm{for} \;  z >  z_{\star, i} \, .
\label{eq:phi0_SFG}
\end{cases}
\ee
All the parameters in Eqs. \eqref{eq:phi_SFG} and \eqref{eq:phi0_SFG} are specified in Table \ref{tab:SFG_par}.

\begin{table}[t]
\centering
\begin{tabular}{l|cccccccc}
\hline\rule{0mm}{5mm}%\\[0.2ex]
~& $\gamma$ & $ \sigma $ & $\log_{10} (L_\star/L_\odot)$ & $\log_{10} (\phi_\star/ \drm{Mpc}^{-3})$ & $k_L$ & $k_{R1}$ & $k_{R2}$ & $z_\star$\\[1mm]
\hline %\\[0.1mm]
\rule{0mm}{5mm}spiral & $1.0$ & $0.50$  & $9.78$ & $-2.12$ & $4.49$ & $-0.54$ & $-7.13$ & $0.53$\\[1mm]
starburst & $1.0$ & $0.35$ & $11.17$ & $-4.46$ & $1.96$ & $\, 3.79$ & $-1.06$ & 1.1 \\[1mm]
SF-AGN & $1.2$ & $0.40$ & $10.80$ & $-3.20$ & $3.17$ & $\, 0.67$ & $3.17$ & $1.1$ \\[1mm]
\hline
\end{tabular}
\caption{Parameters of the infrared luminosity function for the three sub-classes of SFGs.}
\label{tab:SFG_par}
\end{table}

\begin{figure}[tbp]
\centering 
\includegraphics[width=.9\textwidth]{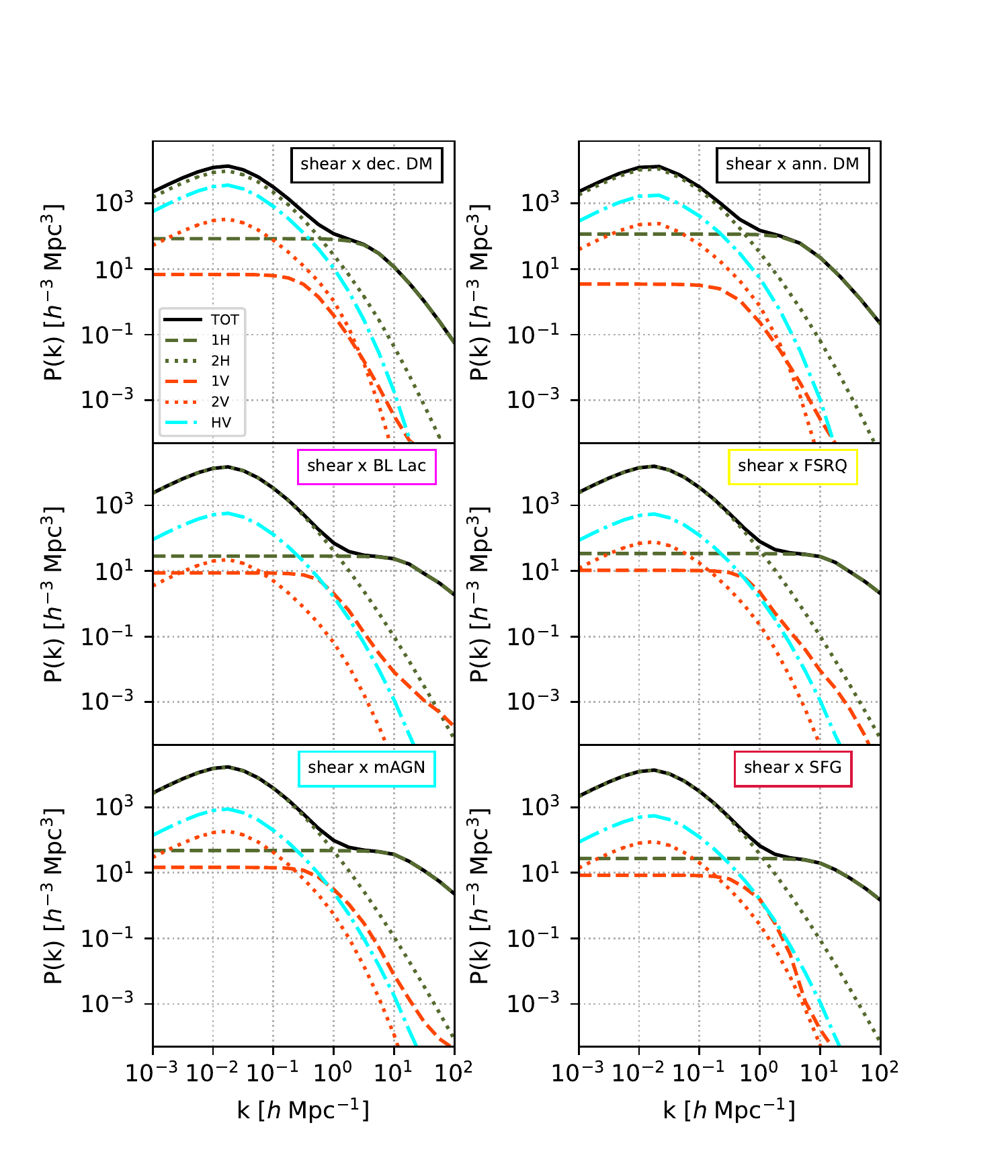}
\hfill
\caption{\label{fig:shearterms} 3D power spectrum of the cross-correlation between cosmic shear and gamma-ray emitters (decaying/annihilating DM and unresolved astrophysical sources), at $z=0.5$, divided into the contributions from a single halo (1H), two halos (2H), a single void (1V), two voids (2V) and an halo and a void (HV).}
\end{figure}
\begin{figure}[tbp]
\centering 
\includegraphics[width=.9\textwidth]{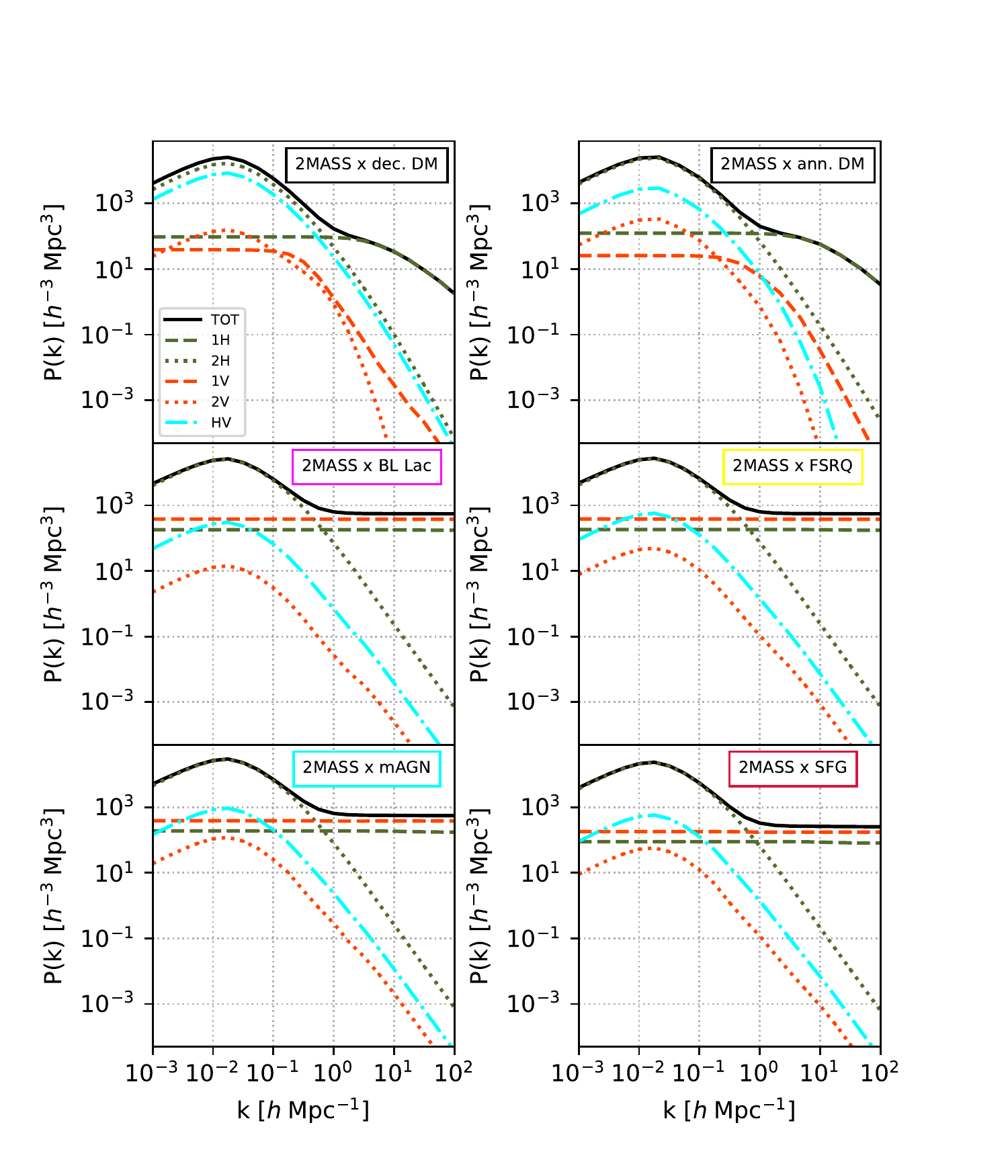}
\hfill
\caption{\label{fig:galterms} The same as Fig. \ref{fig:shearterms} for the cross-correlation of gamma-ray emitters with the 2MASS galaxy catalog.}
\end{figure}

\section{Mass-to-luminosity relations}
\label{app:B}
The astrophysical sources are better characterised by their luminosity, as opposed to the DM halos which are better characterised by the halo mass. Therefore, the power spectra involving the astrophysical components include an integral over the luminosity and the relation $M(L)$ between the luminosity of the astrophysical class and the host DM halo is required. Following \cite{Camera:2014rja}, the mass-to-luminosity relations for blazars (BL Lacs and FSRQs), mAGN and SFGs read
\begin{eqnarray}
M_i(L) &=& 10^{13} \, M_\odot \; \left(\dfrac{M_{\star, i}}{10^{8.8} \, (1+z)^{1.4}} \right)^{0.645} \quad i = \drm{blazar, mAGN} \\[3mm]
M_\drm{SFG}(L) &=& \dfrac{10^{12} \, M_\odot}{(1+z)^{1.61}} \, \left(\dfrac{L}{6.8 \cdot 10^{39} \drm{ erg/s}} \right)^{0.92} \; ,
\end{eqnarray}
with
\begin{eqnarray}
M_{\star, \drm{blazar}} &=& 10^9 \, \left(\dfrac{L}{10^{48} \drm{ erg/s}}\right)^{0.36} \; \\[3mm]
M_{\star, \drm{mAGN}} &=& 4.6 \cdot 10^9 \, \left(\dfrac{L}{10^{48}\drm{ erg/s}}\right)^{0.16} \; .
\end{eqnarray}

\section{Terms of the 3D power spectrum and variations on the void profile}\label{app:C}

When computing the 3D power spectrum within the HVM, three additional terms, with respect to the HM, arise: the correlation between points within the same void (1Void term), within two different voids (2Void term) and between points belonging to a halo and a void, respectively (Halo-Void term). In Figs. \ref{fig:shearterms} and \ref{fig:galterms} we show the 3D power spectrum of the cross-correlation between gamma-ray emitters (decaying/annihilating DM, BL Lacs, FSRQs, mAGN and SFGs) and either cosmic shear or the 2MASS galaxy catalog, respectively (calculated at $z=0.5$). For each panel of the figures, we report the contributions from the terms of Eqs. \eqref{eq:1H}-\eqref{eq:HV}. In general, the linear power on large scales is well recovered and the transition between linear and non-linear scales (around $k\sim1\;h\,\text{Mpc}^{-1}$) is clearly visible in all cases.

Note how the 1Halo and 1Void terms in the cross-correlation with galaxies (Fig. \ref{fig:galterms}) act as a shot noise, since the integrand of Eqs. \eqref{eq:1H} and \eqref{eq:1V} are fairly constant with the scale $k$, for the relevant source functions.

\begin{figure}[tbp]
\centering 
\includegraphics[width=.9\textwidth]{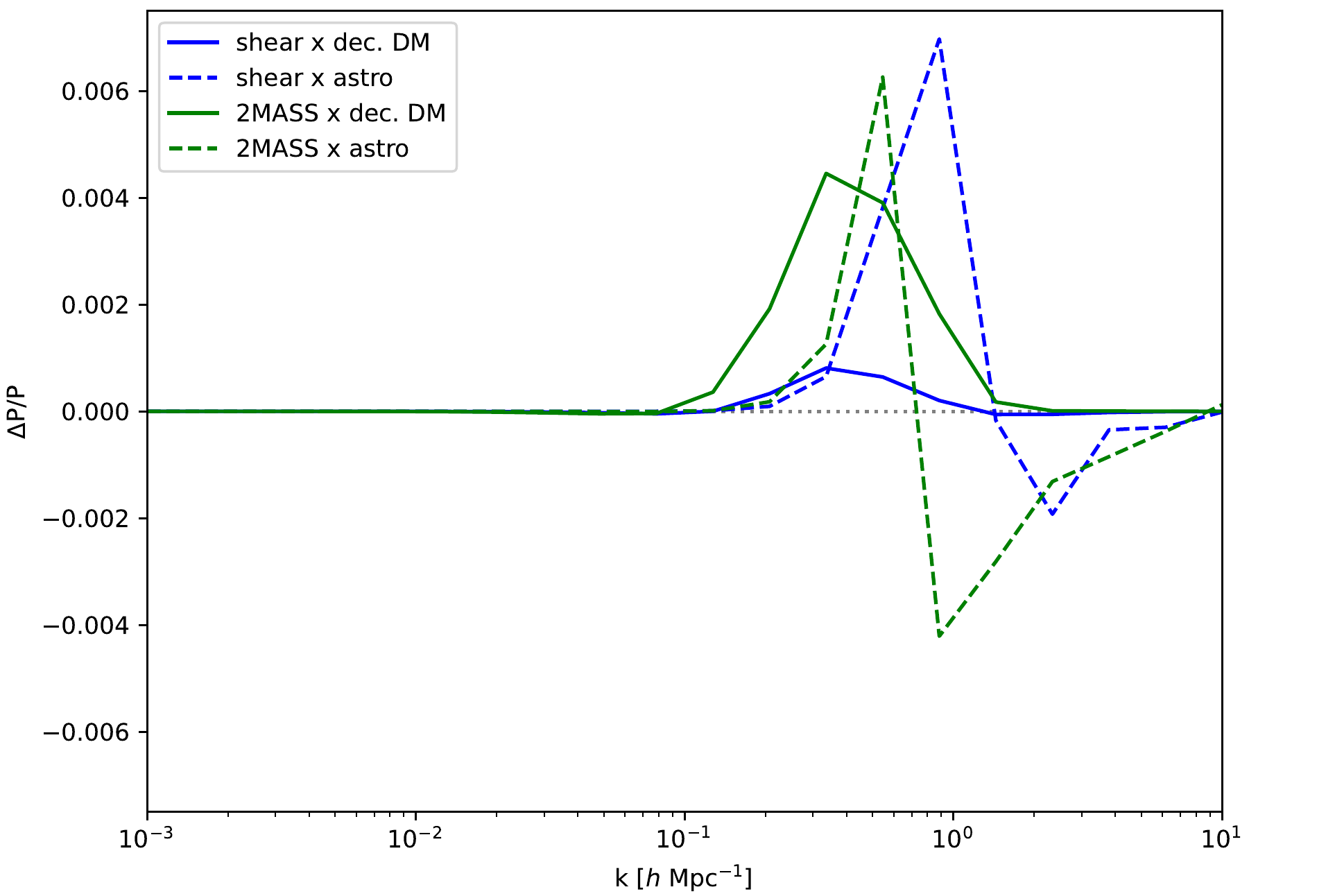}
\hfill
\caption{\label{fig:tanh} Relative variations on the 3D power spectrum upon using the HSW profile \cite{hamaus2014} or the tanh-profile \cite{voivodic2020}. We show them for the cross-correlation of the 2MASS galaxy catalog (green) or the cosmic shear (blue) with the decaying dark matter signal (solid) and the total unresolved astrophysical signal (dashed). Differences are confined well below the few per mille level.}
\end{figure}

The 3D power spectrum depends on the considered structure's profile for all cross-correlations used in this work (see Sec. \ref{sec:PS}). However, while the halo density profile has been studied for a long time and the NFW profile is typically considered the standard choice, the void density profile is debated in the literature and strongly depends on the used void finder (as discussed in Sec. \ref{sec:profile}). Here, we considered both voids with a radius-dependent central density, showing compensation walls, and voids with empty centers and no compensation walls. The former are well reproduced by the Hamaus-Sutter-Wandelt profile \cite{hamaus2014} of Eq. \eqref{eq:HSW} and the latter by the profile proposed in Voivodic et al. \cite{voivodic2020}, shown in Eq. \eqref{eq:tanh}. Despite having different properties, Fig. \ref{fig:tanh} shows how the choice of the void profile leads to sub-percent variations of the 3D power spectrum, for all cross-correlations.

\section{DM and non-DM parameters' correlation in the Fisher analysis}\label{app:D}

\begin{figure}[tbp]
\centering 
\includegraphics[width=.95\textwidth]{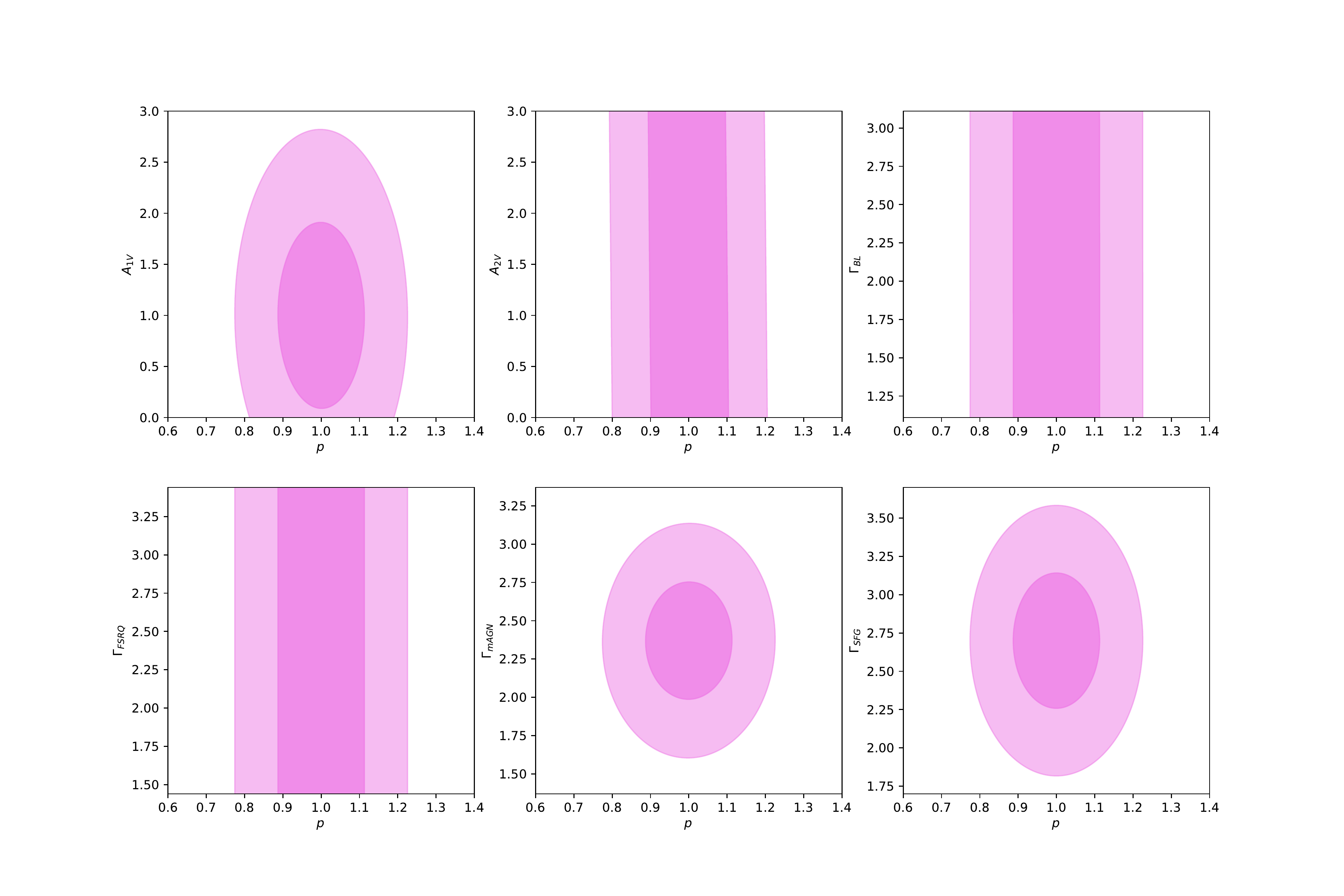}
\hfill
\caption{\label{fig:contour} Bivariate correlations between the DM free parameter of the Fisher matrix analysis $p$ and the six non-DM parameters. All are referred to a WIMP with $m_\text{DM} = 100$ GeV.}
\end{figure}

In Sec. \ref{sec:results} we showed the forecasted bound on the DM decay lifetime, computed through a Fisher matrix analysis with seven parameters: one normalization for the DM signal, two normalizations for the astrophysical signal and the four spectral indexes of unresolved astrophysical sources, accounting for energy-dependent variations of the $C_\ell$'s. The Fisher matrix in Eq. \eqref{eq:fisher} can be seen as the inverse covariance of the model's free parameters. Thus, it can be used to perform a multivariate study on the degeneration of the parameter. In particular, we can consider the six submatrixes related to the correlations between the DM parameter $p$ and each of the non-DM parameters $A_\text{1V}$, $A_\text{2V}$, $\Gamma_\text{BL}$, $\Gamma_\text{FSRQ}$, $\Gamma_\text{mAGN}$ and $\Gamma_\text{SFG}$. In Fig. \ref{fig:contour} we show the bivariate correlations for a WIMP with a mass of 100 GeV. In general, no strong correlation is present and the non-DM parameters have little impact on the constraining power of $p$ (and consequently on that of the DM decay lifetime).

\bibliographystyle{JHEP}
\bibliography{ref}

\providecommand{\href}[2]{#2}\begingroup\raggedright\begin{thebibliography}{100}

\bibitem{camera2013}
S.~Camera, M.~Fornasa, N.~Fornengo and M.~Regis, \emph{{A Novel Approach in the
  Weakly Interacting Massive Particle Quest: Cross-correlation of Gamma-Ray
  Anisotropies and Cosmic Shear}},
  \href{https://doi.org/10.1088/2041-8205/771/1/L5}{\emph{Astrophys. J. Lett.}
  {\bfseries 771} (2013) L5} [\href{https://arxiv.org/abs/1212.5018}{{\ttfamily
  1212.5018}}].

\bibitem{Camera:2014rja}
S.~Camera, M.~Fornasa, N.~Fornengo and M.~Regis, \emph{{Tomographic-spectral
  approach for dark matter detection in the cross-correlation between cosmic
  shear and diffuse $\gamma$-ray emission}},
  \href{https://doi.org/10.1088/1475-7516/2015/06/029}{\emph{JCAP} {\bfseries
  06} (2015) 029} [\href{https://arxiv.org/abs/1411.4651}{{\ttfamily
  1411.4651}}].

\bibitem{fornengo2014}
N.~Fornengo and M.~Regis, \emph{{Particle dark matter searches in the
  anisotropic sky}},
  \href{https://doi.org/10.3389/fphy.2014.00006}{\emph{Front. Physics}
  {\bfseries 2} (2014) 6} [\href{https://arxiv.org/abs/1312.4835}{{\ttfamily
  1312.4835}}].

\bibitem{xia2015}
J.-Q.~Xia, A.~Cuoco, E.~Branchini and M.~Viel, \emph{{Tomography of the
  Fermi-lat $\gamma$-ray Diffuse Extragalactic Signal via Cross Correlations
  With Galaxy Catalogs}},
  \href{https://doi.org/10.1088/0067-0049/217/1/15}{\emph{Astrophys. J. Suppl.}
  {\bfseries 217} (2015) 15}
  [\href{https://arxiv.org/abs/1503.05918}{{\ttfamily 1503.05918}}].

\bibitem{cuoco2015}
A.~Cuoco, J.-Q.~Xia, M.~Regis, E.~Branchini, N.~Fornengo and M.~Viel,
  \emph{{Dark Matter Searches in the Gamma-ray Extragalactic Background via
  Cross-correlations With Galaxy Catalogs}},
  \href{https://doi.org/10.1088/0067-0049/221/2/29}{\emph{Astrophys. J. Suppl.}
  {\bfseries 221} (2015) 29}
  [\href{https://arxiv.org/abs/1506.01030}{{\ttfamily 1506.01030}}].

\bibitem{fornasa2016}
M.~Fornasa et~al., \emph{{Angular power spectrum of the diffuse gamma-ray
  emission as measured by the Fermi Large Area Telescope and constraints on its
  dark matter interpretation}},
  \href{https://doi.org/10.1103/PhysRevD.94.123005}{\emph{Phys. Rev. D}
  {\bfseries 94} (2016) 123005}
  [\href{https://arxiv.org/abs/1608.07289}{{\ttfamily 1608.07289}}].

\bibitem{regis2015}
M.~Regis, J.-Q.~Xia, A.~Cuoco, E.~Branchini, N.~Fornengo and M.~Viel,
  \emph{{Particle dark matter searches outside the Local Group}},
  \href{https://doi.org/10.1103/PhysRevLett.114.241301}{\emph{Phys. Rev. Lett.}
  {\bfseries 114} (2015) 241301}
  [\href{https://arxiv.org/abs/1503.05922}{{\ttfamily 1503.05922}}].

\bibitem{xia2011}
J.-Q.~Xia, A.~Cuoco, E.~Branchini, M.~Fornasa and M.~Viel, \emph{{A
  cross-correlation study of the Fermi-LAT $\gamma$-ray diffuse extragalactic
  signal}}, \href{https://doi.org/10.1111/j.1365-2966.2011.19200.x}{\emph{Mon.
  Not. Roy. Astron. Soc.} {\bfseries 416} (2011) 2247}
  [\href{https://arxiv.org/abs/1103.4861}{{\ttfamily 1103.4861}}].

\bibitem{shirasaki2015}
M.~Shirasaki, S.~Horiuchi and N.~Yoshida, \emph{{Cross-Correlation of the
  Extragalactic Gamma-ray Background with Luminous Red Galaxies}},
  \href{https://doi.org/10.1103/PhysRevD.92.123540}{\emph{Phys. Rev. D}
  {\bfseries 92} (2015) 123540}
  [\href{https://arxiv.org/abs/1511.07092}{{\ttfamily 1511.07092}}].

\bibitem{cuoco2017}
A.~Cuoco, M.~Bilicki, J.-Q.~Xia and E.~Branchini, \emph{{Tomographic imaging of
  the Fermi-LAT gamma-ray sky through cross-correlations: A wider and deeper
  look}}, \href{https://doi.org/10.3847/1538-4365/aa8553}{\emph{Astrophys. J.
  Suppl.} {\bfseries 232} (2017) 10}
  [\href{https://arxiv.org/abs/1709.01940}{{\ttfamily 1709.01940}}].

\bibitem{ammazzalorso2018}
S.~Ammazzalorso, N.~Fornengo, S.~Horiuchi and M.~Regis, \emph{{Characterizing
  the local gamma-ray Universe via angular cross-correlations}},
  \href{https://doi.org/10.1103/PhysRevD.98.103007}{\emph{Phys. Rev. D}
  {\bfseries 98} (2018) 103007}
  [\href{https://arxiv.org/abs/1808.09225}{{\ttfamily 1808.09225}}].

\bibitem{hashimoto2020}
D.~Hashimoto, O.~Macias, A.J.~Nishizawa, K.~Hayashi, M.~Takada, M.~Shirasaki
  et~al., \emph{{Constraining dark matter annihilation with HSC Low Surface
  Brightness Galaxies}},
  \href{https://doi.org/10.1088/1475-7516/2020/01/059}{\emph{JCAP} {\bfseries
  01} (2020) 059} [\href{https://arxiv.org/abs/1906.06701}{{\ttfamily
  1906.06701}}].

\bibitem{branchini2017}
E.~Branchini, S.~Camera, A.~Cuoco, N.~Fornengo, M.~Regis, M.~Viel et~al.,
  \emph{{Cross-correlating the $\gamma$-ray sky with Catalogs of Galaxy
  Clusters}}, \href{https://doi.org/10.3847/1538-4365/228/1/8}{\emph{Astrophys.
  J. Suppl.} {\bfseries 228} (2017) 8}
  [\href{https://arxiv.org/abs/1612.05788}{{\ttfamily 1612.05788}}].

\bibitem{hashimoto2019}
D.~Hashimoto, A.J.~Nishizawa, M.~Shirasaki, O.~Macias, S.~Horiuchi, H.~Tashiro
  et~al., \emph{{Measurement of redshift dependent cross correlation of HSC
  clusters and Fermi $\gamma$ rays}},
  \href{https://arxiv.org/abs/1805.08139}{{\ttfamily 1805.08139}}.

\bibitem{colavincenzo2020}
M.~Colavincenzo, X.~Tan, S.~Ammazzalorso, S.~Camera, M.~Regis, J.-Q.~Xia
  et~al., \emph{{Searching for gamma-ray emission from galaxy clusters at low
  redshift}}, \href{https://doi.org/10.1093/mnras/stz3263}{\emph{Mon. Not. Roy.
  Astron. Soc.} {\bfseries 491} (2020) 3225}
  [\href{https://arxiv.org/abs/1907.05264}{{\ttfamily 1907.05264}}].

\bibitem{tan2020}
X.~Tan, M.~Colavincenzo and S.~Ammazzalorso, \emph{{Bounds on WIMP dark matter
  from galaxy clusters at low redshift}},
  \href{https://doi.org/10.1093/mnras/staa1127}{\emph{Mon. Not. Roy. Astron.
  Soc.} {\bfseries 495} (2020) 114}
  [\href{https://arxiv.org/abs/1907.06905}{{\ttfamily 1907.06905}}].

\bibitem{fornengo2015}
N.~Fornengo, L.~Perotto, M.~Regis and S.~Camera, \emph{{Evidence of
  Cross-correlation between the CMB Lensing and the $\Gamma$-ray sky}},
  \href{https://doi.org/10.1088/2041-8205/802/1/L1}{\emph{Astrophys. J. Lett.}
  {\bfseries 802} (2015) L1} [\href{https://arxiv.org/abs/1410.4997}{{\ttfamily
  1410.4997}}].

\bibitem{feng2017}
C.~Feng, A.~Cooray and B.~Keating, \emph{{Planck Lensing and Cosmic Infrared
  Background Cross-Correlation with Fermi-LAT: Tracing Dark Matter Signals in
  the Gamma-Ray Background}},
  \href{https://doi.org/10.3847/1538-4357/836/1/127}{\emph{Astrophys. J.}
  {\bfseries 836} (2017) 127}
  [\href{https://arxiv.org/abs/1608.04351}{{\ttfamily 1608.04351}}].

\bibitem{shirasaki2014}
M.~Shirasaki, S.~Horiuchi and N.~Yoshida, \emph{{Cross-Correlation of Cosmic
  Shear and Extragalactic Gamma-ray Background: Constraints on the Dark Matter
  Annihilation Cross-Section}},
  \href{https://doi.org/10.1103/PhysRevD.90.063502}{\emph{Phys. Rev. D}
  {\bfseries 90} (2014) 063502}
  [\href{https://arxiv.org/abs/1404.5503}{{\ttfamily 1404.5503}}].

\bibitem{troster2017}
T.~Tr\"oster et~al., \emph{{Cross-correlation of weak lensing and gamma rays:
  implications for the nature of dark matter}},
  \href{https://doi.org/10.1093/mnras/stx365}{\emph{Mon. Not. Roy. Astron.
  Soc.} {\bfseries 467} (2017) 2706}
  [\href{https://arxiv.org/abs/1611.03554}{{\ttfamily 1611.03554}}].

\bibitem{shirasaki2016}
M.~Shirasaki, O.~Macias, S.~Horiuchi, S.~Shirai and N.~Yoshida,
  \emph{{Cosmological constraints on dark matter annihilation and decay:
  Cross-correlation analysis of the extragalactic $\gamma$-ray background and
  cosmic shear}}, \href{https://doi.org/10.1103/PhysRevD.94.063522}{\emph{Phys.
  Rev. D} {\bfseries 94} (2016) 063522}
  [\href{https://arxiv.org/abs/1607.02187}{{\ttfamily 1607.02187}}].

\bibitem{shirasaki2018}
M.~Shirasaki, O.~Macias, S.~Horiuchi, N.~Yoshida, C.-H.~Lee and A.J.~Nishizawa,
  \emph{{Correlation of extragalactic \ensuremath{\gamma} rays with cosmic
  matter density distributions from weak gravitational lensing}},
  \href{https://doi.org/10.1103/PhysRevD.97.123015}{\emph{Phys. Rev. D}
  {\bfseries 97} (2018) 123015}
  [\href{https://arxiv.org/abs/1802.10257}{{\ttfamily 1802.10257}}].

\bibitem{ammazzalorso2020}
{\scshape DES} collaboration, \emph{{Detection of Cross-Correlation between
  Gravitational Lensing and $\gamma$ Rays}},
  \href{https://doi.org/10.1103/PhysRevLett.124.101102}{\emph{Phys. Rev. Lett.}
  {\bfseries 124} (2020) 101102}
  [\href{https://arxiv.org/abs/1907.13484}{{\ttfamily 1907.13484}}].

\bibitem{zandanel2015}
F.~Zandanel, C.~Weniger and S.~Ando, \emph{{The role of the eROSITA all-sky
  survey in searches for sterile neutrino dark matter}},
  \href{https://doi.org/10.1088/1475-7516/2015/09/060}{\emph{JCAP} {\bfseries
  09} (2015) 060} [\href{https://arxiv.org/abs/1505.07829}{{\ttfamily
  1505.07829}}].

\bibitem{Caputo:2019djj}
A.~Caputo, M.~Regis and M.~Taoso, \emph{{Searching for Sterile Neutrino with
  X-ray Intensity Mapping}},
  \href{https://doi.org/10.1088/1475-7516/2020/03/001}{\emph{JCAP} {\bfseries
  03} (2020) 001} [\href{https://arxiv.org/abs/1911.09120}{{\ttfamily
  1911.09120}}].

\bibitem{serra2014}
P.~Serra, G.~Lagache, O.~Dor\'e, A.~Pullen and M.~White,
  \emph{{Cross-correlation of cosmic far-infrared background anisotropies with
  large scale structures}},
  \href{https://doi.org/10.1051/0004-6361/201423958}{\emph{Astron. Astrophys.}
  {\bfseries 570} (2014) A98}
  [\href{https://arxiv.org/abs/1404.1933}{{\ttfamily 1404.1933}}].

\bibitem{Gong:2015hke}
Y.~Gong, A.~Cooray, K.~Mitchell-Wynne, X.~Chen, M.~Zemcov and J.~Smidt,
  \emph{{Axion decay and anisotropy of near-IR extragalactic background
  light}}, \href{https://doi.org/10.3847/0004-637X/825/2/104}{\emph{Astrophys.
  J.} {\bfseries 825} (2016) 104}
  [\href{https://arxiv.org/abs/1511.01577}{{\ttfamily 1511.01577}}].

\bibitem{Caputo:2020msf}
A.~Caputo, A.~Vittino, N.~Fornengo, M.~Regis and M.~Taoso, \emph{{Searching for
  axion-like particle decay in the near-infrared background: an updated
  analysis}}, \href{https://doi.org/10.1088/1475-7516/2021/05/046}{\emph{JCAP}
  {\bfseries 05} (2021) 046}
  [\href{https://arxiv.org/abs/2012.09179}{{\ttfamily 2012.09179}}].

\bibitem{pinetti2019}
E.~Pinetti, S.~Camera, N.~Fornengo and M.~Regis, \emph{{Synergies across the
  spectrum for particle dark matter indirect detection: how HI intensity
  mapping meets gamma rays}},
  \href{https://doi.org/10.1088/1475-7516/2020/07/044}{\emph{JCAP} {\bfseries
  07} (2020) 044} [\href{https://arxiv.org/abs/1911.04989}{{\ttfamily
  1911.04989}}].

\bibitem{Pan:2012}
D.C.~Pan, M.S.~Vogeley, F.~Hoyle, Y.-Y.~Choi and C.~Park, \emph{{Cosmic Voids
  in Sloan Digital Sky Survey Data Release 7}},
  \href{https://doi.org/10.1111/j.1365-2966.2011.20197.x}{\emph{Mon. Not. Roy.
  Astron. Soc.} {\bfseries 421} (2012) 926}
  [\href{https://arxiv.org/abs/1103.4156}{{\ttfamily 1103.4156}}].

\bibitem{Douglass:2022ebt}
K.A.~Douglass, D.~Veyrat and S.~BenZvi, \emph{{Updated void catalogs of the
  SDSS DR7 main sample}},  \href{https://arxiv.org/abs/2202.01226}{{\ttfamily
  2202.01226}}.

\bibitem{Mao:2016onb}
Q.~Mao et~al., \emph{{A Cosmic Void Catalog of SDSS DR12 BOSS Galaxies}},
  \href{https://doi.org/10.3847/1538-4357/835/2/161}{\emph{Astrophys. J.}
  {\bfseries 835} (2017) 161}
  [\href{https://arxiv.org/abs/1602.02771}{{\ttfamily 1602.02771}}].

\bibitem{voivodic2020}
R.~Voivodic, H.~Rubira and M.~Lima, \emph{{The Halo Void (Dust) Model of Large
  Scale Structure}},
  \href{https://doi.org/10.1088/1475-7516/2020/10/033}{\emph{JCAP} {\bfseries
  10} (2020) 033} [\href{https://arxiv.org/abs/2003.06411}{{\ttfamily
  2003.06411}}].

\bibitem{planck2018}
{\scshape Planck} collaboration, \emph{{Planck 2018 results. VI. Cosmological
  parameters}},
  \href{https://doi.org/10.1051/0004-6361/201833910}{\emph{Astron. Astrophys.}
  {\bfseries 641} (2020) A6}
  [\href{https://arxiv.org/abs/1807.06209}{{\ttfamily 1807.06209}}].

\bibitem{limber1953}
D.~Limber, \emph{{The analysis of counts of the extragalactic nebulae in terms
  of a fluctuating density field}},
  \href{https://doi.org/10.1086/145672}{\emph{Astrophys. J.} {\bfseries 134}
  (1953) A6}.

\bibitem{limber1992}
N.~Kaiser, \emph{{Weak gravitational lensing of distant galaxies}},
  \href{https://doi.org/10.1086/171151}{\emph{Astrophys. J.} {\bfseries 388}
  (1992) 272}.

\bibitem{limber1998}
N.~Kaiser, \emph{{Weak lensing and cosmology}},
  \href{https://doi.org/10.1086/305515}{\emph{Astrophys. J.} {\bfseries 498}
  (1998) 26} [\href{https://arxiv.org/abs/astro-ph/9610120}{{\ttfamily
  astro-ph/9610120}}].

\bibitem{cooray2002}
A.~Cooray and R.K.~Sheth, \emph{{Halo Models of Large Scale Structure}},
  \href{https://doi.org/10.1016/S0370-1573(02)00276-4}{\emph{Phys. Rept.}
  {\bfseries 372} (2002) 1}
  [\href{https://arxiv.org/abs/astro-ph/0206508}{{\ttfamily
  astro-ph/0206508}}].

\bibitem{bernardeau2002}
F.~Bernardeau, S.~Colombi, E.~Gaztanaga and R.~Scoccimarro, \emph{{Large scale
  structure of the universe and cosmological perturbation theory}},
  \href{https://doi.org/10.1016/S0370-1573(02)00135-7}{\emph{Phys. Rept.}
  {\bfseries 367} (2002) 1}
  [\href{https://arxiv.org/abs/astro-ph/0112551}{{\ttfamily
  astro-ph/0112551}}].

\bibitem{baumann2012}
D.~Baumann, A.~Nicolis, L.~Senatore and M.~Zaldarriaga, \emph{{Cosmological
  Non-Linearities as an Effective Fluid}},
  \href{https://doi.org/10.1088/1475-7516/2012/07/051}{\emph{JCAP} {\bfseries
  07} (2012) 051} [\href{https://arxiv.org/abs/1004.2488}{{\ttfamily
  1004.2488}}].

\bibitem{carrasco2012}
J.J.M.~Carrasco, M.P.~Hertzberg and L.~Senatore, \emph{{The Effective Field
  Theory of Cosmological Large Scale Structures}},
  \href{https://doi.org/10.1007/JHEP09(2012)082}{\emph{JHEP} {\bfseries 09}
  (2012) 082} [\href{https://arxiv.org/abs/1206.2926}{{\ttfamily 1206.2926}}].

\bibitem{carrasco2014}
J.J.M.~Carrasco, S.~Foreman, D.~Green and L.~Senatore, \emph{{The Effective
  Field Theory of Large Scale Structures at Two Loops}},
  \href{https://doi.org/10.1088/1475-7516/2014/07/057}{\emph{JCAP} {\bfseries
  07} (2014) 057} [\href{https://arxiv.org/abs/1310.0464}{{\ttfamily
  1310.0464}}].

\bibitem{konstandin2019}
T.~Konstandin, R.A.~Porto and H.~Rubira, \emph{{The effective field theory of
  large scale structure at three loops}},
  \href{https://doi.org/10.1088/1475-7516/2019/11/027}{\emph{JCAP} {\bfseries
  11} (2019) 027} [\href{https://arxiv.org/abs/1906.00997}{{\ttfamily
  1906.00997}}].

\bibitem{teyssier2002}
R.~Teyssier, \emph{{Cosmological hydrodynamics with adaptive mesh refinement: a
  new high resolution code called ramses}},
  \href{https://doi.org/10.1051/0004-6361:20011817}{\emph{Astron. Astrophys.}
  {\bfseries 385} (2002) 337}
  [\href{https://arxiv.org/abs/astro-ph/0111367}{{\ttfamily
  astro-ph/0111367}}].

\bibitem{springel2005}
V.~Springel, \emph{{The Cosmological simulation code GADGET-2}},
  \href{https://doi.org/10.1111/j.1365-2966.2005.09655.x}{\emph{Mon. Not. Roy.
  Astron. Soc.} {\bfseries 364} (2005) 1105}
  [\href{https://arxiv.org/abs/astro-ph/0505010}{{\ttfamily
  astro-ph/0505010}}].

\bibitem{moore1999}
B.~Moore, T.R.~Quinn, F.~Governato, J.~Stadel and G.~Lake, \emph{{Cold collapse
  and the core catastrophe}},
  \href{https://doi.org/10.1046/j.1365-8711.1999.03039.x}{\emph{Mon. Not. Roy.
  Astron. Soc.} {\bfseries 310} (1999) 1147}
  [\href{https://arxiv.org/abs/astro-ph/9903164}{{\ttfamily
  astro-ph/9903164}}].

\bibitem{navarro1996}
J.F.~Navarro, C.S.~Frenk and S.D.M.~White, \emph{{The Structure of cold dark
  matter halos}}, \href{https://doi.org/10.1086/177173}{\emph{Astrophys. J.}
  {\bfseries 462} (1996) 563}
  [\href{https://arxiv.org/abs/astro-ph/9508025}{{\ttfamily
  astro-ph/9508025}}].

\bibitem{evrard2002}
{\scshape VIRGO} collaboration, \emph{{Galaxy clusters in Hubble volume
  simulations: Cosmological constraints from sky survey populations}},
  \href{https://doi.org/10.1086/340551}{\emph{Astrophys. J.} {\bfseries 573}
  (2002) 7} [\href{https://arxiv.org/abs/astro-ph/0110246}{{\ttfamily
  astro-ph/0110246}}].

\bibitem{angulo2014}
R.E.~Angulo, C.M.~Baugh, C.S.~Frenk and C.G.~Lacey, \emph{{Extending the halo
  mass resolution of $N$-body simulations}},
  \href{https://doi.org/10.1093/mnras/stu1084}{\emph{Mon. Not. Roy. Astron.
  Soc.} {\bfseries 442} (2014) 3256}
  [\href{https://arxiv.org/abs/1310.3880}{{\ttfamily 1310.3880}}].

\bibitem{LSST}
{\scshape LSST} collaboration, \emph{{LSST: from Science Drivers to Reference
  Design and Anticipated Data Products}},
  \href{https://doi.org/10.3847/1538-4357/ab042c}{\emph{Astrophys. J.}
  {\bfseries 873} (2019) 111}
  [\href{https://arxiv.org/abs/0805.2366}{{\ttfamily 0805.2366}}].

\bibitem{DES}
{\scshape DES} collaboration, \emph{{The Dark Energy Survey}},
  \href{https://arxiv.org/abs/astro-ph/0510346}{{\ttfamily astro-ph/0510346}}.

\bibitem{Euclid}
{\scshape Euclid Theory Working Group} collaboration, \emph{{Cosmology and
  fundamental physics with the Euclid satellite}},
  \href{https://doi.org/10.12942/lrr-2013-6}{\emph{Living Rev. Rel.} {\bfseries
  16} (2013) 6} [\href{https://arxiv.org/abs/1206.1225}{{\ttfamily
  1206.1225}}].

\bibitem{sunyaev1980}
R.A.~Sunyaev and Y.B.~Zeldovich, \emph{{The Velocity of clusters of galaxies
  relative to the microwave background. The Possibility of its measurement}},
  {\emph{Mon. Not. Roy. Astron. Soc.} {\bfseries 190} (1980) 413}.

\bibitem{orlowski2021}
{\scshape ACT} collaboration, \emph{{Atacama Cosmology Telescope measurements
  of a large sample of candidates from the Massive and Distant Clusters of WISE
  Survey - Sunyaev-Zeldovich effect confirmation of MaDCoWS candidates using
  ACT}}, \href{https://doi.org/10.1051/0004-6361/202141200}{\emph{Astron.
  Astrophys.} {\bfseries 653} (2021) A135}
  [\href{https://arxiv.org/abs/2105.00068}{{\ttfamily 2105.00068}}].

\bibitem{sachs1967}
R.K.~Sachs and A.M.~Wolfe, \emph{{Perturbations of a cosmological model and
  angular variations of the microwave background}},
  \href{https://doi.org/10.1007/s10714-007-0448-9}{\emph{Astrophys. J.}
  {\bfseries 147} (1967) 73}.

\bibitem{lemarchand2019}
N.~Lemarchand, J.~Grain, G.~Hurier, F.~Lacasa and A.~Fert\'e, \emph{{Secondary
  CMB anisotropies from magnetized haloes - I. Power spectra of the Faraday
  rotation angle and conversion rate}},
  \href{https://doi.org/10.1051/0004-6361/201834485}{\emph{Astron. Astrophys.}
  {\bfseries 630} (2019) A149}
  [\href{https://arxiv.org/abs/1810.09221}{{\ttfamily 1810.09221}}].

\bibitem{mead2015}
A.~Mead, J.~Peacock, C.~Heymans, S.~Joudaki and A.~Heavens, \emph{{An accurate
  halo model for fitting non-linear cosmological power spectra and baryonic
  feedback models}}, \href{https://doi.org/10.1093/mnras/stv2036}{\emph{Mon.
  Not. Roy. Astron. Soc.} {\bfseries 454} (2015) 1958}
  [\href{https://arxiv.org/abs/1505.07833}{{\ttfamily 1505.07833}}].

\bibitem{schmidt2016}
F.~Schmidt, \emph{{Towards a self-consistent halo model for the nonlinear
  large-scale structure}},
  \href{https://doi.org/10.1103/PhysRevD.93.063512}{\emph{Phys. Rev. D}
  {\bfseries 93} (2016) 063512}
  [\href{https://arxiv.org/abs/1511.02231}{{\ttfamily 1511.02231}}].

\bibitem{chen2019}
A.Y.~Chen and N.~Afshordi, \emph{{Amending the halo model to satisfy
  cosmological conservation laws}},
  \href{https://doi.org/10.1103/PhysRevD.101.103522}{\emph{Phys. Rev. D}
  {\bfseries 101} (2020) 103522}
  [\href{https://arxiv.org/abs/1912.04872}{{\ttfamily 1912.04872}}].

\bibitem{valageas2011}
P.~Valageas and T.~Nishimichi, \emph{{Combining perturbation theories with halo
  models for the matter bispectrum}},
  \href{https://doi.org/10.1051/0004-6361/201116638}{\emph{Astron. Astrophys.}
  {\bfseries 532} (2011) A4} [\href{https://arxiv.org/abs/1102.0641}{{\ttfamily
  1102.0641}}].

\bibitem{biswas2010}
R.~Biswas, E.~Alizadeh and B.D.~Wandelt, \emph{{Voids as a Precision Probe of
  Dark Energy}}, \href{https://doi.org/10.1103/PhysRevD.82.023002}{\emph{Phys.
  Rev. D} {\bfseries 82} (2010) 023002}
  [\href{https://arxiv.org/abs/1002.0014}{{\ttfamily 1002.0014}}].

\bibitem{pollina2016}
G.~Pollina, M.~Baldi, F.~Marulli and L.~Moscardini, \emph{{Cosmic voids in
  coupled dark energy cosmologies: the impact of halo bias}},
  \href{https://doi.org/10.1093/mnras/stv2503}{\emph{Mon. Not. Roy. Astron.
  Soc.} {\bfseries 455} (2016) 3075}
  [\href{https://arxiv.org/abs/1506.08831}{{\ttfamily 1506.08831}}].

\bibitem{sahlen2016}
M.~Sahl\'en, I.n.~Zubeld\'\i{}a and J.~Silk, \emph{{Cluster\textendash{}void
  Degeneracy Breaking: Dark Energy, Planck, and the Largest Cluster and Void}},
  \href{https://doi.org/10.3847/2041-8205/820/1/L7}{\emph{Astrophys. J. Lett.}
  {\bfseries 820} (2016) L7}
  [\href{https://arxiv.org/abs/1511.04075}{{\ttfamily 1511.04075}}].

\bibitem{pisani2015}
A.~Pisani, P.M.~Sutter, N.~Hamaus, E.~Alizadeh, R.~Biswas, B.D.~Wandelt et~al.,
  \emph{{Counting voids to probe dark energy}},
  \href{https://doi.org/10.1103/PhysRevD.92.083531}{\emph{Phys. Rev. D}
  {\bfseries 92} (2015) 083531}
  [\href{https://arxiv.org/abs/1503.07690}{{\ttfamily 1503.07690}}].

\bibitem{barreira2015}
A.~Barreira, M.~Cautun, B.~Li, C.~Baugh and S.~Pascoli, \emph{{Weak lensing by
  voids in modified lensing potentials}},
  \href{https://doi.org/10.1088/1475-7516/2015/08/028}{\emph{JCAP} {\bfseries
  08} (2015) 028} [\href{https://arxiv.org/abs/1505.05809}{{\ttfamily
  1505.05809}}].

\bibitem{cai2015}
Y.-C.~Cai, N.~Padilla and B.~Li, \emph{{Testing Gravity using Cosmic Voids}},
  \href{https://doi.org/10.1093/mnras/stv777}{\emph{Mon. Not. Roy. Astron.
  Soc.} {\bfseries 451} (2015) 1036}
  [\href{https://arxiv.org/abs/1410.1510}{{\ttfamily 1410.1510}}].

\bibitem{perico2019}
E.L.D.~Perico, R.~Voivodic, M.~Lima and D.F.~Mota, \emph{{Cosmic voids in
  modified gravity scenarios}},
  \href{https://doi.org/10.1051/0004-6361/201935949}{\emph{Astron. Astrophys.}
  {\bfseries 632} (2019) A52}
  [\href{https://arxiv.org/abs/1905.12450}{{\ttfamily 1905.12450}}].

\bibitem{voivodic2017}
R.~Voivodic, M.~Lima, C.~Llinares and D.F.~Mota, \emph{{Modelling Void
  Abundance in Modified Gravity}},
  \href{https://doi.org/10.1103/PhysRevD.95.024018}{\emph{Phys. Rev. D}
  {\bfseries 95} (2017) 024018}
  [\href{https://arxiv.org/abs/1609.02544}{{\ttfamily 1609.02544}}].

\bibitem{damico2011}
G.~D'Amico, M.~Musso, J.~Norena and A.~Paranjape, \emph{{Excursion Sets and
  Non-Gaussian Void Statistics}},
  \href{https://doi.org/10.1103/PhysRevD.83.023521}{\emph{Phys. Rev. D}
  {\bfseries 83} (2011) 023521}
  [\href{https://arxiv.org/abs/1011.1229}{{\ttfamily 1011.1229}}].

\bibitem{chan2019}
K.C.~Chan, N.~Hamaus and M.~Biagetti, \emph{{Constraint of Void Bias on
  Primordial non-Gaussianity}},
  \href{https://doi.org/10.1103/PhysRevD.99.121304}{\emph{Phys. Rev. D}
  {\bfseries 99} (2019) 121304}
  [\href{https://arxiv.org/abs/1812.04024}{{\ttfamily 1812.04024}}].

\bibitem{song2009}
H.~Song and J.~Lee, \emph{{The Mass Distribution of SDSS Galaxy Groups in Void
  Regions and Its Implication on the Primordial non-Gaussianity}},
  \href{https://doi.org/10.1088/0004-637X/701/1/L25}{\emph{Astrophys. J. Lett.}
  {\bfseries 701} (2009) L25}
  [\href{https://arxiv.org/abs/0811.1339}{{\ttfamily 0811.1339}}].

\bibitem{cai2016}
Y.-C.~Cai, A.~Taylor, J.A.~Peacock and N.~Padilla, \emph{{Redshift-space
  distortions around voids}},
  \href{https://doi.org/10.1093/mnras/stw1809}{\emph{Mon. Not. Roy. Astron.
  Soc.} {\bfseries 462} (2016) 2465}
  [\href{https://arxiv.org/abs/1603.05184}{{\ttfamily 1603.05184}}].

\bibitem{chantavat2016}
T.~Chantavat, U.~Sawangwit, P.M.~Sutter and B.D.~Wandelt, \emph{{Cosmological
  parameter constraints from CMB lensing with cosmic voids}},
  \href{https://doi.org/10.1103/PhysRevD.93.043523}{\emph{Phys. Rev. D}
  {\bfseries 93} (2016) 043523}
  [\href{https://arxiv.org/abs/1409.3364}{{\ttfamily 1409.3364}}].

\bibitem{hamaus2014}
N.~Hamaus, B.D.~Wandelt, P.M.~Sutter, G.~Lavaux and M.S.~Warren,
  \emph{{Cosmology with Void-Galaxy Correlations}},
  \href{https://doi.org/10.1103/PhysRevLett.112.041304}{\emph{Phys. Rev. Lett.}
  {\bfseries 112} (2014) 041304}
  [\href{https://arxiv.org/abs/1307.2571}{{\ttfamily 1307.2571}}].

\bibitem{hamaus2015}
N.~Hamaus, P.M.~Sutter, G.~Lavaux and B.D.~Wandelt, \emph{{Probing cosmology
  and gravity with redshift-space distortions around voids}},
  \href{https://doi.org/10.1088/1475-7516/2015/11/036}{\emph{JCAP} {\bfseries
  11} (2015) 036} [\href{https://arxiv.org/abs/1507.04363}{{\ttfamily
  1507.04363}}].

\bibitem{hamaus2016}
N.~Hamaus, A.~Pisani, P.M.~Sutter, G.~Lavaux, S.~Escoffier, B.D.~Wandelt
  et~al., \emph{{Constraints on Cosmology and Gravity from the Dynamics of
  Voids}}, \href{https://doi.org/10.1103/PhysRevLett.117.091302}{\emph{Phys.
  Rev. Lett.} {\bfseries 117} (2016) 091302}
  [\href{https://arxiv.org/abs/1602.01784}{{\ttfamily 1602.01784}}].

\bibitem{chuang2017}
C.-H.~Chuang, F.-S.~Kitaura, Y.~Liang, A.~Font-Ribera, C.~Zhao, P.~McDonald
  et~al., \emph{{Linear redshift space distortions for cosmic voids based on
  galaxies in redshift space}},
  \href{https://doi.org/10.1103/PhysRevD.95.063528}{\emph{Phys. Rev. D}
  {\bfseries 95} (2017) 063528}
  [\href{https://arxiv.org/abs/1605.05352}{{\ttfamily 1605.05352}}].

\bibitem{lavaux2012}
G.~Lavaux and B.D.~Wandelt, \emph{{Precision cosmography with stacked voids}},
  \href{https://doi.org/10.1088/0004-637X/754/2/109}{\emph{Astrophys. J.}
  {\bfseries 754} (2012) 109}
  [\href{https://arxiv.org/abs/1110.0345}{{\ttfamily 1110.0345}}].

\bibitem{bond1991}
J.R.~Bond, S.~Cole, G.~Efstathiou and N.~Kaiser, \emph{{Excursion set mass
  functions for hierarchical Gaussian fluctuations}},
  \href{https://doi.org/10.1086/170520}{\emph{Astrophys. J.} {\bfseries 379}
  (1991) 440}.

\bibitem{sheth2004}
R.K.~Sheth and R.~van~de Weygaert, \emph{{A Hierarchy of voids: Much ado about
  nothing}}, \href{https://doi.org/10.1111/j.1365-2966.2004.07661.x}{\emph{Mon.
  Not. Roy. Astron. Soc.} {\bfseries 350} (2004) 517}
  [\href{https://arxiv.org/abs/astro-ph/0311260}{{\ttfamily
  astro-ph/0311260}}].

\bibitem{simone2011}
A.~De~Simone, M.~Maggiore and A.~Riotto, \emph{{Excursion Set Theory for
  generic moving barriers and non-Gaussian initial conditions}},
  \href{https://doi.org/10.1111/j.1365-2966.2010.18078.x}{\emph{Mon. Not. Roy.
  Astron. Soc.} {\bfseries 412} (2011) 2587}
  [\href{https://arxiv.org/abs/1007.1903}{{\ttfamily 1007.1903}}].

\bibitem{jennings2013}
E.~Jennings, Y.~Li and W.~Hu, \emph{{The abundance of voids and the excursion
  set formalism}}, \href{https://doi.org/10.1093/mnras/stt1169}{\emph{Mon. Not.
  Roy. Astron. Soc.} {\bfseries 434} (2013) 2167}
  [\href{https://arxiv.org/abs/1304.6087}{{\ttfamily 1304.6087}}].

\bibitem{hamaus2014b}
N.~Hamaus, P.M.~Sutter and B.D.~Wandelt, \emph{{Universal Density Profile for
  Cosmic Voids}},
  \href{https://doi.org/10.1103/PhysRevLett.112.251302}{\emph{Phys. Rev. Lett.}
  {\bfseries 112} (2014) 251302}
  [\href{https://arxiv.org/abs/1403.5499}{{\ttfamily 1403.5499}}].

\bibitem{duffy2008}
A.R.~Duffy, J.~Schaye, S.T.~Kay and C.~Dalla~Vecchia, \emph{{Dark matter halo
  concentrations in the Wilkinson Microwave Anisotropy Probe year 5
  cosmology}},
  \href{https://doi.org/10.1111/j.1745-3933.2008.00537.x}{\emph{Mon. Not. Roy.
  Astron. Soc.} {\bfseries 390} (2008) L64}
  [\href{https://arxiv.org/abs/0804.2486}{{\ttfamily 0804.2486}}].

\bibitem{sheth1999}
R.K.~Sheth and G.~Tormen, \emph{{Large scale bias and the peak background
  split}}, \href{https://doi.org/10.1046/j.1365-8711.1999.02692.x}{\emph{Mon.
  Not. Roy. Astron. Soc.} {\bfseries 308} (1999) 119}
  [\href{https://arxiv.org/abs/astro-ph/9901122}{{\ttfamily
  astro-ph/9901122}}].

\bibitem{press1974}
W.H.~Press and P.~Schechter, \emph{{Formation of galaxies and clusters of
  galaxies by selfsimilar gravitational condensation}},
  \href{https://doi.org/10.1086/152650}{\emph{Astrophys. J.} {\bfseries 187}
  (1974) 425}.

\bibitem{sheth2001}
R.K.~Sheth, H.J.~Mo and G.~Tormen, \emph{{Ellipsoidal collapse and an improved
  model for the number and spatial distribution of dark matter haloes}},
  \href{https://doi.org/10.1046/j.1365-8711.2001.04006.x}{\emph{Mon. Not. Roy.
  Astron. Soc.} {\bfseries 323} (2001) 1}
  [\href{https://arxiv.org/abs/astro-ph/9907024}{{\ttfamily
  astro-ph/9907024}}].

\bibitem{berlind2002}
A.A.~Berlind and D.H.~Weinberg, \emph{{The Halo occupation distribution:
  Towards an empirical determination of the relation between galaxies and
  mass}}, \href{https://doi.org/10.1086/341469}{\emph{Astrophys. J.} {\bfseries
  575} (2002) 587} [\href{https://arxiv.org/abs/astro-ph/0109001}{{\ttfamily
  astro-ph/0109001}}].

\bibitem{berlind2003}
A.A.~Berlind, D.H.~Weinberg, A.J.~Benson, C.M.~Baugh, S.~Cole, R.~Dave et~al.,
  \emph{{The Halo occupation distribution and the physics of galaxy
  formation}}, \href{https://doi.org/10.1086/376517}{\emph{Astrophys. J.}
  {\bfseries 593} (2003) 1}
  [\href{https://arxiv.org/abs/astro-ph/0212357}{{\ttfamily
  astro-ph/0212357}}].

\bibitem{zheng2007}
Z.~Zheng, A.L.~Coil and I.~Zehavi, \emph{{Galaxy Evolution from Halo Occupation
  Distribution Modeling of DEEP2 and SDSS Galaxy Clustering}},
  \href{https://doi.org/10.1086/521074}{\emph{Astrophys. J.} {\bfseries 667}
  (2007) 760} [\href{https://arxiv.org/abs/astro-ph/0703457}{{\ttfamily
  astro-ph/0703457}}].

\bibitem{zheng2005}
Z.~Zheng, A.A.~Berlind, D.H.~Weinberg, A.J.~Benson, C.M.~Baugh, S.~Cole et~al.,
  \emph{{Theoretical models of the halo occupation distribution: Separating
  central and satellite galaxies}},
  \href{https://doi.org/10.1086/466510}{\emph{Astrophys. J.} {\bfseries 633}
  (2005) 791} [\href{https://arxiv.org/abs/astro-ph/0408564}{{\ttfamily
  astro-ph/0408564}}].

\bibitem{2MASS}
{\scshape 2MASS} collaboration, \emph{{The Two Micron All Sky Survey (2MASS)}},
  \href{https://doi.org/10.1086/498708}{\emph{Astron. J.} {\bfseries 131}
  (2006) 1163}.

\bibitem{zehavi2005}
{\scshape SDSS} collaboration, \emph{{The Luminosity and color dependence of
  the galaxy correlation function}},
  \href{https://doi.org/10.1086/431891}{\emph{Astrophys. J.} {\bfseries 630}
  (2005) 1} [\href{https://arxiv.org/abs/astro-ph/0408569}{{\ttfamily
  astro-ph/0408569}}].

\bibitem{patiri2006}
S.G.~Patiri, F.~Prada, J.~Holtzman, A.~Klypin and J.~Betancort-Rijo, \emph{{The
  Properties of Galaxies in Voids}},
  \href{https://doi.org/10.1111/j.1365-2966.2006.10975.x}{\emph{Mon. Not. Roy.
  Astron. Soc.} {\bfseries 372} (2006) 1710}
  [\href{https://arxiv.org/abs/astro-ph/0605703}{{\ttfamily
  astro-ph/0605703}}].

\bibitem{constantin2008}
A.~Constantin, F.~Hoyle and M.S.~Vogeley, \emph{{Active Galactic Nuclei in Void
  Regions}}, \href{https://doi.org/10.1086/524310}{\emph{Astrophys. J.}
  {\bfseries 673} (2008) 715}
  [\href{https://arxiv.org/abs/0710.1631}{{\ttfamily 0710.1631}}].

\bibitem{furniss2015}
A.~Furniss, P.M.~Sutter, J.R.~Primack and A.~Dom\'\i{}nguez, \emph{{A
  Correlation Between Hard Gamma-ray Sources and Cosmic Voids Along the Line of
  Sight}}, \href{https://doi.org/10.1093/mnras/stu2196}{\emph{Mon. Not. Roy.
  Astron. Soc.} {\bfseries 446} (2015) 2267}
  [\href{https://arxiv.org/abs/1407.6370}{{\ttfamily 1407.6370}}].

\bibitem{ajello2014}
M.~Ajello et~al., \emph{{The Cosmic Evolution of Fermi BL Lacertae Objects}},
  \href{https://doi.org/10.1088/0004-637X/780/1/73}{\emph{Astrophys. J.}
  {\bfseries 780} (2014) 73} [\href{https://arxiv.org/abs/1310.0006}{{\ttfamily
  1310.0006}}].

\bibitem{ajello2012}
M.~Ajello et~al., \emph{{The Luminosity Function of Fermi-detected
  Flat-Spectrum Radio Quasars}},
  \href{https://doi.org/10.1088/0004-637X/751/2/108}{\emph{Astrophys. J.}
  {\bfseries 751} (2012) 108}
  [\href{https://arxiv.org/abs/1110.3787}{{\ttfamily 1110.3787}}].

\bibitem{willot2001}
C.J.~Willott, S.~Rawlings, K.M.~Blundell, M.~Lacy and S.A.~Eales, \emph{{The
  radio luminosity function from the low-frequency 3crr, 6ce \& 7crs complete
  samples}}, \href{https://doi.org/10.1046/j.1365-8711.2001.04101.x}{\emph{Mon.
  Not. Roy. Astron. Soc.} {\bfseries 322} (2001) 536}
  [\href{https://arxiv.org/abs/astro-ph/0010419}{{\ttfamily
  astro-ph/0010419}}].

\bibitem{mauro2014}
M.~Di~Mauro, F.~Calore, F.~Donato, M.~Ajello and L.~Latronico, \emph{{Diffuse
  $\gamma$-ray emission from misaligned active galactic nuclei}},
  \href{https://doi.org/10.1088/0004-637X/780/2/161}{\emph{Astrophys. J.}
  {\bfseries 780} (2014) 161}
  [\href{https://arxiv.org/abs/1304.0908}{{\ttfamily 1304.0908}}].

\bibitem{gruppioni2013}
C.~Gruppioni et~al., \emph{{The Herschel PEP/HerMES Luminosity Function. I:
  Probing the Evolution of PACS selected Galaxies to z\textasciitilde{}4}},
  \href{https://doi.org/10.1093/mnras/stt308}{\emph{Mon. Not. Roy. Astron.
  Soc.} {\bfseries 432} (2013) 23}
  [\href{https://arxiv.org/abs/1302.5209}{{\ttfamily 1302.5209}}].

\bibitem{bartelmann2010}
M.~Bartelmann, \emph{{Gravitational Lensing}},
  \href{https://doi.org/10.1088/0264-9381/27/23/233001}{\emph{Class. Quant.
  Grav.} {\bfseries 27} (2010) 233001}
  [\href{https://arxiv.org/abs/1010.3829}{{\ttfamily 1010.3829}}].

\bibitem{euclid2020}
{\scshape Euclid} collaboration, \emph{{Euclid preparation: VII. Forecast
  validation for Euclid cosmological probes}},
  \href{https://doi.org/10.1051/0004-6361/202038071}{\emph{Astron. Astrophys.}
  {\bfseries 642} (2020) A191}
  [\href{https://arxiv.org/abs/1910.09273}{{\ttfamily 1910.09273}}].

\bibitem{fornengo2014b}
N.~Fornengo, L.~Maccione and A.~Vittino, \emph{{Constraints on particle dark
  matter from cosmic-ray antiprotons}},
  \href{https://doi.org/10.1088/1475-7516/2014/04/003}{\emph{JCAP} {\bfseries
  04} (2014) 003} [\href{https://arxiv.org/abs/1312.3579}{{\ttfamily
  1312.3579}}].

\bibitem{Blanco:2018esa}
C.~Blanco and D.~Hooper, \emph{{Constraints on Decaying Dark Matter from the
  Isotropic Gamma-Ray Background}},
  \href{https://doi.org/10.1088/1475-7516/2019/03/019}{\emph{JCAP} {\bfseries
  03} (2019) 019} [\href{https://arxiv.org/abs/1811.05988}{{\ttfamily
  1811.05988}}].

\bibitem{cembranos2011}
J.A.R.~Cembranos, A.~de~la Cruz-Dombriz, A.~Dobado, R.A.~Lineros and
  A.L.~Maroto, \emph{{Photon spectra from WIMP annihilation}},
  \href{https://doi.org/10.1103/PhysRevD.83.083507}{\emph{Phys. Rev. D}
  {\bfseries 83} (2011) 083507}
  [\href{https://arxiv.org/abs/1009.4936}{{\ttfamily 1009.4936}}].

\bibitem{razzaque2009}
S.~Razzaque, C.D.~Dermer and J.D.~Finke, \emph{{The stellar contribution to the
  extra-galactic background light and absorption of TeV gamma-rays}},
  \href{https://doi.org/10.1088/0004-637X/697/1/483}{\emph{Astrophys. J.}
  {\bfseries 697} (2009) 483}
  [\href{https://arxiv.org/abs/0807.4294}{{\ttfamily 0807.4294}}].

\bibitem{ackermann2018}
{\scshape Fermi-LAT} collaboration, \emph{{Unresolved Gamma-Ray Sky through its
  Angular Power Spectrum}},
  \href{https://doi.org/10.1103/PhysRevLett.121.241101}{\emph{Phys. Rev. Lett.}
  {\bfseries 121} (2018) 241101}
  [\href{https://arxiv.org/abs/1812.02079}{{\ttfamily 1812.02079}}].

\bibitem{ackermann2015}
{\scshape Fermi-LAT} collaboration, \emph{{The spectrum of isotropic diffuse
  gamma-ray emission between 100 MeV and 820 GeV}},
  \href{https://doi.org/10.1088/0004-637X/799/1/86}{\emph{Astrophys. J.}
  {\bfseries 799} (2015) 86} [\href{https://arxiv.org/abs/1410.3696}{{\ttfamily
  1410.3696}}].

\bibitem{ackermann2012}
{\scshape Fermi-LAT} collaboration, \emph{{Anisotropies in the diffuse
  gamma-ray background measured by the Fermi LAT}},
  \href{https://doi.org/10.1103/PhysRevD.85.083007}{\emph{Phys. Rev. D}
  {\bfseries 85} (2012) 083007}
  [\href{https://arxiv.org/abs/1202.2856}{{\ttfamily 1202.2856}}].

\bibitem{Topchiev:2017xfp}
N.~Topchiev et~al., \emph{{High-energy gamma-ray studying with GAMMA-400}},
  \href{https://doi.org/10.22323/1.301.0802}{\emph{PoS} {\bfseries ICRC2017}
  (2018) 802} [\href{https://arxiv.org/abs/1707.04882}{{\ttfamily
  1707.04882}}].

\bibitem{DESb}
{\scshape DES} collaboration, \emph{{Dark Energy Survey Year 3 results:
  Cosmological constraints from galaxy clustering and weak lensing}},
  \href{https://doi.org/10.1103/PhysRevD.105.023520}{\emph{Phys. Rev. D}
  {\bfseries 105} (2022) 023520}
  [\href{https://arxiv.org/abs/2105.13549}{{\ttfamily 2105.13549}}].

\bibitem{DESc}
{\scshape DES} collaboration, \emph{{Dark Energy Survey Year 3 results:
  Cosmological constraints from galaxy clustering and weak lensing}},
  \href{https://doi.org/10.1103/PhysRevD.105.023520}{\emph{Phys. Rev. D}
  {\bfseries 105} (2022) 023520}
  [\href{https://arxiv.org/abs/2105.13549}{{\ttfamily 2105.13549}}].

\bibitem{Donato:2008yx}
F.~Donato, N.~Fornengo and D.~Maurin, \emph{{Antideuteron fluxes from dark
  matter annihilation in diffusion models}},
  \href{https://doi.org/10.1103/PhysRevD.78.043506}{\emph{Phys. Rev. D}
  {\bfseries 78} (2008) 043506}
  [\href{https://arxiv.org/abs/0803.2640}{{\ttfamily 0803.2640}}].

\bibitem{Aramaki:2015pii}
T.~Aramaki et~al., \emph{{Review of the theoretical and experimental status of
  dark matter identification with cosmic-ray antideuterons}},
  \href{https://doi.org/10.1016/j.physrep.2016.01.002}{\emph{Phys. Rept.}
  {\bfseries 618} (2016) 1} [\href{https://arxiv.org/abs/1505.07785}{{\ttfamily
  1505.07785}}].

\bibitem{Jeffrey:1961}
H.~Jeffreys, \emph{Theory of Probability}, Oxford Classic Texts in the Physical
  Sciences, Oxford University Press,, Oxford, U.K. (1939).

\bibitem{Vogeley:1996xu}
M.S.~Vogeley and A.S.~Szalay, \emph{{Eigenmode analysis of galaxy redshift
  surveys I. theory and methods}},
  \href{https://doi.org/10.1086/177399}{\emph{Astrophys. J.} {\bfseries 465}
  (1996) 34} [\href{https://arxiv.org/abs/astro-ph/9601185}{{\ttfamily
  astro-ph/9601185}}].

\bibitem{Tegmark:1996bz}
M.~Tegmark, A.~Taylor and A.~Heavens, \emph{{Karhunen-Loeve eigenvalue problems
  in cosmology: How should we tackle large data sets?}},
  \href{https://doi.org/10.1086/303939}{\emph{Astrophys. J.} {\bfseries 480}
  (1997) 22} [\href{https://arxiv.org/abs/astro-ph/9603021}{{\ttfamily
  astro-ph/9603021}}].

\bibitem{Lara:2004ee}
L.~Lara, G.~Giovannini, W.D.~Cotton, L.~Feretti, J.M.~Marcaide, I.~Marquez
  et~al., \emph{{A New sample of large angular size radio galaxies. 3.
  Statistics and evolution of the grown population}},
  \href{https://doi.org/10.1051/0004-6361:20035676}{\emph{Astron. Astrophys.}
  {\bfseries 421} (2004) 899}
  [\href{https://arxiv.org/abs/astro-ph/0404373}{{\ttfamily
  astro-ph/0404373}}].

\bibitem{Inoue:2011bm}
Y.~Inoue, \emph{{Contribution of the Gamma-ray Loud Radio Galaxies Core
  Emissions to the Cosmic MeV and GeV Gamma-Ray Background Radiation}},
  \href{https://doi.org/10.1088/0004-637X/733/1/66}{\emph{Astrophys. J.}
  {\bfseries 733} (2011) 66} [\href{https://arxiv.org/abs/1103.3946}{{\ttfamily
  1103.3946}}].

\bibitem{Fermi-LAT:2012nqz}
{\scshape Fermi-LAT} collaboration, \emph{{GeV Observations of Star-forming
  Galaxies with \textbackslash{}textit{Fermi} LAT}},
  \href{https://doi.org/10.1088/0004-637X/755/2/164}{\emph{Astrophys. J.}
  {\bfseries 755} (2012) 164}
  [\href{https://arxiv.org/abs/1206.1346}{{\ttfamily 1206.1346}}].

\end{thebibliography}\endgroup

\end{document}